\documentclass[lineno]{jfm}

\usepackage{graphicx}
\usepackage{newtxtext}
\usepackage{newtxmath}
\usepackage{natbib}
\usepackage{hyperref}
\usepackage{soul}
\usepackage{xcolor}
\hypersetup{
    colorlinks = true,
    urlcolor   = blue,
    citecolor  = black,
}

\newcommand{\RomanNumeralCaps}[1]

\usepackage{amsmath}

\usepackage[caption=false]{subfig}
\usepackage{bbm}
\usepackage{tabularx}
\usepackage{comment}

\newcommand{\mycomment}[1]{}
\newcommand{\reviewerone}[1]{\color{black} #1 \color{black}}
\newcommand{\reviewertwo}[1]{\color{black} #1 \color{black}}
\newcommand{\reviewerthree}[1]{\color{black} #1 \color{black}}
\newcommand{\reviewers}[1]{\color{black} #1 \color{black}}

\shorttitle{Growth and breakdown of resolvent modes in channel flow}
\shortauthor{E.~Ballouz, S.~T.~M.~Dawson and H.~J.~Bae}

\title{Transient growth and nonlinear breakdown of wavelet-based resolvent modes in turbulent channel flow}

\author{Eric~Ballouz\aff{1}
  \corresp{\email{eballouz@caltech.edu}},
  Scott~T.~M.~Dawson\aff{2},
 \and H.~Jane~Bae\aff{3}}

\affiliation{\aff{1} Mechanical and Civil Engineering, California Institute of Technology,
Pasadena, CA 91125, USA
\aff{2} Mechanical, Materials, and Aerospace Engineering, Illinois Institute of Technology, Chicago, IL 60616, USA
\aff{3} Graduate Aerospace Laboratories, California Institute of Technology,
Pasadena, CA 91125, USA}

\begin{document}

\maketitle
\nolinenumbers
\begin{abstract}
In this work, we study the effectiveness of the time-localised principal resolvent forcing mode at actuating the near wall cycle of turbulence.  
This mode is restricted to a wavelet pulse and computed from a singular value decomposition of the windowed wavelet-based resolvent operator \citep{ballouz2024wavelet} such that it produces the largest amplification via the linearised Navier-Stokes equations. 
We then inject this time-localised mode into the turbulent minimal flow unit at different intensities, and measure the deviation of the system's response from the optimal resolvent response mode. 
Using the most energetic spatial wave numbers for the minimal flow unit -- \emph{i.e.} constant in the streamwise direction and once-periodic in the spanwise direction -- the forcing mode takes the shape of streamwise rolls and produces a response mode in the form of streamwise streaks that transiently grow and decay.
Though other works such as \citet{bae2021nonlinear} demonstrate the importance of principal resolvent forcing modes to buffer layer turbulence, none instantaneously track their time-dependent interaction with the turbulence, which is made possible by their formulation in a wavelet basis.
For initial times and close to the wall, the turbulent minimal flow unit matches the principal response mode well, but due to nonlinear effects, the response across all forcing intensities decays prematurely with a higher forcing intensity leading to faster energy decay. 
Nevertheless, the principal resolvent forcing mode does lead to significant energy amplification and is more effective than a randomly-generated forcing structure and the second suboptimal resolvent forcing mode at amplifying the near-wall streaks. 
We compute the nonlinear energy transfer to secondary modes and observe that the breakdown of the actuated mode proceeds similarly across all forcing intensities: in the near-wall region, the induced streak forks into a structure twice-periodic in the spanwise direction; in the outer region, the streak breaks up into a structure that is once-periodic in the streamwise direction. In both regions, spanwise oscillations account for the dominant share of nonlinear energy transfer.

\end{abstract}

\begin{keywords}

\end{keywords}
\nolinenumbers
\section{Introduction}

Near-wall turbulence is organised into streamwise rolls, and alternating low- and high-speed streamwise streaks \citep{klebanoff1962three, kline1967structure, blackwelder1979streamwise, smithmetzler1983characteristics,  johansson1987generation}. These coherent structures are well-documented and their characterization is the subject of multiple works \citep{bakewell1967viscous, landahl1980note, butler1993optimal, chernyshenko2005mechanism, delalamo2006linear}. Additionally, many studies point to a quasi-periodic cycle, wherein the streamwise streaks are amplified by streamwise vortices, meander, then break down, which subsequently regenerates new quasi-streamwise vortices \citep{kim1971production, robinson1991coherent, hamilton1995regeneration, waleffe1997self,panton2001overview, adrian2007hairpin, smits2011high, jimenez2018coherent}. The cycle, also known as the self-sustaining process, can be more clearly observed in a minimal flow unit \citep{jimenez1991minimal}, where the domain is artificially restricted in the streamwise and spanwise directions in order to exclude the dynamics of the outer region of the channel. %Despite the restriction, the study of the minimal flow unit reveals that the near-wall cycle can self-sustain even when motions at larger scales are inhibited \citep{jimenez1999autonomous}.
Recent methods based on graph-theoretic approaches \citep{elnahhas2024dynamics} have provided new evidence of the self-sustaining process and revealed consistent patterns of
energy exchange between rolls and streaks in both the minimal flow unit and larger channels.

Although nonlinear mechanisms play a role in the self-sustaining process, a lot of attention has been given to linear mechanisms and instabilities as the drivers of this process \citep{panton2001overview, jimenez2013linear,lozano2021cause}. 
One example is the Orr mechanism \citep{orr1907stability, jimenez2013linear}, in which the mean shear near the wall tilts velocity perturbations forward in the streamwise direction and stretches vertical scales, intensifying the wall-normal velocity perturbations. 
Another example is lift-up \citep{hwang2010self}, which occurs when wall-normal velocity perturbations transport slow-moving fluid near the wall away into the faster flow field farther away from the wall.
Works such as \citet{delalamo2006linear} and  \citet{pujals2009note} show that, even after removing the nonlinear term from the perturbation equations, linear transient growth via the mean shear generates the dominant (streaky) structures in wall-bounded turbulence. The linearised system additionally accounts for much of the energy spectra and reproduces the self-similar profile in the logarithmic region. 
Similarly, \citet{lozano2021cause} show through numerical experiments that the minimal flow unit can sustain turbulence without the nonlinear feedback between the velocity fluctuations and the mean velocity profile, except when the Orr-mechanism or push-over (momentum transfer from the spanwise mean shear into the streamwise velocity perturbation) are suppressed. The authors thus argue for linear transient growth as a prominent mechanism for transferring energy from the mean flow to turbulent fluctuations.
In addition to the traditional near-wall streaks, 
smaller scales in the minimal channel are also capable of significant transient growth via purely linear mechanisms \citep{markeviciute2024threshold}. This is shown using a linearisation of the Navier-Stokes equations about a base flow composed of the mean turbulent channel flow profile with an added near-wall streak. The transient growth of the small scales is found to be especially driven by spanwise gradients, further underlining the ability of the linear push-over mechanism to amplify perturbations. In this work, using methods from \citet{cho2018scale, symon2021energy, ding2025mode}, we lend special attention to the role of the spanwise self-advection term in transferring energy across scales.

The linear amplification process linking streamwise vortices and streamwise streaks has also been fruitfully studied through the lens of resolvent analysis \citep{mckeon2010critical,moarref2014foundation,mckeon2017engine}. In resolvent analysis, the Navier-Stokes equations are reframed as a linear dynamical system for the velocity fluctuations; \reviewerthree{the nonlinear term, along with other exogenous inputs to the system, are represented as an external forcing term acting on this system.} The goal is then to solve for the spatial structure of the (nonlinear) forcing that generates the response (velocity) with the largest linear energy amplification. 
Resolvent analysis is traditionally applied to systems exhibiting spatial and temporal homogeneity. The linearised Navier-Stokes are first Fourier transformed in time and the homogeneous spatial directions prior to resolvent analysis which thus reveals the linear amplification properties of individual wave numbers and frequencies. In the context of wall-bounded turbulent flows, the equations are Fourier-transformed in the streamwise and spanwise directions, and time, and the targeted length scales are the traditional streamwise and spanwise streak spacings in the buffer layer \citep{mckeon2010critical, mckeon2017engine, moarref2014foundation, bae2021nonlinear}. Despite using a linearisation of the equations of motion, traditional resolvent analysis identifies streamwise rolls as the most perturbing structures and streamwise streaks as the most amplified structures. \reviewers{Resolvent analysis thus provides a partial explanation for why rolls and streaks are observed in the turbulent channel, though the linearly obtained modes fail to reproduce second-order turbulent statistics with quantitative accuracy, especially for higher Reynolds numbers. 
Indeed, resolvent analysis ignores the feedback of the output modes into the nonlinear forcing term.  One way to reincorporate this effect into resolvent models is through the inclusion of a linear eddy viscosity model in the linearised equations of motion \citep{morra2019relevance, morra2021colour, symon2023eddy}. Another useful correction includes weighting the resolvent operator with the statistical properties of the nonlinear fluctuation terms \citep{zare2017colour, morra2019relevance, morra2021colour}; the forcing terms, which are considered uncorrelated (``white") across space and time in the traditional formulation of resolvent analysis, are reweighted to reflect the observed coherence of the nonlinear terms in the turbulent system.}

Beyond simply using resolvent analysis as a way to study the self-sustaining process, resolvent modes have been used as a tractable way of tackling the control of turbulent flows. The resolvent response modes are often used as models for the fully turbulent flow. These modes can be cheaply computed for a variety of wave numbers and frequencies to identify the structures that undergo the largest kinetic energy amplification under the linearised dynamics and are thus expected to figure prominently in the turbulent flow. In \citet{toedtli2019predicting}, resolvent modes were used to model the response of turbulent channel flow \reviewertwo{at $\Rey_\tau \approx 180$} to a varying-phase opposition control. Though the Reynolds stresses of the model did not accurately capture the profile from direct numerical simulations, the change in the profiles and drag were well represented by the resolvent model. This suggests that the cheap computation of resolvent modes can be used in lieu of direct numerical simulations to identify forcing frequencies and length scales that can enact desirable changes in the turbulent flow.  \citet{liu2021unsteady} have also applied this framework to reduce pressure fluctuations along a cavity in supersonic flow \reviewertwo{($\Rey = 10,000$, computed using the cavity depth and the free stream velocity)}. In \citet{yeh2019resolvent}, resolvent analysis of a flow over an airfoil \reviewertwo{($\Rey = 23,000$, based on chord length)} is also used to identify a forcing frequency and length scale that enhance momentum mixing and reduce flow separation. A similar framework is used in \citet{lin2023flow} for a plunging cylinder \reviewertwo{($\Rey = 500$, based on cylinder diameter)} to reduce lift fluctuations. The assumption underpinning the above approach is that resolvent forcing modes, though only optimal for the linearised equations, are nevertheless efficient at actuating the fully coupled system.

In \citet{bae2021nonlinear}, linearly identified resolvent modes were indeed found to play a role in the transfer of energy to coherent near-wall turbulent perturbations, even within a fully nonlinear turbulent flow \reviewertwo{at $\Rey_\tau = 186$}, lending credibility to the assumption underpinning the works of \citet{yeh2019resolvent, liu2021unsteady, lin2023flow}. Via a modified simulation in which the contribution of the leading resolvent forcing mode is subtracted from the nonlinear term at every time step, the streak-regeneration process is interrupted and buffer layer turbulence in the minimal flow unit is greatly suppressed. % ADD STUFF
%. 

Resolvent forcing and response modes have shown promise in emulating the behaviour of turbulent flows under forcing, and this is linked to the role the forcing modes themselves play in amplifying near-wall turbulence in simulated flows. 
%%%%%%%%%%% TIME %%%%%%%%%%
In a more physical scenario, however, the flow would only be forced intermittently, either by spontaneous events arising in the unforced flow or by externally imposed actuation. It would be valuable to study how a turbulent flow instantaneously reacts to such forcing, but since the flow in \citet{bae2021nonlinear} is altered at each time step, its time-dependent reaction is difficult to measure.
We thus emphasise the importance of computing optimal time-localised forcing modes and transient response mode. Traditional formulations of resolvent analysis\reviewertwo{---in which the linearised equations of motion are Fourier-transformed in time, or Laplace-transformed in time and the real part of the Laplace variable is taken to zero to allow transients to decay---}are incapable of representing such scenarios. 
%The resolvent operator is traditionally constructed once the linearised Navier-Stokes equations are Fourier-transformed in time, and 
The resulting resolvent modes are Fourier modes in time that lack transient growth information. 
In this work, we use a version of resolvent analysis formulated in a wavelet basis \citep{ballouz2023wavelet, ballouz2024transient, ballouz2024wavelet}. The individual wavelets can capture information localised in a particular time interval. \reviewertwo{Additionally, in contrast to the Laplace transform which prescribes the same decay rate for the forcing and response modes,} wavelet-based resolvent analysis, \reviewers{particularly its windowed version,} allows one \reviewers{to compute optimal forcing and response modes that are separated in time or frequency}, \emph{e.g.} optimal pulse-like forcing modes and their corresponding transient response trajectories that can extend over the entire time domain. \reviewers{A discretised and wavelet-transform signal in time is a concatenation of increasingly subsampled signals, each discretised over time shift. By windowing for a particular element of the transform in either the response or the forcing, the user can easily select the degree to which the original signal is subsampled, which roughly determines the frequency range (scale), as well as the interval in time (shift) in which the user wishes to constrain the forcing or response. More on this topic can be found in \S\ref{sec:resolvent}. }%\reviewertwo{Wavelets are also}

% optimal linear growth
We note that maximally growing transient trajectories for the linearised Navier-Stokes equations have long been computed under the optimal transient growth framework, and used to study turbulent flows \citep{butler1993optimal, schmid2000stability, pujals2009note, jimenez2013linear, encinar2020momentum}. This framework finds the optimal initial condition that leads to a maximally energetic state at a chosen final time. While optimal transient growth has also been successful at producing rolls and streaks as the optimal perturbation and response structures, respectively, important differences exist between this method and wavelet-based resolvent analysis. Optimal transient growth is sensitive to the choice of final time.
More importantly, as is expounded in \S\ref{sec:resolvent}, optimal transient growth and wavelet-based resolvent analysis use different measures of optimality: the former maximises the kinetic energy ratio between the initial condition and the solution at the chosen terminal time, while the latter maximises the integrated kinetic energy of the response over the entire time domain. 
Maximising the integrated kinetic energy may better capture structures that tend to persist in time, rather than spike and decay rapidly.  
% cause and effect
Moreover, wavelet-based resolvent analysis can identify optimal forcing terms arising at various points of the time domain, and not just its origin. For example, when applied to turbulent oscillating Stokes flows \citep{ballouz2024transient}, the computed forcing and response modes coincide with the times in the cycle when the streamwise root-mean-square velocity peaks. The forcing modes usually precede the response modes. Another version of this spatio-temporal resolvent analysis computes time-sparse modes and similarly yields time-localised optimal forcing modes that precede their transiently varying responses 
\citep{lopez2023sparsity, lopez2024spacetime}. Both of these results suggest that time-localised resolvent analysis correctly extracts cause-and-effect relations between the computed modes. 

The transient growth of any structure within a fully turbulent flow is modulated by the myriad nonlinear interactions not considered when using methods that rely on the linearised equations of motion.
Therefore, with the objective of controlling near-wall turbulence in mind, we wish to determine whether significant transient growth can be achieved in a turbulent flow via the injection of a resolvent forcing mode into the simulation of a turbulent flow.
The transient resolvent response mode corresponding to the injected forcing will allow us to measure the instantaneous discrepancy between the actuated turbulent and linearised flows, and probe the efficacy of the resolvent forcing modes at actuating nonlinear flows.
We aim to identify the time scales during which the turbulent system responds similarly to the optimal linear response mode and beyond which nonlinear effects distort the effect of the resolvent forcing mode. 
We study the mechanisms that erode the effects of the injected mode, especially the nonlinear interactions that transfer energy from the actuated scale to secondary ones and force the turbulent flow to deviate from the optimal linear response. The system we use is the minimal flow unit \reviewertwo{at $\Rey_\tau = 186$}. 

%%%%%%%%%%%%%
This paper is organised as follows. In \S\ref{sec:base_flow}, we compute the base flow for the minimal flow unit at $\Rey_\tau = 186$, which we use in \S\ref{sec:resolvent} to formulate the resolvent operator in a time-localised wavelet basis.
We compute resolvent modes as in \citet{ballouz2024wavelet}, making sure to constrain the forcing to a wavelet-shaped pulse. This yields a forcing mode in the shape of streamwise rolls that is compactly supported in time, in addition to an optimal streak-like response that grows transiently before decaying. The justification for the choice of spatial wavenumbers and wavelets is given in \S\ref{parameters}. We then solve the fully nonlinear forced Navier-Stokes equations for the minimal flow unit at $Re_\tau = 186$, using the time-localised wavelet-based resolvent forcing mode as our forcing term. This step is detailed in \S\ref{dns}. We track the evolution of this resolvent forcing mode as it generates and amplifies streamwise streaks, and compute relevant turbulent statistics, which we present in \S\ref{results}. 
In \S\ref{sec:NLT}, we focus on the nonlinear energy transfer from the induced streak to secondary modes.
%A preliminary version of this work was published in \citet{ballouz2023transient}. Additions include a more detailed formulation of wavelet-based resolvent analysis (\S\ref{sec:resolvent}), an analysis of nonlinear energy transfer from the induced streak to secondary modes (\S\ref{NLT}), and a quasi-linear model of this process (\S\ref{quasilinear}). 
Concluding remarks are given in \S\ref{conclusion}.

\section{Methods}\label{sec:methods}

%\subsection{System and discretization}\label{discretization}

In this work, we consider the flow in the minimal flow unit of size $L_1 \times L_2 \times L_3 = 1.72\delta \times 2\delta \times 0.86 \delta$, where $\delta$ is the channel half-height, and $x_1$, $x_2$ and $x_3$ are the streamwise, wall-normal and spanwise directions, respectively. We denote the velocity fluctuation field by $\boldsymbol u = [u_1, u_2, u_3]^T$, where $u_1$, $u_2$ and $u_3$ are the streamwise, wall-normal, and spanwise components respectively. The system is characterised by the friction Reynolds number $Re_\tau = \delta u_\tau / \nu \approx 186$, where $u_\tau$ is the friction velocity, and $\nu$ is the kinematic viscosity. The flow is periodic in the streamwise and spanwise directions, and the no-slip and no-penetration conditions hold at the walls of the channel.

\subsection{Base flow} \label{sec:base_flow}
For the direct numerical simulations (DNS) in this work, we discretise the streamwise and spanwise directions uniformly using $N_1 = N_3 = 32$ grid points, which results in streamwise and spanwise grid spacings of $\Delta x_1^+ \approx 10$ and $\Delta x_3^+ \approx 5$. In the wall-normal direction, the grid is of size $N_2 =  128$ and stretched according to a hyperbolic tangent distribution, which results in a wall-normal spacing of $\min(\Delta x_2^+) \approx 0.17$ near the wall and $\max(\Delta x_2^+) \approx 7.6$ at the centreline. Here, the superscript $+$ denotes wall units normalised with $u_\tau$ and $\nu$. We discretise the incompressible Navier-Stokes equations with a staggered, second-order accurate, central finite difference method in space \citep{orlandi2000fluid}, and a fractional step method is used to compute pressure \citep{kim1985application}. Time-advancement is performed with an explicit third-order-accurate Runge-Kutta method \citep{wray1990minimal}. The DNS code has been validated in previous studies of turbulent channel flows \citep{lozano2016turbulent, bae2018turbulence, bae2019}. Using this discretisation, we obtain a mean streamwise velocity profile $U_1(x_2)$ by averaging DNS results of the unforced system in the homogeneous directions and time. This mean profile is used in the subsequent sections.

%%%
\subsection{Spatio-temporal resolvent modes}\label{sec:resolvent}

In this section , we describe how we compute the time-localised resolvent forcing modes and their corresponding transient responses for the minimal flow unit \citep{ballouz2024transient, ballouz2024wavelet}. 
We first formulate the incompressible Navier-Stokes equations for the velocity fluctuations about a mean turbulent flow field $\boldsymbol{U} = (U_1, U_2, U_3)$:
%$\boldsymbol{U}(x_2) = (U(x_2), 0, 0)$:

\begin{equation}\label{eq:NS}
    \frac{\partial{u_i}}{\partial t} + U_j\frac{\partial{u_i}}{\partial{x_j}} + u_j\frac{\partial{U_i}}{\partial{x_j}} = -\frac{\partial{p}}{\partial{x_i}} + \frac{1}{\Rey} \frac{\partial^2u_i}{\partial{x_j}\partial{x_j}} + f_i,\quad \frac{\partial{u_i}}{\partial{x_i}}= 0.
\end{equation}
Here $p$ represents the pressure fluctuation, and $f_i$ the nonlinear term in the $x_i$-momentum equation. The base flow is obtained from \S\ref{sec:base_flow}, and satisfies $U_2 = U_3 = 0$ while the mean streamwise component $U_1$ is constant in time.
Equations \eqref{eq:NS} are Fourier-transformed in the $x_1$- and $x_3$-directions and discretised in $x_2$ and time. 
The time domain $[0, T]$ is periodic\reviewertwo{---\emph{i.e.} when constructing the resolvent operator for the discretised system, periodic boundary conditions are used for the temporal derivative matrix; this is further discussed below---} and we choose $T = 22 \; \delta / u_\tau$, which is long enough to allow the resolvent modes to decay to zero. The discretisation in time is uniform, with a grid size of $N_t = 128$ corresponding to a spacing of $\Delta t \approx 0.17 \; \delta / u_\tau$.
The wall-normal discretisation is the same as in the DNS (\S\ref{sec:base_flow}). 

The discretised equations are further wavelet-transformed in time by premultiplying them by a discrete wavelet transform operator $\mathsfbi W$. For an arbitrary square-integrable function $g(t)$, the wavelet expansion on a dyadic time grid of size $N_t$\reviewerone{---the grid is uniform and $N_t$ is a power of 2---}is defined as:
\begin{equation}\label{eq:waveletExpansion}
    g(t) = \sum_{\ell = 1}^L \sum_{m = 0}^{N_t/2^\ell-1} \frac{1}{\sqrt{2^\ell}}\tilde g^{(w)}(\ell, m) \eta \left (\frac{t}{2^\ell} - m \right) + \sum_{m = 0}^{N_t/2^L-1}  \frac{1}{\sqrt{2^L}} \tilde g^{(s)}(m) \zeta \left (\frac{t}{2^L} - m  \right ),
\end{equation}
where $\eta(t)$ and $\zeta(t)$ denote the wavelet and scaling functions, respectively \citep{mallat2001wavelet,najmi2012wavelets}, and $L$ satisfies $2^L \leq N_t$ and represents the largest scale captured by the wavelet expansion. \reviewers{The dilations and shifts of $\eta(t)$ act as high-pass filters and coefficients $\tilde g^{(w)}(\ell, m) $ capture the high-frequency information of $g(t)$; while $\zeta(t)$ acts as a low-pass filter and coefficients $\tilde g^{(s)}(m)$ capture the remaining low-frequency information of the function.}
The matrix $\mathsfbi W$ approximates this wavelet expansion: given $\boldsymbol{g}$ -- the discretisation of $g(t)$ in time -- its wavelet transform $\boldsymbol{\tilde g}$ is computed as
\begin{equation}
    \boldsymbol{\tilde g} := \begin{bmatrix}
        \boldsymbol{\tilde g^{(s)}}
        \\ \boldsymbol{\tilde g^{(w)}}
    \end{bmatrix} = \mathsfbi W \boldsymbol{g},
\end{equation}
where the elements of
$\boldsymbol{\tilde g^{(s)}}$ and $\boldsymbol{\tilde g^{(w)}}$ are respectively $\tilde g^{(w)}(\ell, m)$ and $\tilde g^{(s)}(m)$ for all $\ell$ and $m$. The matrix $\mathsfbi W$ is of size $N_t \times N_t$, and the transform $\boldsymbol{\tilde g}$ is of size $N_t$.
The projection onto $\eta (t/2^\ell - m)$ or $\zeta(t/2^L - m ) $ roughly captures a portion of the frequency content of $g$ determined by $\ell$, centred in a time interval determined by $m$. Larger $\ell$ corresponds to a narrower band of frequencies closer to zero, while larger $m$ corresponds to later times. Each of $\eta$ and $\zeta$ capture difference portions of the frequency spectrum. 

The choice of wavelet-scaling-function pair is not unique and determines the properties of the transform operator $\mathsfbi W$. In this work, we use a single-level Daubechies-8 wavelet transform \citep{daubechies1992ten}. The Daubechies wavelets and their corresponding scaling functions are compactly supported in time and form an orthonormal basis, resulting in a sparse banded and unitary operator $\mathsfbi W$ \citep{mallat2001wavelet, najmi2012wavelets, ballouz2024wavelet}. 

%%%%%%%%%%%%%%
For a given streamwise and spanwise wavenumber pair $(2\pi k_1/L_1, 2\pi k_3/L_3)$, where $k_1, \; k_3 \in \mathbb Z $, we obtain the following discretised and transformed equations
\begin{equation}\label{NS_transformed}
    \widetilde{\mathsfbi D_t} \widetilde {\mathsfbi u}_i + \widetilde {\mathsfbi U}_j \widetilde{\mathsfbi D_j}\widetilde {\mathsfbi u}_i + \widetilde {\mathsfbi{dU}_{ij}} \widetilde {\mathsfbi u}_j = -\widetilde{\mathsfbi D_i} \widetilde {\mathsfbi p} + \frac{1}{\Rey} \widetilde{\mathsfbi D^2_{jj}} \widetilde {\mathsfbi u}_i + \widetilde {\mathsfbi f}_i,\quad \widetilde{\mathsfbi D_i} \widetilde {\mathsfbi u}_i= 0.
\end{equation}
Here, $\tilde {\mathsfbi u}_i$, $\tilde {\mathsfbi p}_i$ and $\tilde {\mathsfbi f}_i$ denote the discretised and transformed velocity, pressure and forcing, respectively. These transformed quantities are functions of wall-normal position $x_2$, and the wavelet scale and shift parameters $\ell$ and $m$, which respectively represent the time interval and frequency support of the wavelet mode \citep{ballouz2024wavelet}. 
The transformed spatial derivative operators are defined as follows: $\widetilde {\mathsfbi D_1} = \text{i} \breve k_1 \mathsfbi I$, where $\mathsfbi I$ is the identity matrix of size $(N_2N_t) \times (N_2N_t)$, $\widetilde {\mathsfbi D_2} = \mathsfbi D_2$, which denotes a block diagonal second-order-accurate central finite difference operator on the staggered $x_2$-- grid, $\widetilde {\mathsfbi D_3} = \text{i} \breve k_3 \mathsfbi I$ and $\widetilde{\mathsfbi D^2_{jj}} = -\breve k_1^2 \mathsfbi I + \mathsfbi D^2_{2} - \breve k_3^2 \mathsfbi I$, where $\mathsfbi D^2_{2}$ denotes a second-order accurate second-order finite difference operator on the staggered $x_2$-- grid. The wavenumbers $\breve k_i$ are the modified wavenumbers for the discretisation scheme used in the DNS, and are defined as $\breve k_1 := 2\delta  \mathrm{sin}( \delta \Delta x_1 \pi k_1/L_1 )/\Delta x_1 $ and $\breve k_3 := 2 \delta \mathrm{sin}(\delta \Delta x_3 \pi k_3/L_3)/\Delta x_3 $. \reviewerone{Here, the use of the modified wavenumbers ensures that the modes satisfy the conservation laws---namely continuity---on the simulation grid and for the discretisation scheme used in the DNS; this is particularly important for the forcing modes, which will be injected into the DNS of the minimal flow unit.}

The modified time derivative operator is defined as $\widetilde{\mathsfbi D_t} = \mathsfbi W \mathsfbi D_t  \mathsfbi W^{-1}$, \reviewers{where $\mathsfbi W$ now denotes the discrete wavelet transform in time for the $(N_tN_2)\times (N_tN_2)$--dimensional system (\emph{i.e.} the transform defined previously for a one-dimensional discrete time signal, pre-multiplying the $N_2\times N_2$--dimensional identity matrix $\mathsfbi I$ via a Kronecker product)}, and $\mathsfbi D_t$ is a second-order-accurate central finite difference matrix in time. 
Though not shown, using a fourth-order accurate finite difference operator does not strongly affect the resolvent modes.  We also define the mean flow term $\widetilde {\mathsfbi U}_j := \mathsfbi W \mathsfbi U_j \mathsfbi W^{-1}$ and the mean shear term $\widetilde {\mathsfbi {dU}}_{ij} := \mathsfbi W \mathsfbi {dU}_{ij} \mathsfbi W^{-1} $, where $\mathsfbi U_j $ and $\mathsfbi {dU}_{ij} $ are diagonal matrices with diagonal terms corresponding to $U_j$ and $dU_i/dx_j$ at each $x_2$ and time grid point, respectively.

We can rearrange equation (\ref{NS_transformed}) as
\begin{equation}\label{inout}
    \begin{bmatrix}
        \widetilde {\mathsfbi u}_1(x_2, \ell, m) \\
         \widetilde {\mathsfbi u}_2(x_2, \ell, m) \\ 
          \widetilde {\mathsfbi u}_3(x_2, \ell, m) \\
        \widetilde{\mathsfbi p}(x_2, \ell, m)
    \end{bmatrix} = \widetilde{\mathsfbi H}^{(k_1, k_3)} \begin{bmatrix}
        \widetilde {\mathsfbi f}_1(x_2, \ell, m) \\
         \widetilde {\mathsfbi f}_2(x_2, \ell, m) \\ 
          \widetilde {\mathsfbi f}_3(x_2, \ell, m) \\ 0
    \end{bmatrix},
\end{equation}
where the linear operator $\widetilde{\mathsfbi H}^{(k_1, k_3)}$ is the resolvent operator in this formulation. The superscript $(k_1, k_3)$ indicates the choice of Fourier parameters for the transformation of equation \eqref{eq:NS}, and the functional dependence on $x_2$, $\ell$ and $m$ denotes the discretisation of \eqref{eq:NS} over these quantities. 
Since we use the same $x_2$--grid as the DNS that produced the mean flow and shear profiles, and taking $N_t$ as the temporal resolution, we note that $\widetilde{\mathsfbi H}^{(k_1, k_3)}$ is a $(4N_2 N_t) \times (4 N_2 N_t)$ matrix. This formulation of the resolvent operator targets \reviewertwo{all the Fourier frequencies resolvent by the temporal grid}, rather than one as in traditional resolvent analysis.

We introduce the additional step of constraining the forcing along a wavelet-shaped pulse of any desired scale $\ell$ and shift $m$ using a temporal windowing matrix $\mathsfbi B$ \citep{jeun2016input, kojima2020, ballouz2023wavelet, ballouz2024transient}.  
Typically, the unconstrained resolvent modes for channel flow are Fourier modes in time centred at a chosen critical layer \citep{mckeon2010critical,bae2021nonlinear, ballouz2024wavelet}. In order to compute an optimal transiently growing response trajectory, we must constrain the forcing in time. %\citep{ballouz2024wavelet}. 
The windowing matrix $\mathsfbi B$ takes the form

\begin{equation}
    \mathsfbi{B} = \text{diag}\big(\mathbbm{1}(\ell = \ell_d)\mathbbm{1}(m = m_d)\big),
\end{equation}
so that we are restricting the forcing to a wavelet or scaling function, centred at a desired time and frequency determined by $\ell_d$ and $m_d$, respectively. For a discussion on the choice of $\ell_d$ and $m_d$, see \S\ref{parameters}.
We then take the singular value decomposition (SVD) of the combined operator 
\begin{equation}
    \widetilde{\mathsfbi H}^{(k_1, k_3)} \mathsfbi B = \sum_{j} \sigma_j \boldsymbol{\tilde \psi}_j(x_2, \ell, m)  \boldsymbol{\tilde \phi}_j^{H}(x_2, \ell, m),
\end{equation}
where $(\cdot)^H$ denotes the conjugate transpose. We index the singular values $\{\sigma_i\}_{i=1}^{\infty}$ such that $\sigma_1 \geq \sigma_2 \geq ... \geq 0$. The right and left singular vectors $\{\boldsymbol{\tilde \phi_j}\}_{j=1}^{\infty}$ and $\{\boldsymbol{\tilde \psi_j}\}_{j=1}^{\infty}$ respectively define orthonormal bases for the spaces containing the nonlinear term (forcing) and the velocity and pressure fluctuations (response). 
For the SVD, we choose the inner product to be the seminorm representing action\reviewerone{---or time-integrated energy---}which we define for an arbitrary vector  $\boldsymbol{\tilde b} =  [\tilde b_1, \tilde b_2, \tilde b_3, \tilde b_p]^T$ to be
\begin{equation}
    \|\boldsymbol{\tilde b} \|^2 = \frac{u_\tau}{\delta} \frac{1}{L_1(2\delta)L_3}\int_0^{L_1}\int_0^{2\delta} \int_{0}^{L_3}\int_{0}^{T} (|b_1|^2 + |b_2|^2 + |b_3|^2 )\,\mathrm{d}t\,\mathrm{d}x_3\,\mathrm{d}x_2 \,\mathrm{d}x_1,
\end{equation}
where $\boldsymbol{b} = [b_1, b_2, b_3, b_p]^T$ is the inverse wavelet- and Fourier-transform of $\boldsymbol{\tilde b}$ \citep{ballouz2024wavelet}. 
The kinetic energy amplification factor is given by the square of the singular values. The forcing modes are therefore ordered decreasingly according to the integrated kinetic energy amplification they undergo when acted on by $\widetilde{\mathsfbi H}^{(k_1, k_3)} \mathsfbi B $, and the response modes are the corresponding amplified coherent structures arising from this action. 
Thus, $\boldsymbol{\tilde \phi_1}= [\tilde \phi_{1, 1}, \; \tilde \phi_{1, 2}, \; \tilde \phi_{1, 3}, \;0]^T$ generates the largest linear energy amplification via the windowed resolvent operator, and $\sigma_1 \boldsymbol{\tilde \psi_1} = \sigma_1 [ \tilde \psi_{1, 1}, \; \tilde \psi_{1, 2}, \; \tilde \psi_{1, 3}, \; \tilde \psi_{1, p}]^T$ is the optimally amplified transient velocity and pressure fluctuation \reviewerone{resulting from the application of $\boldsymbol{\tilde \phi_1}$ on the linearised system}. Here, $ \tilde \psi_{j, i}$ and $ \tilde \phi_{j, i}$ correspond to the $i^{th}$ velocity component, and $ \tilde \psi_{j, p}$ and $ \tilde \phi_{j, p}$ refer to the pressure component.
For all $j$, $\boldsymbol{\phi_j}$, the inverse transform of $\boldsymbol{\tilde \phi_j}$, is shaped in time according to the wavelet or scaling function chosen by $\mathsfbi B$. As described in \citet{moarref2014foundation}, the resolvent modes for certain Fourier parameters occur in equivalent pairs of equal singular values due to the symmetry of the channel geometry. The pairs of singular modes each form a singular plane, and for the numerical experiments in \S\ref{dns}, we linearly combine the two equivalent forcing modes (e.g. $\boldsymbol{\tilde \phi_1}$ and $\boldsymbol{\tilde \phi_2}$) to form a forcing vector in the singular plane that acts primarily on the bottom half of the channel. Thus, upon injecting this mode into the DNS of the forced Navier-Stokes equations, only the bottom of half of the channel is subject to the forcing, allowing us to use the \reviewertwo{unforced} top half as a benchmark \reviewertwo{for comparison} \citep{bae2021nonlinear}. Henceforth, $\boldsymbol{\tilde \phi_1}$ refers to the linear combination of the first two equivalent forcing modes that concentrates the forcing in the bottom-half of the channel, renormalised to satisfy $\|\boldsymbol{\tilde \phi_1} \| = 1$, and $\boldsymbol{\tilde \psi_1}$ denotes the corresponding response mode. The modes
$\boldsymbol{\tilde \phi_3}$ and $\boldsymbol{\tilde \psi_3}$ are defined similarly with regards to the third and fourth resolvent forcing and response modes.
We use $\boldsymbol{\breve \psi_i}$ and $\boldsymbol{\breve \phi_i}$ to denote the inverse wavelet transforms in time of $\boldsymbol{\tilde \psi_i}$ and $\boldsymbol{\tilde \phi_i}$, \reviewerthree{or equivalently, the spatial Fourier transforms of $\boldsymbol{ \psi_i}$ and $\boldsymbol{ \phi_i}$, respectively.}

\subsection{Choice of spatial and temporal scales} \label{parameters} 
The minimal flow unit allows us to isolate one buffer layer streak. This streak appears to stretch the entire streamwise length of the unit, which is only large enough to contain one low- and high-speed streak pair in the spanwise direction.
We thus choose to target the Fourier mode given by the streamwise and spanwise wavenumbers of $k_1 = 0$ and $k_3 = 1$, respectively. These length scales also correspond to a peak in the spectral energy content for the minimal flow unit \citep{bae2021nonlinear}. 

\begin{figure} 
    \centering
    \subfloat[]{
        \includegraphics[width=0.44\textwidth]{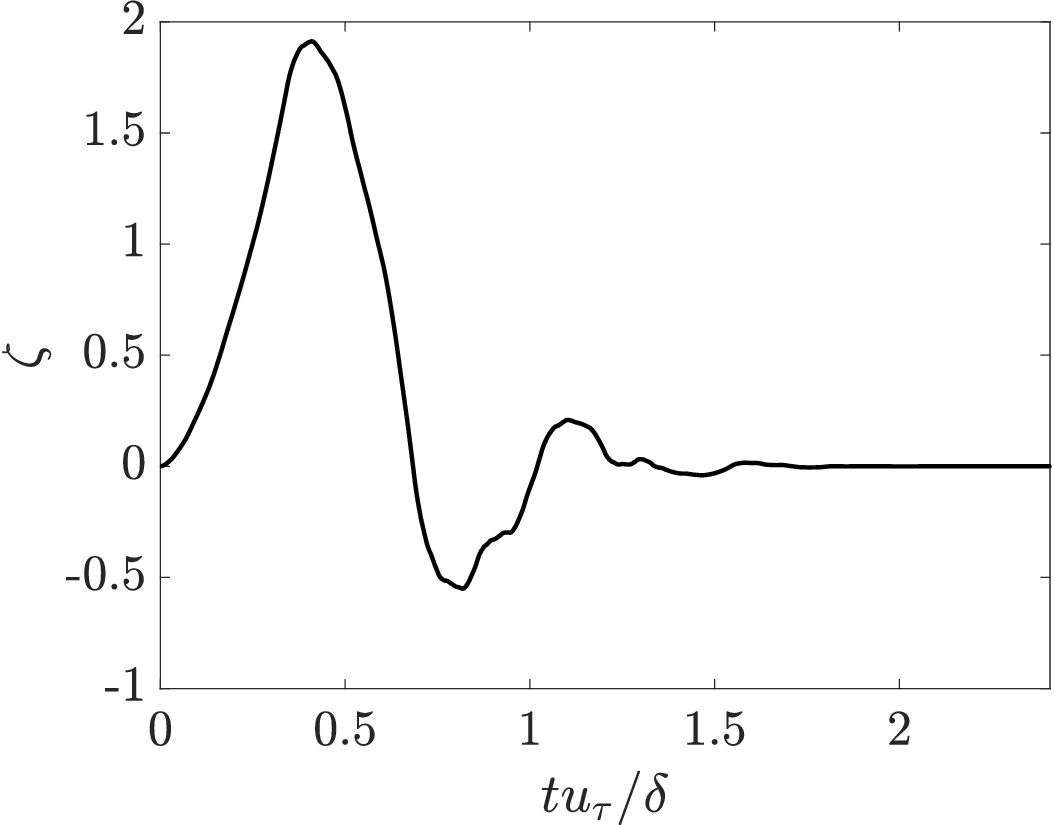}}
    \subfloat[]{
        \includegraphics[width=0.46\textwidth]{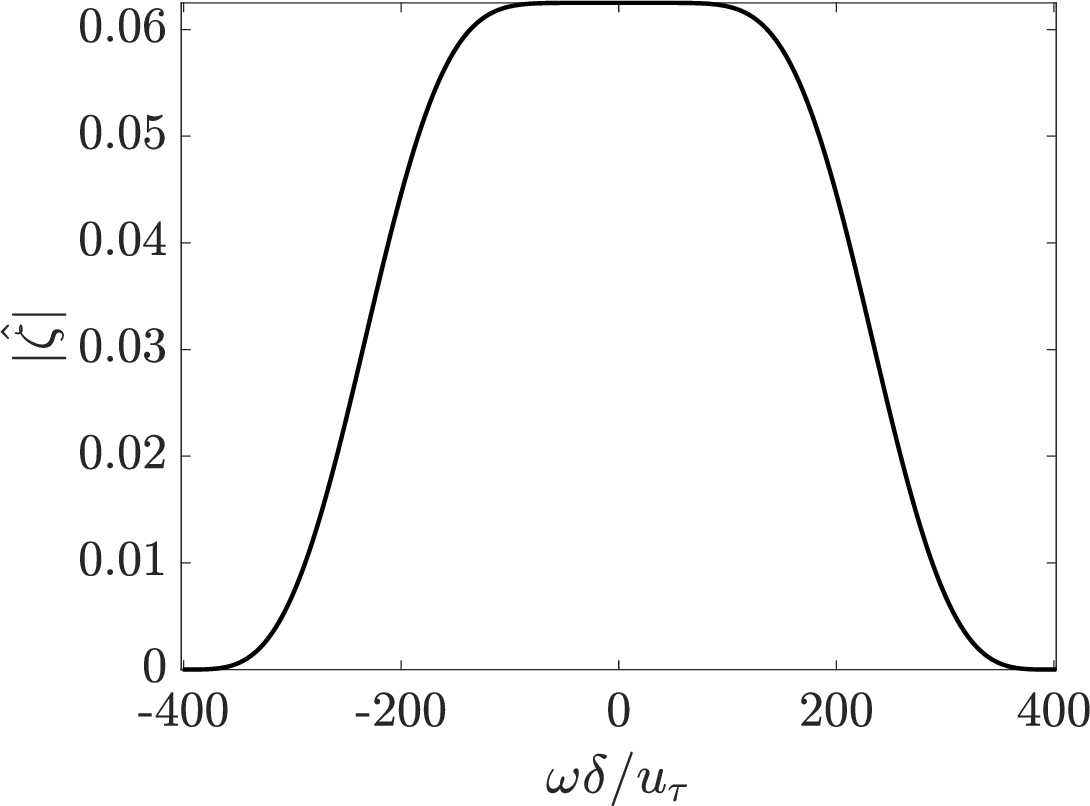}}
    \caption{Daubechies-8 scaling function $\zeta$ in time (a) and frequency (b) domain.
    }
    \label{Daub8}
\end{figure}

%%%% move this up
Traditional resolvent analysis in which the Navier-Stokes equations are Fourier-transformed in time reveals that a temporal frequency of $\omega = 0$ produces the modes with the largest kinetic energy amplification \reviewertwo{for the minimal flow unit at $\Rey_\tau = 186$ }\citep{bae2021nonlinear}. 
To target this frequency, we constrain the forcing term to a Daubechies-8 scaling function of arbitrary shift $m$ by using $\mathsfbi B$ to select the corresponding elements of $\widetilde{\mathsfbi f}$. The scaling function is shown in figure \ref{Daub8}(a), and its Fourier spectrum, shown in figure \ref{Daub8}(b), indeed encompasses the target frequency.
Since the scaling function is compactly-supported in time, wavelet-based resolvent analysis will not be able to target $\omega = 0$ uniquely but will capture a wide range of frequencies: a trade-off exists between time and frequency localisation, and the more precision we require in one domain, the less we preserve in the other \citep{mallat2001wavelet, najmi2012wavelets}. We note that the simulations detailed in \S\ref{dns} resolve temporal wavenumbers up to $\omega \delta / u_\tau \approx 55,000$. The scaling function satisfies $\left ( \int_{-\infty}^{+\infty} |\zeta(t)|^2 \mathrm{d}t \right ) \; u_\tau / \delta = 1$.

%%%%%%%%%%%%%%%

\begin{figure}
    \centering
\subfloat[]{
        \includegraphics[width=0.47\textwidth]{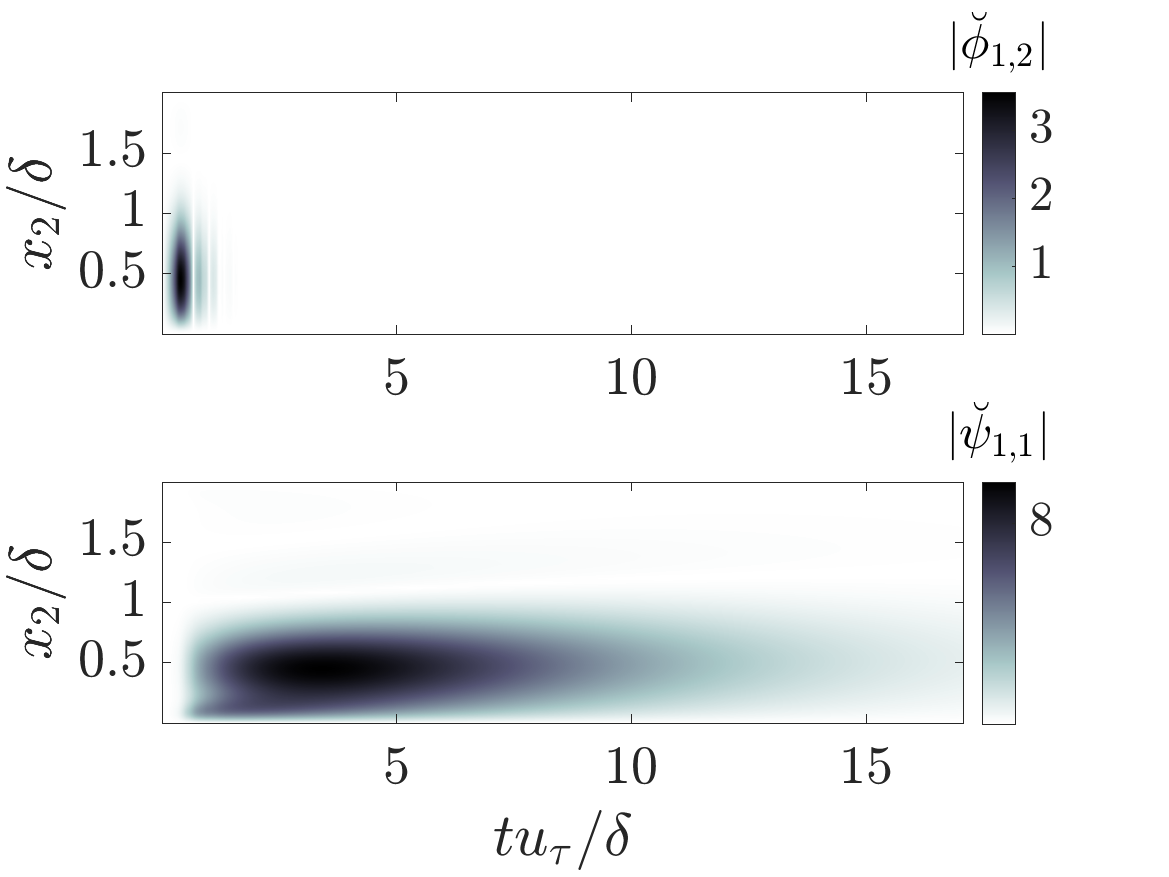}}
\subfloat[]{
        \includegraphics[width=0.43\textwidth]{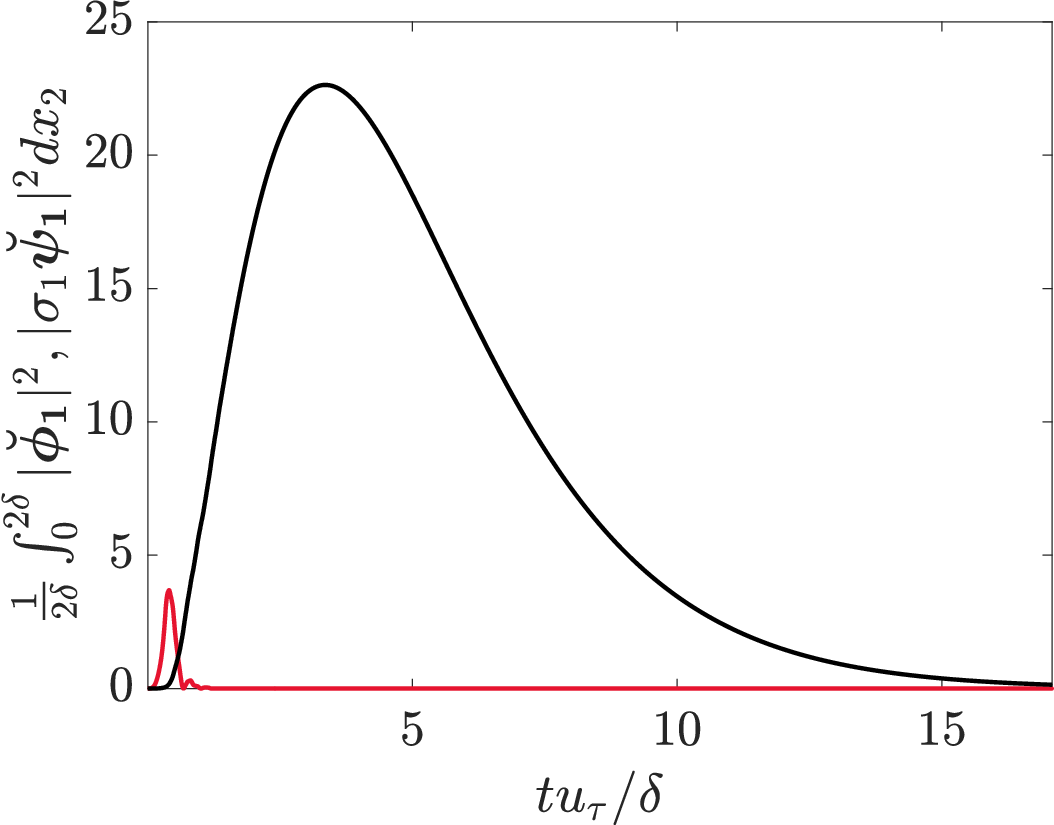}}
        
\subfloat[]{
        \includegraphics[width=0.45\textwidth]{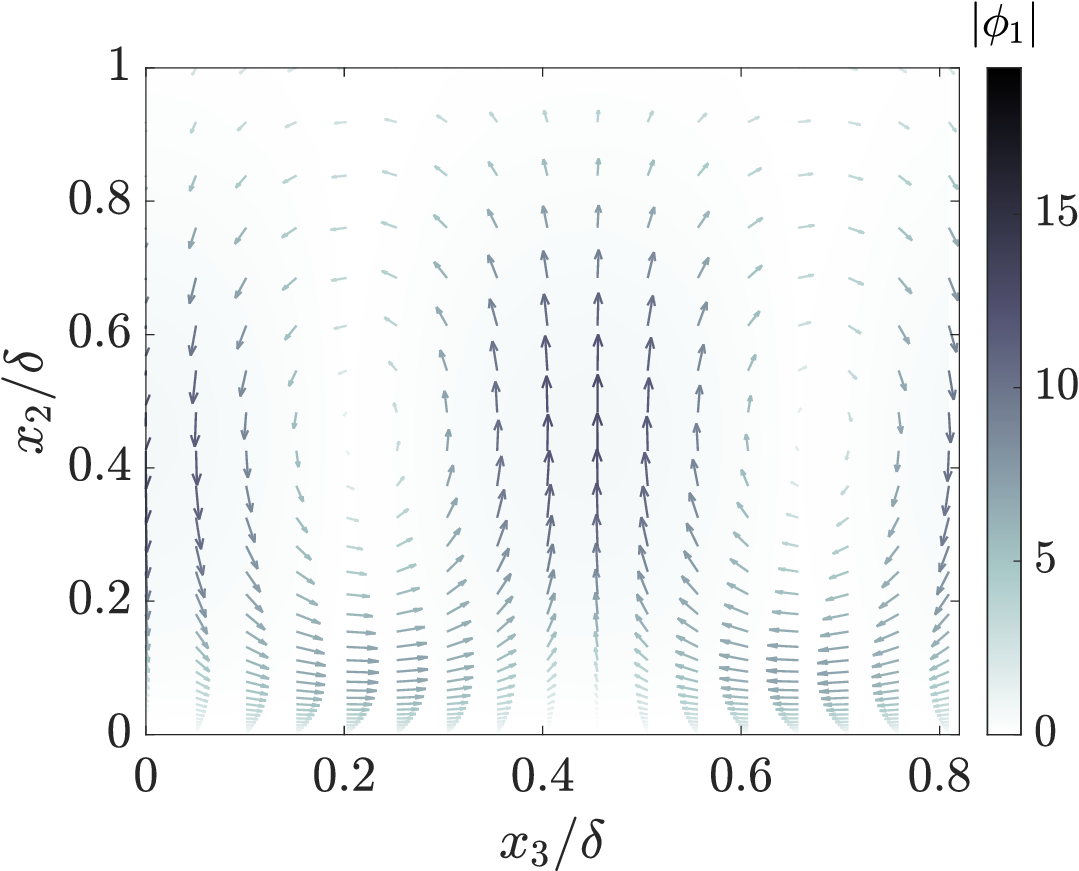}}
\subfloat[]{
        \includegraphics[width=0.45\textwidth]{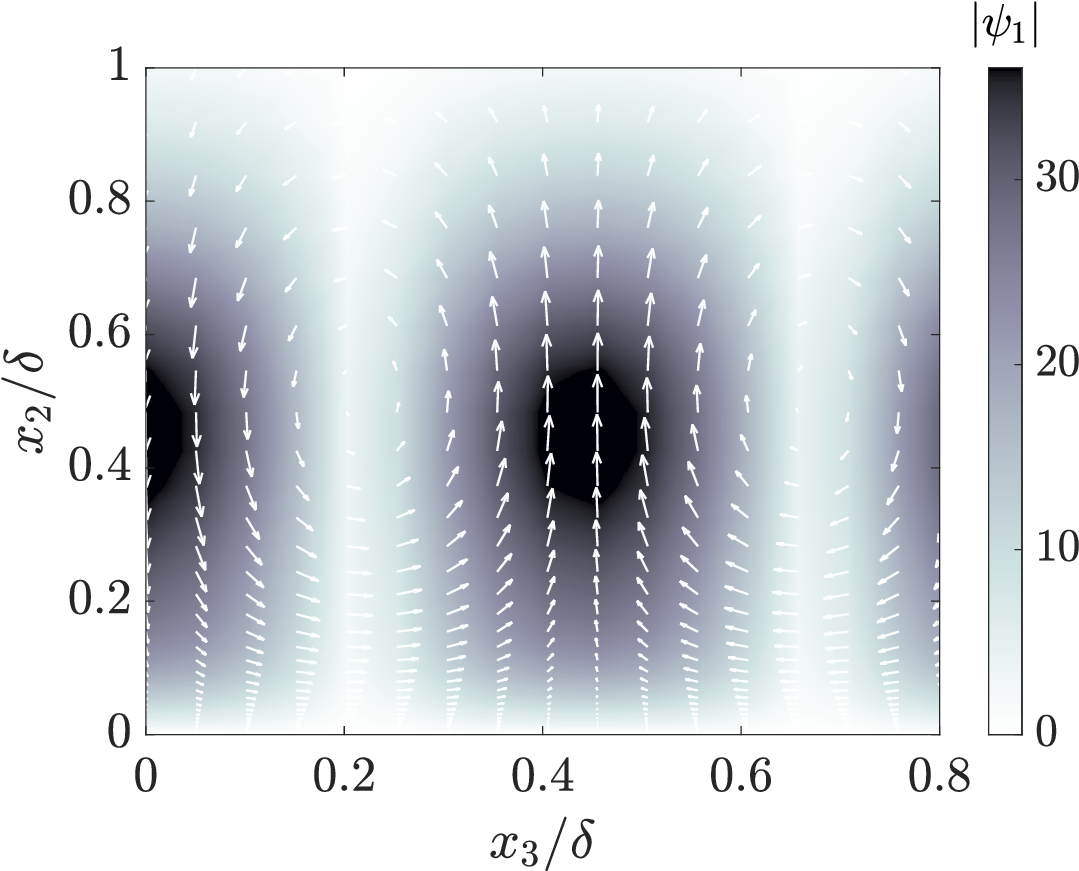}}
    \caption{(a) Magnitudes of the wall-normal component of the principal forcing mode (top) and the streamwise component of the principal response mode (bottom). (b) Integrated energy of the principal forcing (red) and response (black) modes. (c) Principal forcing mode shown at peak amplitude ($t u_\tau /\delta \approx 0.40)$. (d) Principal response modes at peak amplitude ($t u_\tau /\delta \approx 2.41)$. In (c,d), the contours represent the streamwise magnitude; the arrows, which show the direction of the cross flow components, are coloured according to their magnitudes $\sqrt{| \phi_{1,2}|^2 + | \phi_{1,3}|^2}$ in (c), or $\sqrt{| \psi_{1,2}|^2 + |\psi_{1,3}|^2}$ in (d).}
    \label{fig:linearmodes}
\end{figure}

The obtained resolvent modes are shown in figure \ref{fig:linearmodes}. Notably, the response modes exhibit transient energy growth and decay as seen in figure \ref{fig:linearmodes}(a,b). The inverse transforms of the modes are shown in figure \ref{fig:linearmodes}(c,d). The modes share many similarities with the Fourier-based modes computed for $\omega = 0$ in \citet{bae2021nonlinear}: the optimal transient nonlinear forcing mode appears in the shape of streamwise rolls, and the optimal velocity fluctuation response appears as predominantly streamwise streaks with alternating signs of the same magnitude. This supports the extensively examined claim that streamwise streaks can be linearly generated by a linear lift-up mechanism, whereby slower moving fluid close to the wall is swept upwards into the faster moving mean flow farther away from the wall. The streak-shaped response mode grows in intensity before fading, 
showcasing the transient growth that is characteristic of non-normal systems.


\subsection{Forced direct numerical simulations}\label{dns}
%%%%% DESCRIBE THE EXPERIMENT MORE %%%%%%%
The time-localised resolvent forcing mode obtained in the previous section maximises the action for the minimal flow unit under the linearised Navier-Stokes equations. The aim of the work is to study the effect of the forcing mode on the fully nonlinear minimal flow unit. To that end, we aim to inject the time-localised resolvent forcing mode into a DNS of the minimal flow unit. 
The forcing mode will be introduced into an already turbulent state of the minimal flow unit. The turbulent initial condition will contain energy at varying scales, triggering cross-scale interactions that will interfere with the linearly driven transient growth of the injected mode.
We inject the forcing at different magnitudes, all small relative to the energy of the initial condition. We expect the produced energy amplification to vary nonlinearly across the different cases.
To characterise how the turbulent system reacts to actuation of varying strength, we aim to track the ensemble-averaged transient response, and measure the energy growth of the targeted mode along with the nonlinear energy transfer to secondary scales. The optimal resolvent response mode will serve as a benchmark for the achievable energy amplification caused by the injected mode.
Below we present the numerical details of the DNS.

%%%%%%%%%%%%%%%%%%%%%
To initialise the ensemble of the forced simulations, we first perform a DNS of the non-actuated minimal flow unit, fixing the mean profile to be $\boldsymbol{U} = (U_1(x_2), 0, 0)$ (\S\ref{sec:base_flow}), which is used to calculate the resolvent modes. Snapshots from this simulation will serve as the initial conditions to the forced simulations. 
Freezing the mean profile ensures that the DNS mean profile matches the one used to compute the resolvent modes for all time. We do this by initializing the flow to have the desired mean streamwise profile of $U_1$, then by removing the steady-state contribution of the right-hand side of the Navier-Stokes equations. 
The initial snapshots from the fixed-mean simulation are separated by $1 \leq \Delta t  u_\tau / \delta \leq 5.2 $, \reviewertwo{which reduces the correlation between the snapshots}. The ensemble sizes range from 1000 to 4000, in order to ensure statistical convergence. \reviewertwo{Doubling the time between initial snapshots---equivalently, averaging over half the available timeseries data---led to streamwise energy fields that only differed by at most $0.5\%$ to $1.3\%$ at any given wall-normal location and time.}
For each initial condition, we also obtain a corresponding unforced fixed-mean time series. We denote the velocity fluctuations for the unforced simulations by $\boldsymbol{u}_0 (x_1, x_2, x_3, t)$. %, and the nonlinear terms in the momentum equations by $\boldsymbol{g}_0(x_1, x_2, x_3, t)$.

Before injecting the forcing mode into the DNS of the minimal flow unit, the mode is normalised so that $\| \boldsymbol{\tilde{\phi_1}}\|^2 = 1$ and scaled by a complex constant $\kappa$ with magnitude
\begin{equation}
    |\kappa| := \gamma  \left (\frac{\delta}{u_\tau^3} \frac{1}{L_1 (2\delta)L_3}\int_{-\infty}^{+\infty}\int_0^{L_1}\int_0^{2\delta}\int_0^{L_3} \left \vert \frac{\partial \boldsymbol{{u}_0}}{\partial t}\right \vert^2_{t=0} \zeta(t)^2\mathrm{d}x_3 \, \mathrm{d}x_2 \, \mathrm{d}x_1\, \mathrm{d}t \right )^{1/2},
\end{equation} 
where $\gamma \in \{ 1\%, 2\%, 5\%, 10\%\}$ such that the resolvent forcing mode is increasing the initial energy of the right-hand side by $\gamma \%$. 
Thus, $|\kappa|^2$ determines the integrated energy injected into the system by the forcing. \reviewerone{Using $\angle \cdot$ to denote the phase,} we choose 
\begin{equation}
\angle \kappa = \angle \int_0^2 \int_0^{+\infty} \partial_t \boldsymbol{\hat u}_0^{(0, 1)} \boldsymbol{\breve \phi_1}^* dt dx_2 ,
\end{equation}
so that the forcing mode is in phase with the right-hand side of the unforced flow field. 
In the limiting case where the nonlinear interactions are negligible, this provision ensures that the added forcing maximally increases the energy of the target $(0, 1)$--mode.

The forced DNS is the solution to the full incompressible Navier-Stokes equations with the additional right-hand side forcing terms shown below:
\begin{gather}\label{eq: forced_NS}
    \frac{\partial{u_i}}{\partial t} + u_j \frac{\partial u_i}{\partial x_j} + U_j\frac{\partial{u_i}}{\partial{x_j}} + u_j\frac{\partial{U_i}}{\partial{x_j}} = -\frac{\partial{p}}{\partial{x_i}} + \frac{1}{\Rey} \frac{\partial^2u_i}{\partial{x_j}\partial{x_j}} + 2 \Real \left (\kappa \breve \phi_{1, i}(x_2, t) e^{\textrm i \frac{2\pi x_3}{L_3}} \right) + 
    \mathcal F_i, \\\frac{\partial{u_i}}{\partial{x_i}}= 0.
\end{gather}
The term $\boldsymbol{\mathcal F} = (\mathcal F_1, 0, 0)$ enforces the condition that the mean streamwise profile of the forced simulation stay fixed and equal to $U_1$ (\S\ref{sec:base_flow}) as in the unforced simulation, \reviewerone{\emph{i.e.} $\boldsymbol{\mathcal F}$  removes the steady-state ($k_1 = 0$, $k_3 = 0$) contribution of the right-hand side of the Navier-Stokes equations. }
We remind the reader that $\boldsymbol{u} (x_1, x_2, x_3, 0 ) = \boldsymbol{u}_0 (x_1, x_2, x_3, 0 )$ for each ensemble member, where $ \boldsymbol{u}_0 $ is obtained from the unforced DNS. We run the forced DNS for a total time of $T = 5.69 \; \delta / u_\tau$.
% describe the rand term
%We use $\boldsymbol{g}(x_1, x_2, x_3, t)$ to refer to the nonlinear term. 
To test the optimality of $\boldsymbol{\phi_1}$ at forcing the turbulent channel, we repeat the case for $\gamma = 5\%$ using $\boldsymbol{\phi_3}$ and a forcing term with a randomly-generated spatial component $\boldsymbol{\phi}_{\text{\bf{rand}}} = \hat {\boldsymbol{\phi}}_{\text{\bf{rand}}}(x_2) \zeta(t)$, which we normalise and scale the same way ($\Vert {\boldsymbol{\phi}}_{\text{\bf{rand}}} \Vert ^2 = 1$).  

\section{Results and discussion} \label{results}
In this section, the notation $\widehat{(\cdot)}^{(k_1, k_3)}$ denotes the Fourier transform in the streamwise and spanwise directions corresponding to the streamwise and spanwise wave numbers $2\pi k_1/L_1$ and $2\pi k_3/L_3$, respectively. 
We define the deviation operator $\Delta$ as the difference between the forced and unforced simulations, e.g., $\Delta \hat{u}_1^{(0,1)} = \hat u_1^{(0,1)} - \hat u_{0,1}^{(0,1)}$. 
%The velocity deviation field is given by $\boldsymbol{\hat q} = [\Delta \hat u_1^{(0,1)}, \Delta \hat u_2^{(0,1)}, \Delta \hat u_3^{(0,1)}]^T$, and the deviation in the nonlinear terms by $\boldsymbol{\hat f} = [\Delta \hat g_1^{(0,1)}, \Delta \hat g_2^{(0,1)}, \Delta \hat g_3^{(0,1)}]^T$, where $g_i$ is the nonlinear term of the momentum equation in the $i^{th}$ direction. 
We denote the ensemble average by $\overline{ (\cdot) }$.

\subsection{Transient energy growth and decay of streaks in the forced DNS}\label{transientEnergyGrowth}

\begin{figure}
    \centering
    \subfloat[]{
    \includegraphics[width=0.44\textwidth]{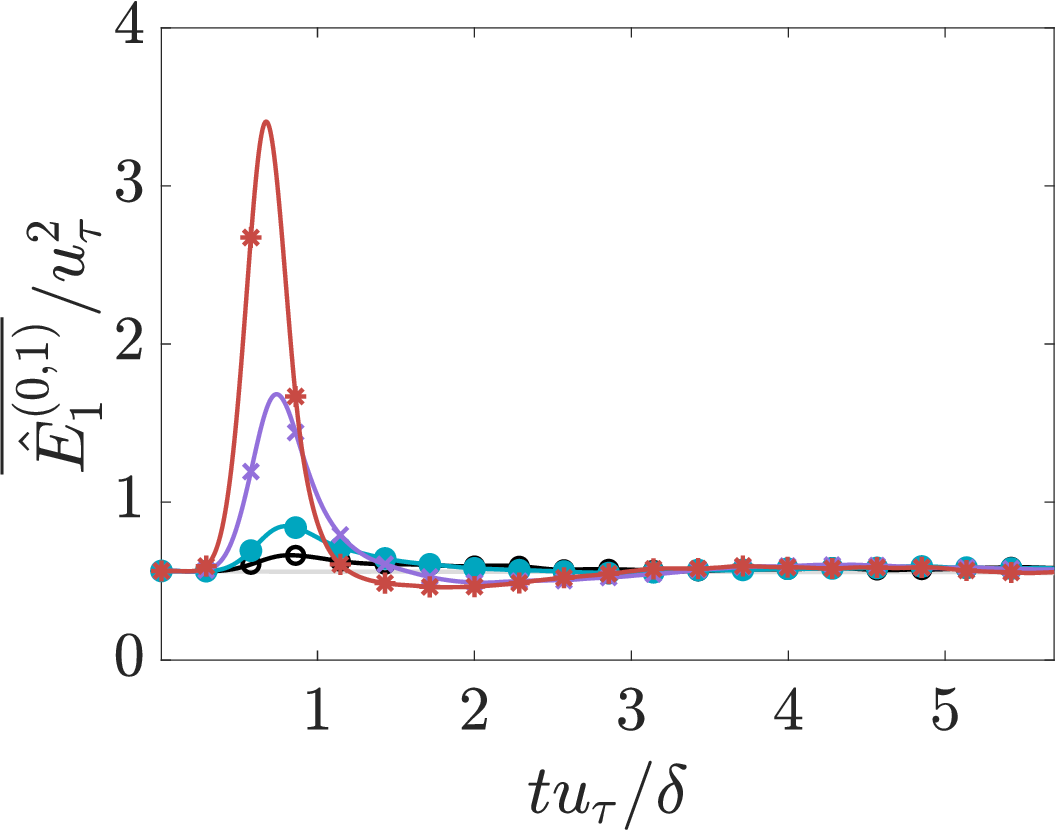}} 
    \subfloat[]{
    \includegraphics[width=0.45\textwidth]{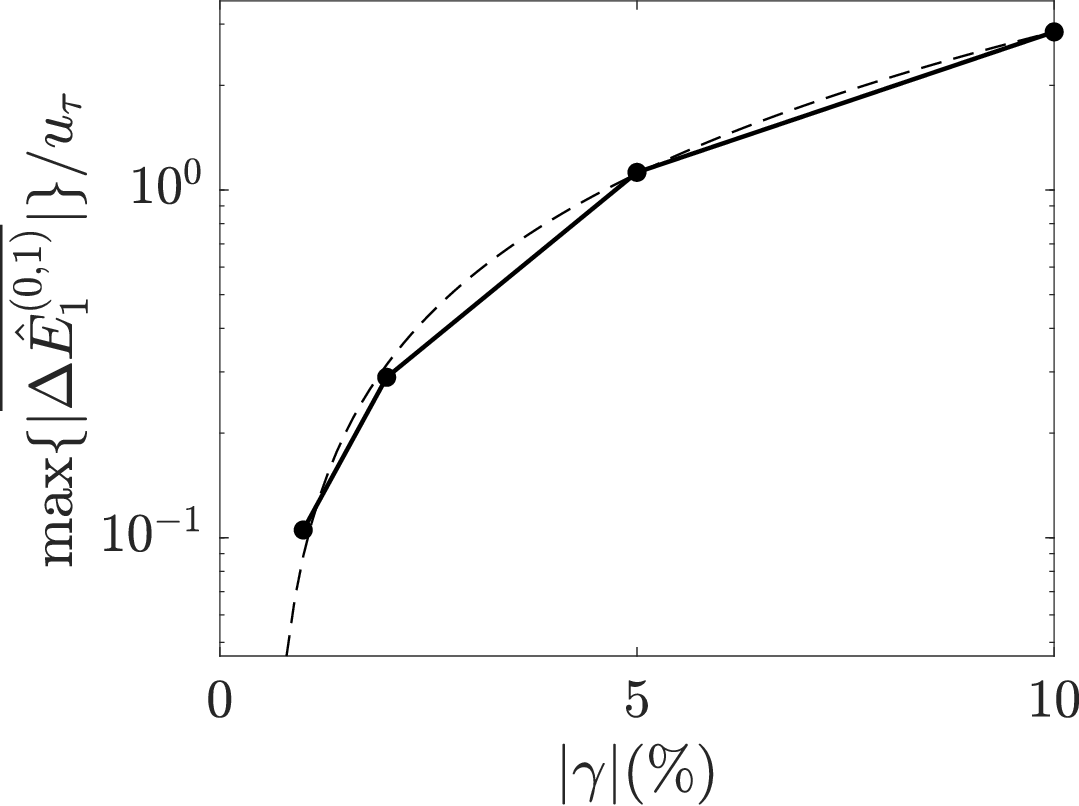}
    } 

    \subfloat[]{
       \includegraphics[width=0.45\textwidth]{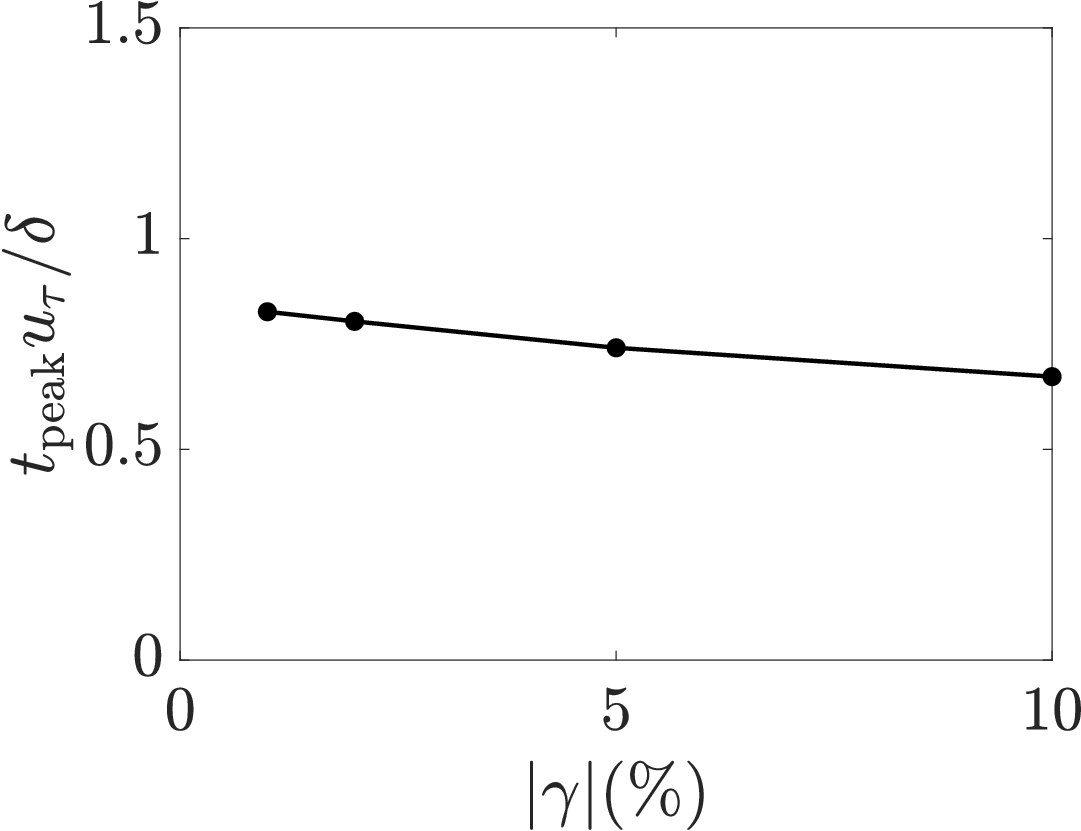}}
    \subfloat[]{
        \includegraphics[width=0.45\textwidth]{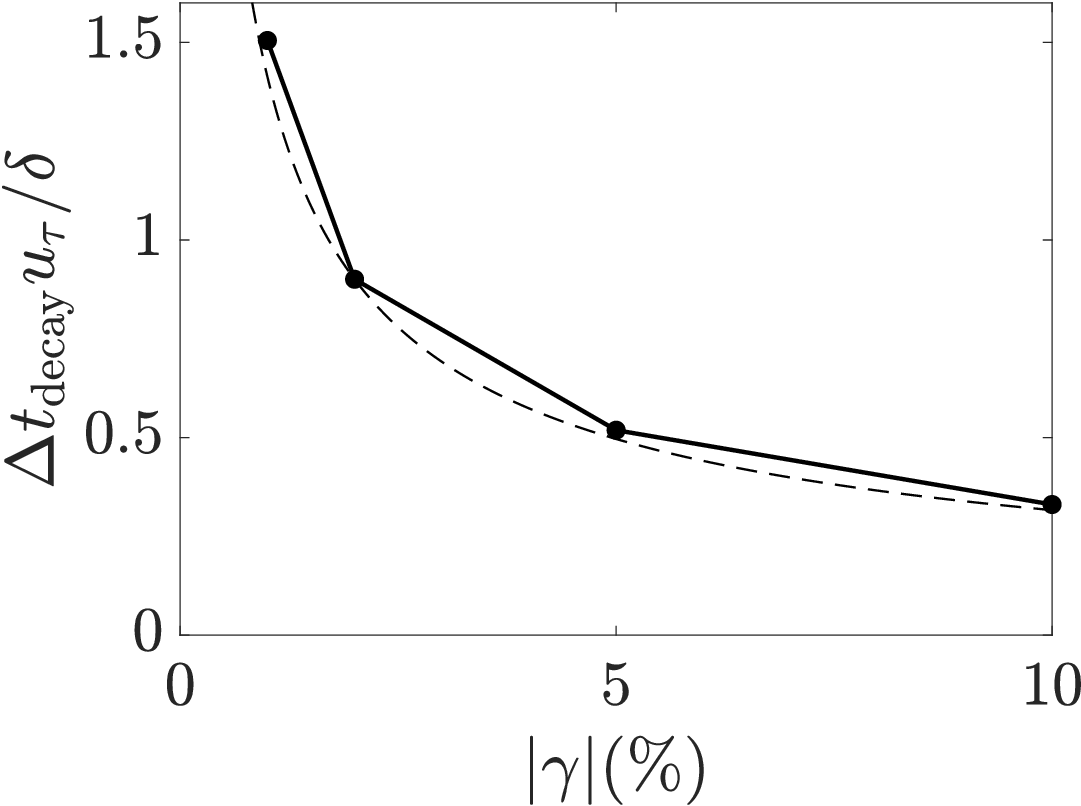}}
    \caption{(a) Average streak energy as a function of time; the cases plotted are $\gamma = 1\%$ (black $\circ$), $\gamma = 2\%$ (cyan $\bullet$), $\gamma = 5\%$ (purple $\times$), and $\gamma = 10\%$ (red $*$). (b) Streak energy peaks, (c) peak times, and (d) decay times as a function of $\gamma$. The dashed line represent the trends (b) $80.48 |\gamma|^{1.44}$, and (d) $|\gamma|^{-0.65}$
    %$\times$ denotes the streak energy in the unforced case. The dashed line in (c) represents the trend $|\gamma|^{-0.65}$. 
    }
    \label{fig:transientStats}
\end{figure}

We define the instantaneous streak energy as 
\begin{equation}
    \hat E_1^{(0,1)}(t) = \left[ \frac{|\hat u_1^{(0,1)}|^2}{2} \right ],
\end{equation}
where $[\cdot] := \int_0^{2\delta} (\cdot) dx_2 / (2\delta)$ denotes the wall-normal average. Figure \ref{fig:transientStats}(a) shows the streak energy contained in the $(0, 1)$-mode as a function of time, for different resolvent-forcing amplitudes. For all cases, the energies grow and peak before decaying and reverting back to non-forced levels.
We observe that the peak energy deviation $\Delta \hat E_1^{(0,1)}$ scales sub-quadratically with the forcing amplitudes and is proportional to $ | \gamma| ^{1.44}$ (figure \ref{fig:transientStats}(b)). For a linear system, the energy peaks would scale quadratically, which indicates that the nonlinearities 
\reviewerone{cause the energy to start decaying before it can reach the peak allowed by optimal linear growth.}%lead to the premature peaking of streak energy.
The stronger the forcing, the faster the streak energy's growth rate, and the faster its decay. 
The peak times, $t_{\text{peak}}$, defined as the times at which the energies reach their maxima, decrease slightly with forcing amplitude, but are relatively constant compared to the decay times, $\Delta t_{\text{decay}}$, which we define as the time it takes for the energy to reduce from the peak to $10\%$ of its peak (figures \ref{fig:transientStats}(c), \ref{fig:transientStats}(d)). Indeed, the differences in decay rate are more dramatic across the different forcing amplitude cases, and scale as $\Delta t_\text{decay} \sim |\gamma|^{-0.65}$.
We note that all fully-coupled simulations decay significantly faster ($\Delta t_\mathrm{decay} \approx 1 \delta / u_\tau$) than the linear response ($\Delta t_\mathrm{decay}\approx 15 \delta / u_\tau$).

\begin{figure}
    \centering
    \includegraphics[width=0.5\linewidth]{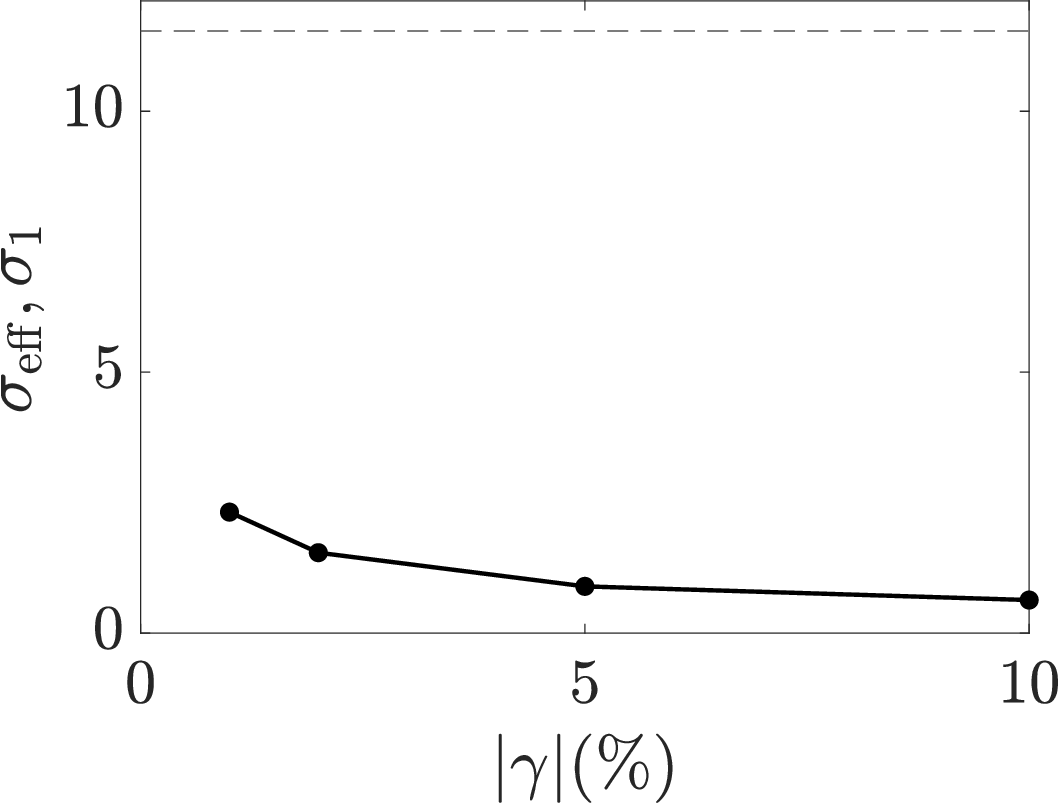}
    \caption{Effective amplification $\sigma_{\text{eff}}$ (solid) and  $\sigma_1$ (dashed).}
    \label{fig:sigmaEffective}
\end{figure}

To measure the proportion of the linearly-amplified energy captured by the forced simulations, we compute an ensemble-averaged forcing efficiency, or effective amplification, which we define to be 
\begin{equation}
    \sigma_\text{eff} = \left(\, \frac{1}{u_\tau \delta}  \int_{0}^{T} \overline{\frac{ \Delta \hat E^{(0,1)}(t)}{|\kappa|^2/2} } \mathrm{d}t \,\right)^{{1}/{2}}.
\end{equation}
This is analogous to $\sigma_1 = \max_{\mathsfbi{\tilde{f}}} \|\tilde{\mathsfbi{H}}^{(0, 1)} \mathsfbi B \mathsfbi{\tilde{f}} \|_2 / \|\mathsfbi{\tilde{f}}\|_2$, where the numerator reflects the energy contained in the velocity perturbation field, and the denominator corresponds to the forcing amplitude. The computed $\sigma_\text{eff}$ is shown in figure \ref{fig:sigmaEffective}. The forcing efficiencies decrease with the intensity of the forcing, and all effective amplifications are lower than $\sigma_1 =  11.54$, \reviewers{indicating that, across forcing amplitudes, the resolvent response mode overpredicts the response of the system to the injected forcing, even for the low-amplitude cases. We expect the agreement to be improved when equations \eqref{eq:NS} are augmented by an eddy viscosity model \citep{zare2017colour, symon2023eddy}.}
Regarding the trend of $\sigma_1$ with forcing amplitude, we see that for smaller resolvent forcing amplitudes, more of the forcing energy is linearly converted into streak energy. 
For higher values of $\gamma$, nonlinear interactions that scale superlinearly curtail the growth of the response mode -- the integrated action of which scales linearly with $\gamma$ in the linearised setting --and hinder the effectiveness of the resolvent forcing mode.

\subsection{Comparison of velocity deviations with linear response}

\begin{figure}
    \centering
    \subfloat[]{
        \includegraphics[width=0.45\textwidth]{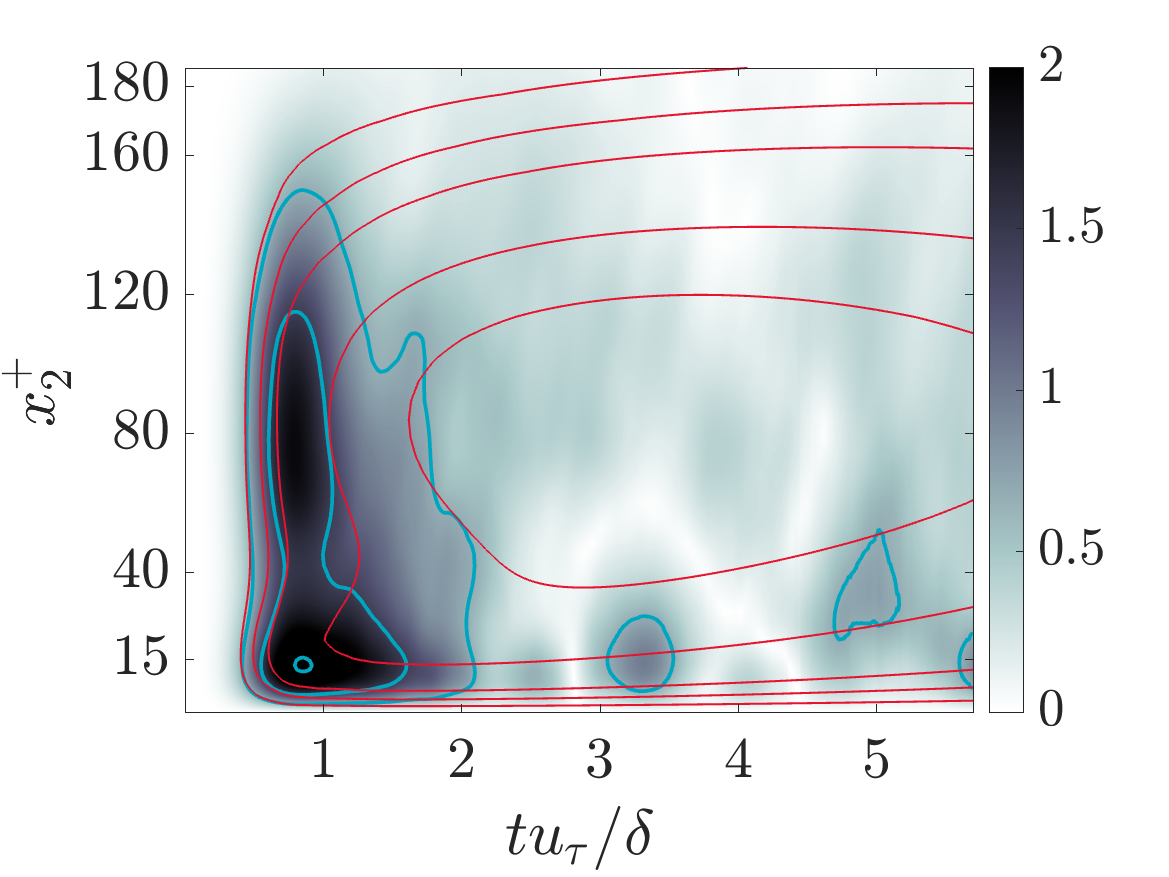}}
    \subfloat[]{
        \includegraphics[width=0.45\textwidth]{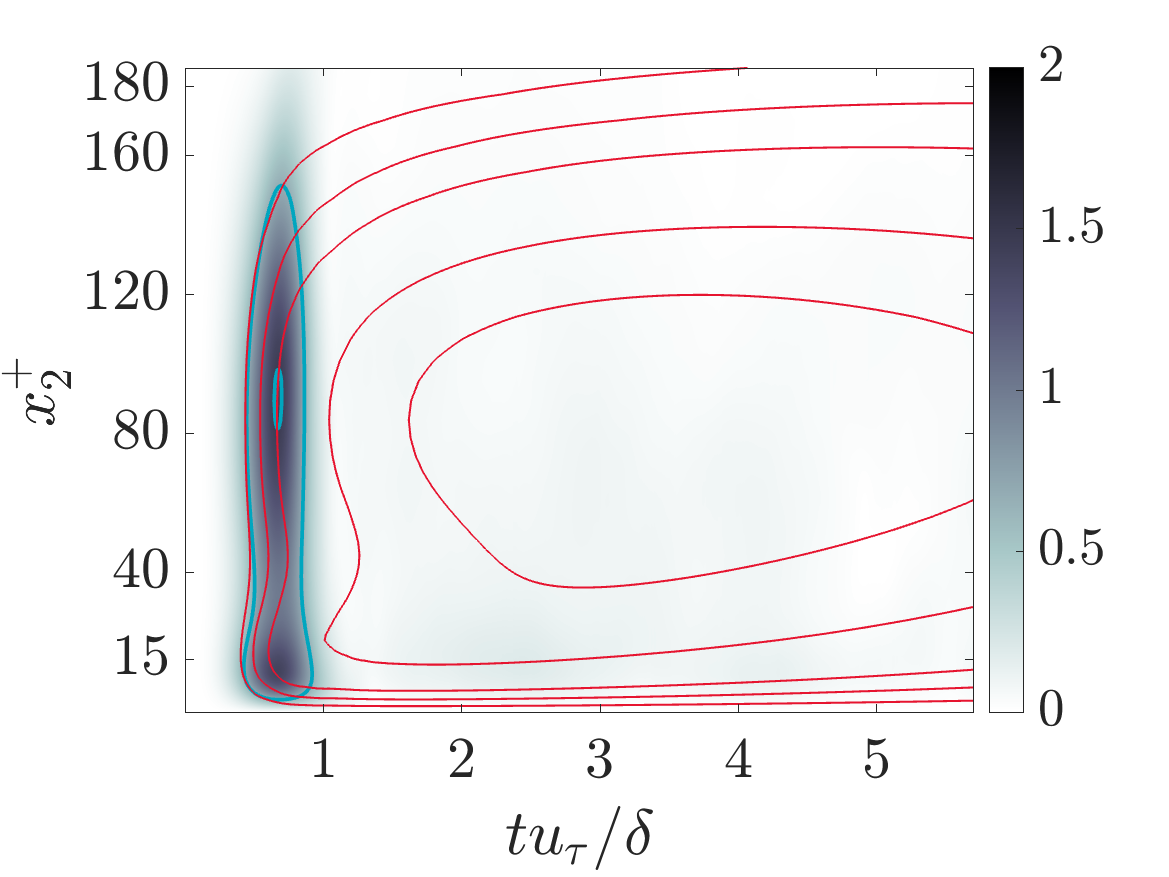}}
    \caption{Average deviation in the streamwise velocity of the $(0,1)$-Fourier mode, $|\overline{\Delta\hat{u}^{(0,1)}/\kappa}|/u_\tau$, for the (a) $\gamma = 1\%$ and (b) $\gamma = 10\%$ cases. The contours correspond to $7\%$, $15\%$, $25\%$, $75\%$ and $90\%$ of the maximum value of $\sigma_1 \boldsymbol{\psi_1}$. The lines represent the forced DNS case (blue), and the resolvent response (red).} 
    \label{fig:uContours}
\end{figure}

\begin{figure}
    \centering
    \subfloat[]{
        \includegraphics[width=0.45\textwidth]{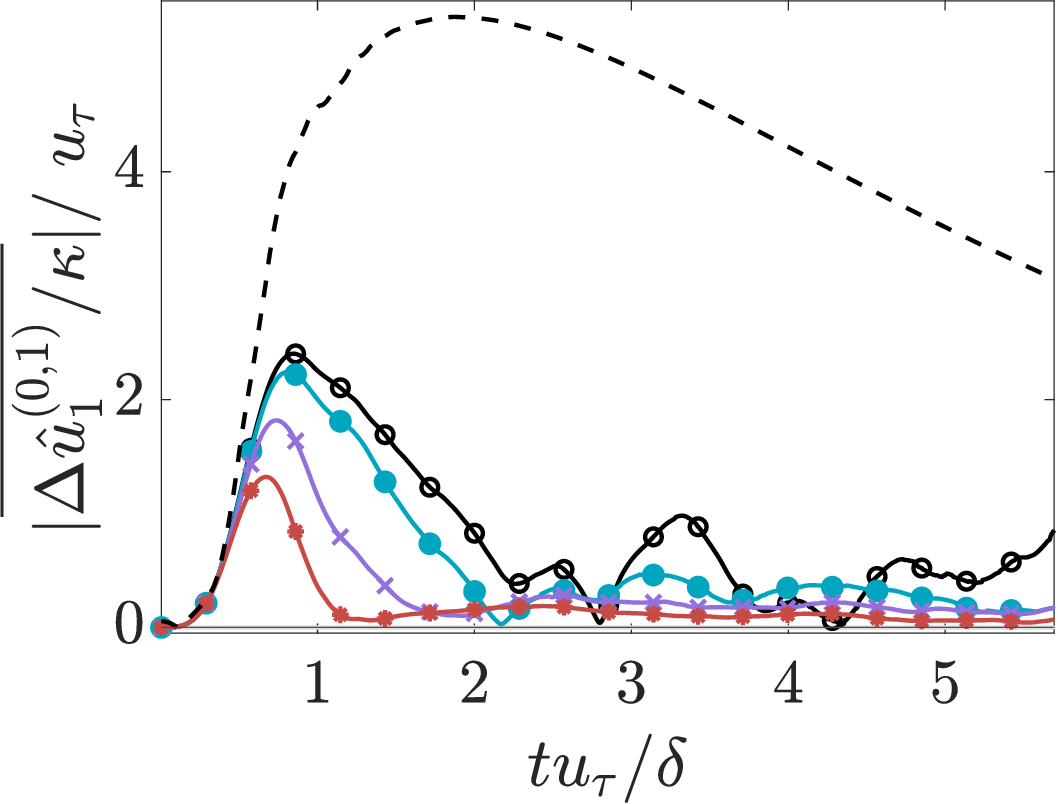}}
     \subfloat[]{
        \includegraphics[width=0.45\textwidth]{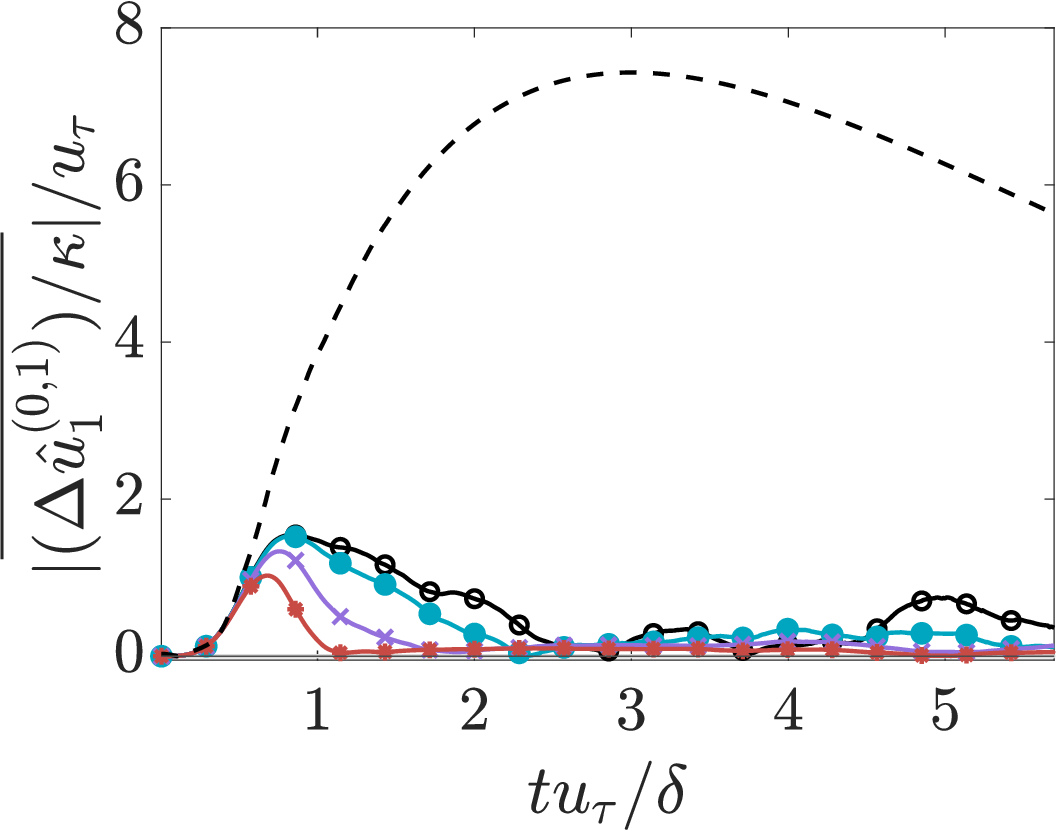}}
    \caption{ 
    Average deviation in the streamwise velocity of the $(0,1)$-Fourier mode at $x_2^+ \approx 16$ (a) and $x_2^+ \approx 39$ (b). The cases plotted are $\gamma = 1\%$ (black $\circ$), $\gamma = 2\%$ (cyan $\bullet$), $\gamma = 5\%$ (purple $\times$), $\gamma = 10\%$ (red $*$), and the linear response mode $\sigma_1 \boldsymbol{\psi_1}$ (black, dashed).}
    \label{absU}
\end{figure}

To visualise the alignment of the velocity fluctuation fields with the linear response mode across all wall normal heights, we plot the contours of $|\overline{\Delta\hat{u}^{(0,1)}/\kappa}|/u_\tau$, \emph{i.e.} the magnitude of the $\hat u_1^{(0,1)}$ deviations normalised by the forcing coefficient $\kappa$, along with the contours of the linear resolvent response modes (figure \ref{fig:uContours}). We note that $\Delta\hat{u}^{(0,1)}$ is divided by the complex value of $\kappa$ prior to ensemble averaging so as to align the phases of the forcing (\S \ref{dns}) across the ensemble members.
At earlier times ($t < 0.7 \delta / u_\tau $), the responses for both the $\gamma = 1\%$ and $\gamma = 10\%$ cases are very similar to the linear mode. The strongly-forced case, however, quickly reverts to the unforced channel flow statistics beyond an eddy turnover time unit. 
In contrast, the lightly forced case of $\gamma = 1\%$ exhibits a longer-lasting velocity deviation, especially in the near wall region ($x_2^+ < 15$).
%% TO DO: add explanation
To more closely investigate how the agreement of the forced simulations and the optimal linear response varies with $x_2^+$, we show $\Delta \hat{u}_1^{(0,1)}$ at two wall-normal heights, along with the linear response at those heights (figure \ref{absU}).
At both wall-normal locations, the initial growth rates are similar to the linear case for all $\gamma$, and we obtain good collapse prior to $t \approx 0.7 \delta /u_\tau$.  \reviewertwo{However, increasing the forcing amplitude causes the $\kappa$--normalised velocity deviation to decay earlier and at a smaller amplitude for both wall-normal heights shown in figure \ref{absU}.}
For a fixed forcing amplitude, we note that $\Delta \hat{u}^{(0,1)}$ diverges from the optimal linear response around the same time at both wall-normal locations plotted in figure \ref{absU}, but 
the agreement between the forced simulations and the optimal linear response turns out better in the near-wall region. 
As $x_2^+$ moves closer to the wall, the growth rate due to linear mechanisms increases, and $\Delta \hat{u}_1^{(0,1)}$ manages to recover more of the linearly-amplified energy before decaying due to nonlinear effects.

\mycomment{
\begin{figure}
    \centering
         \subfloat[]{
        \includegraphics[width=0.45\textwidth]{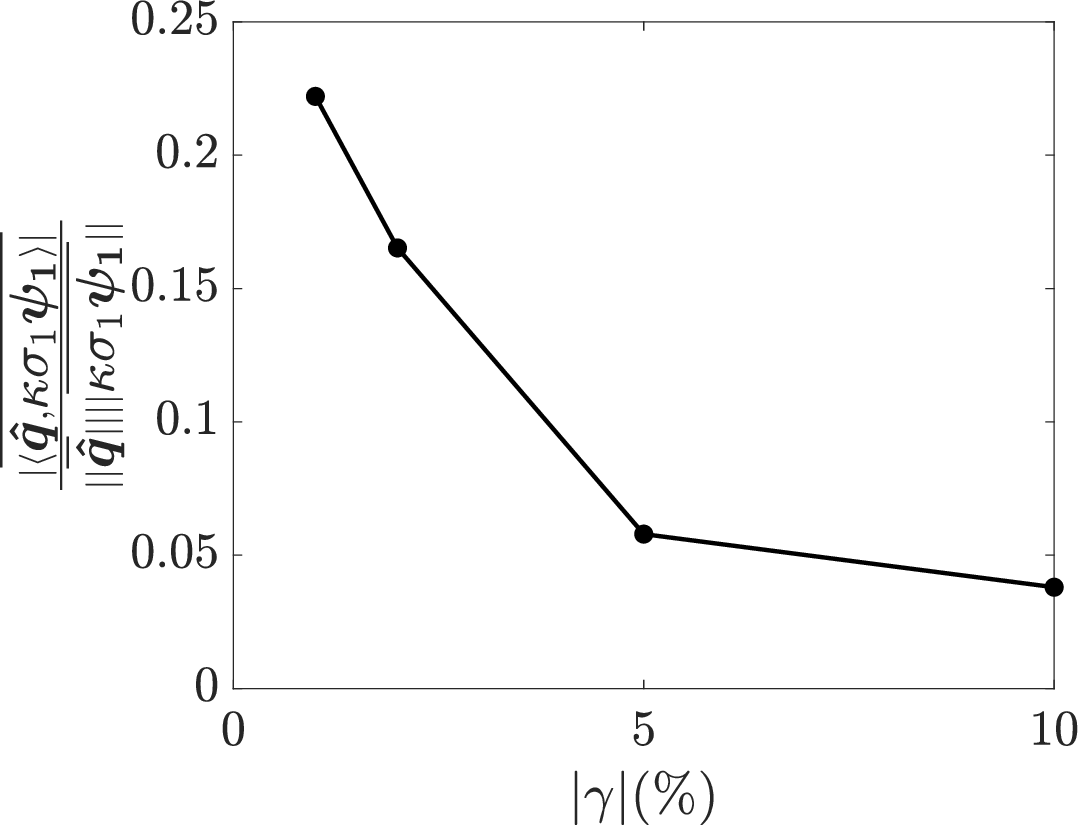}}
        \subfloat[]{
        \includegraphics[width=0.45\textwidth]{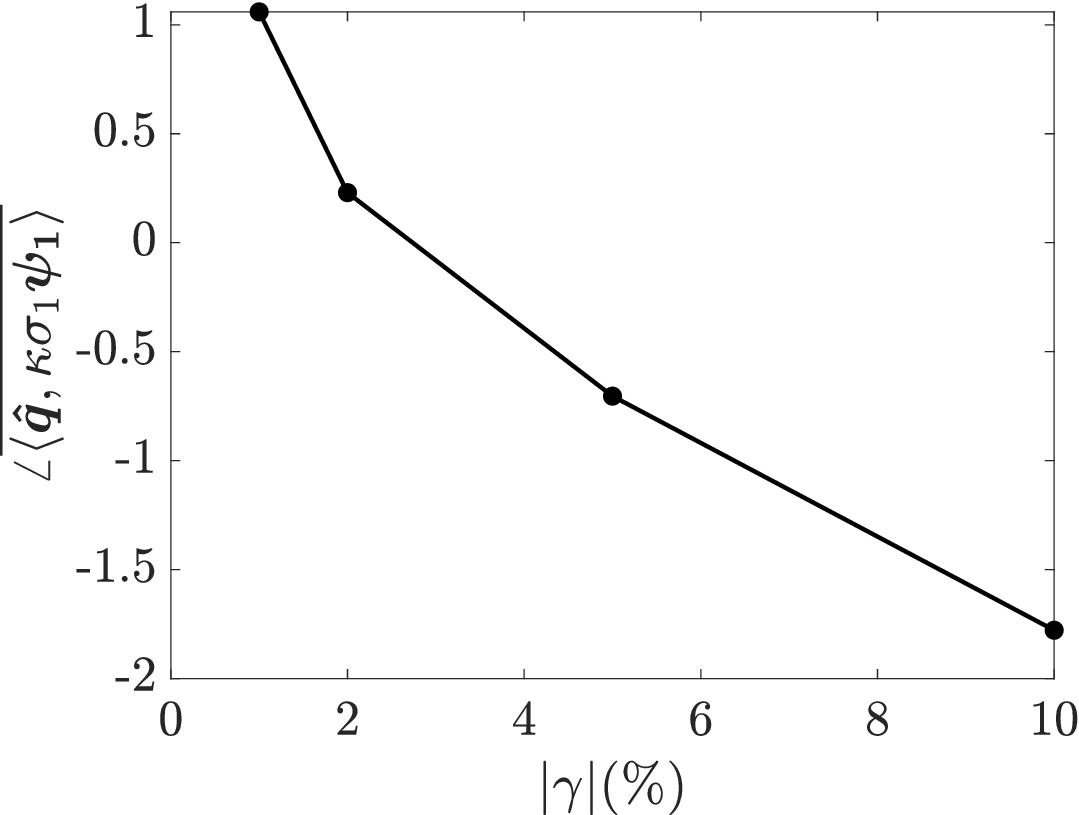}}
        
         \subfloat[]{
        \includegraphics[width=0.45\textwidth]{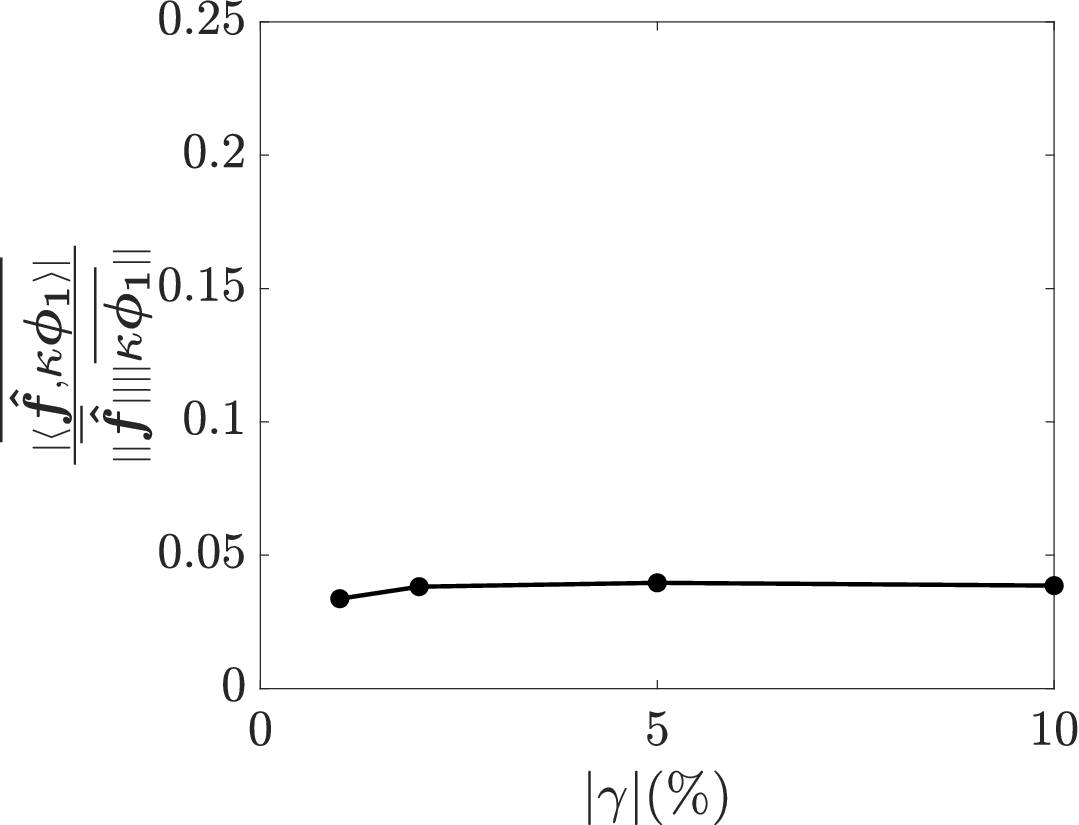}}
         \subfloat[]{
        \includegraphics[width=0.45\textwidth]{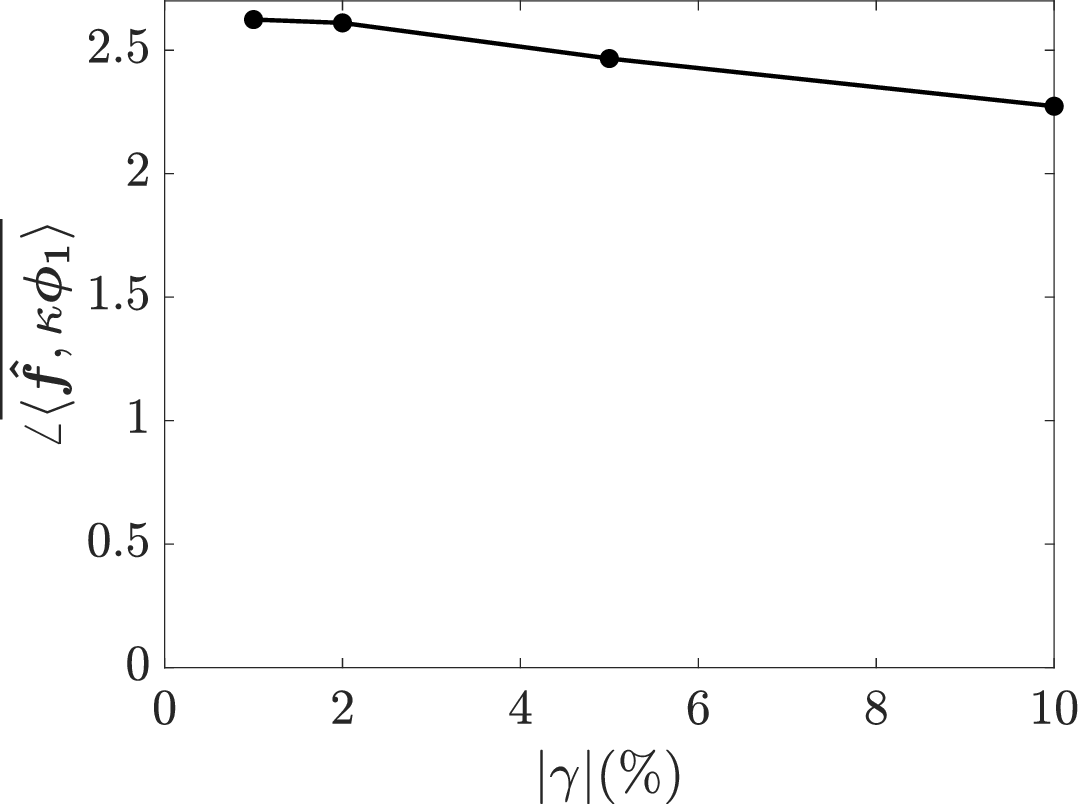}}
    \caption{Magnitude (a) and phase (b) of the average projection of the velocity field deviation onto the principle resolvent response mode $\boldsymbol{\psi_1}$. Magnitude (c) and phase (d) of the average projection of the deviation in the nonlinear terms onto the principle resolvent forcing mode $\boldsymbol{\phi_1}$.}
    \label{fig:responseAlign}
\end{figure}
}

\begin{figure}
    \centering
    \includegraphics[width=0.45\textwidth]{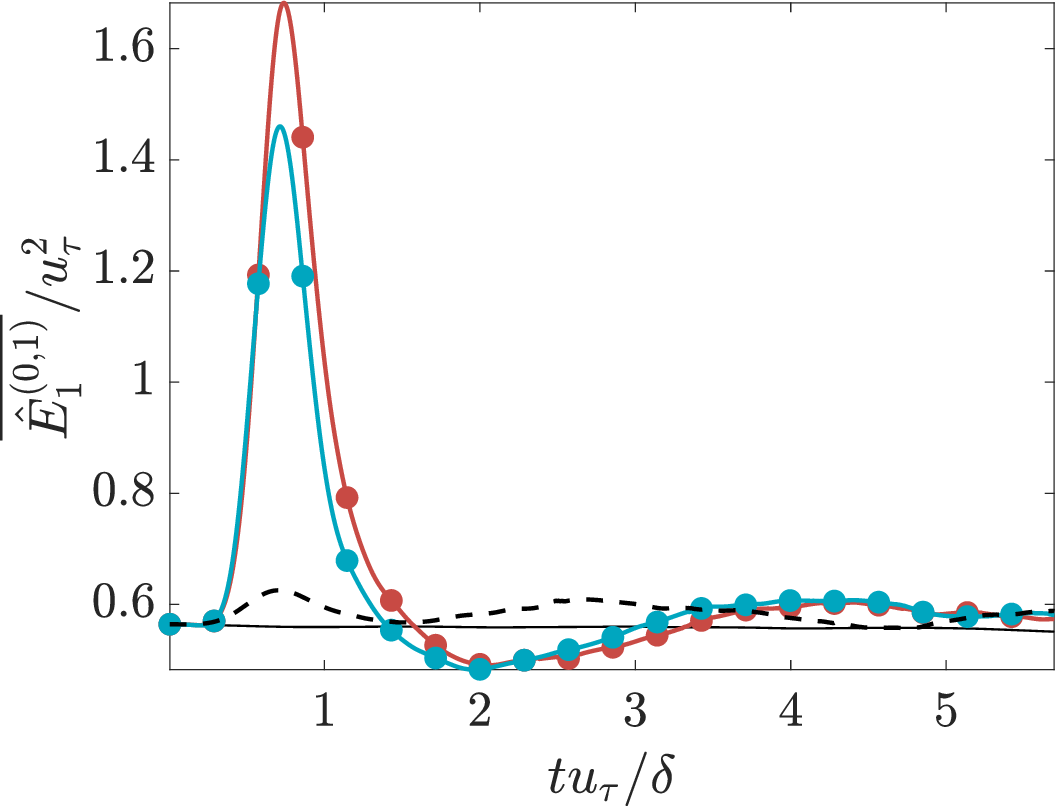}
    \caption{Average streak energy for cases forced by $\boldsymbol{\phi_1}$ (red $*$), $\boldsymbol{\phi_3}$ (cyan $\bullet$), and $\boldsymbol{\phi}_\mathrm{\bf rand}$ (dashed black). The unforced case is shown in black.}
    \label{fig:optimality}
\end{figure}

\begin{figure}
    \centering
    \subfloat[]{
        \includegraphics[width=0.31\textwidth]{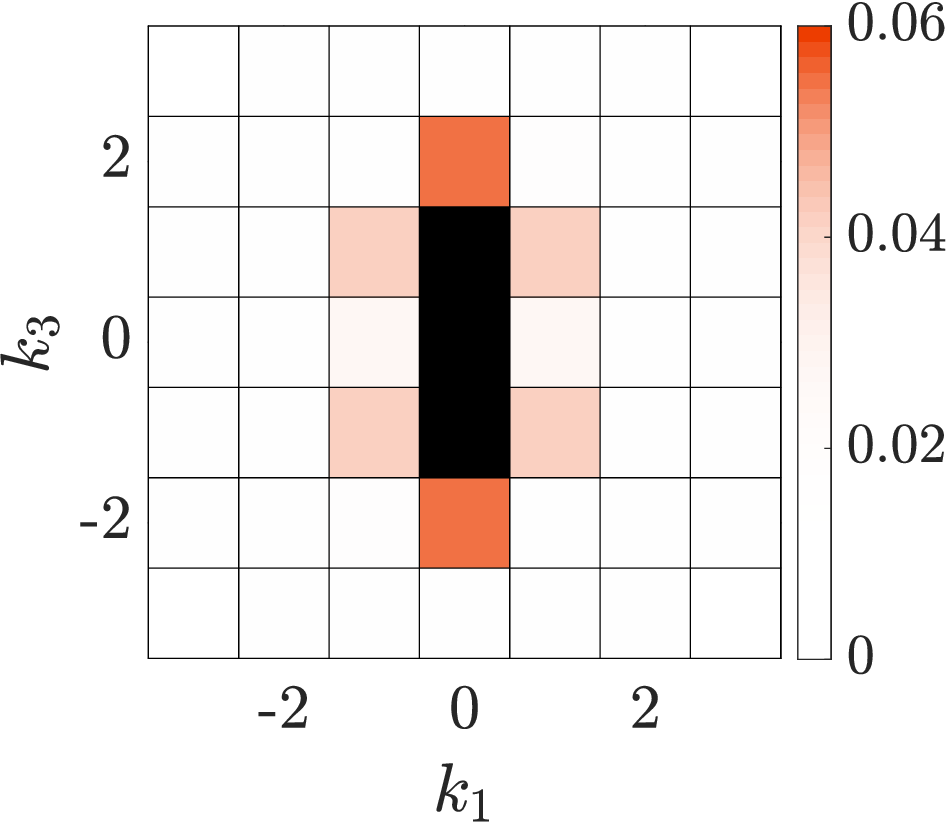}}
    \subfloat[]{
        \includegraphics[width=0.31\textwidth]{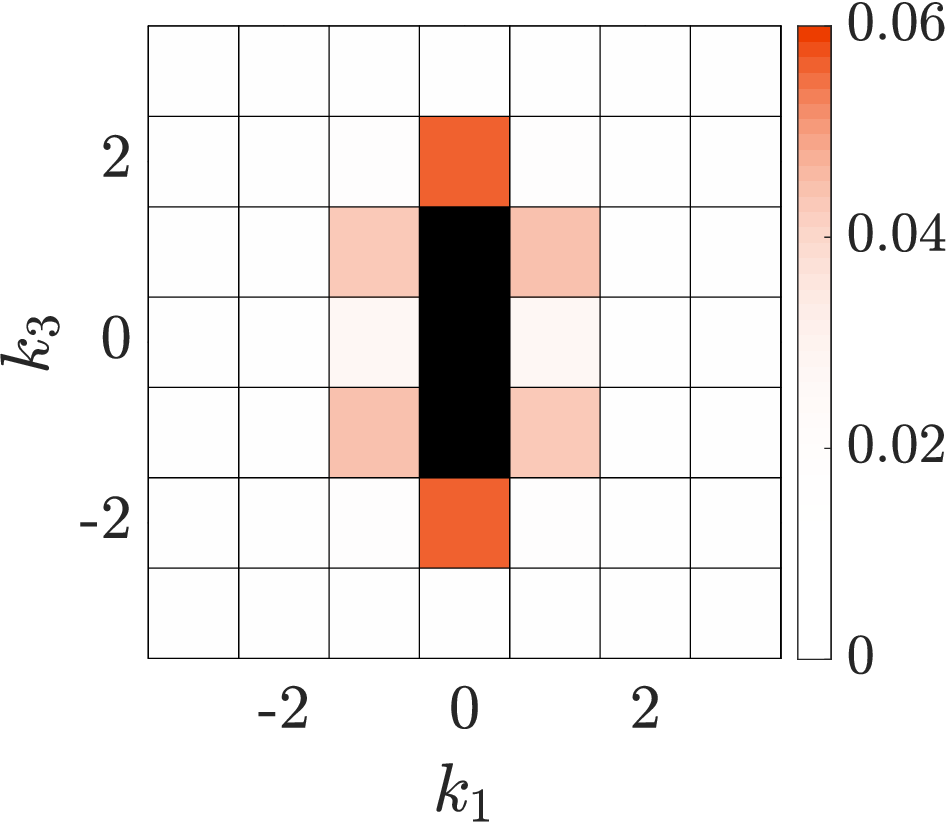}}
    \subfloat[]{
        \includegraphics[width=0.31\textwidth]{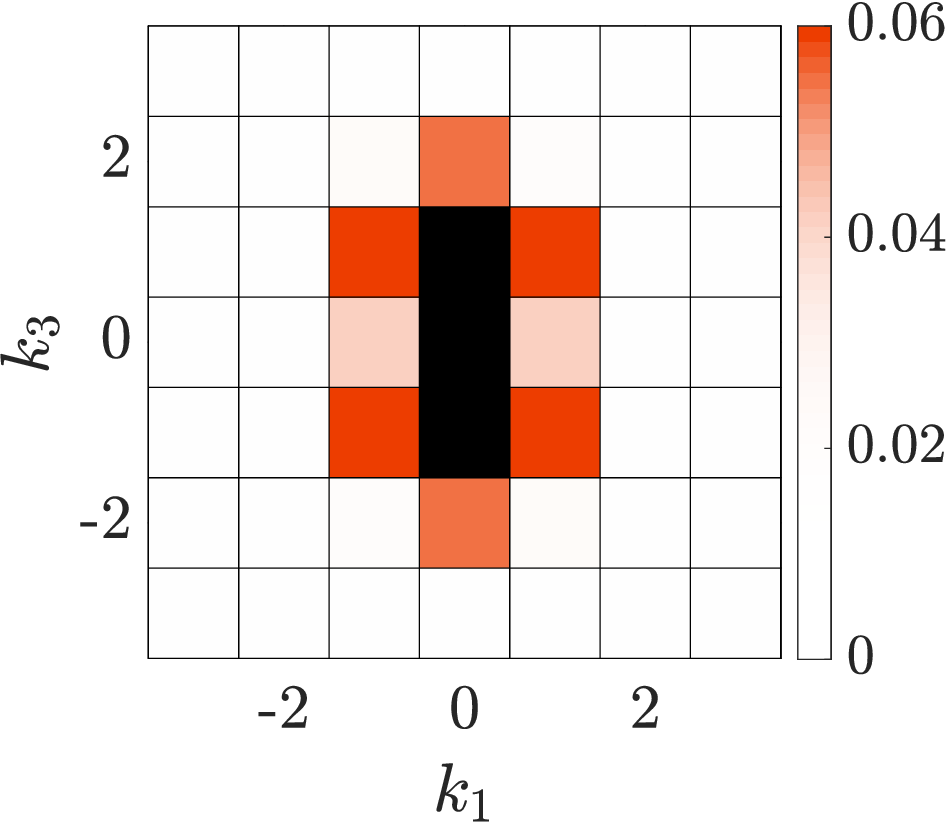}}
        
    \subfloat[]{
        \includegraphics[width=0.31\textwidth]{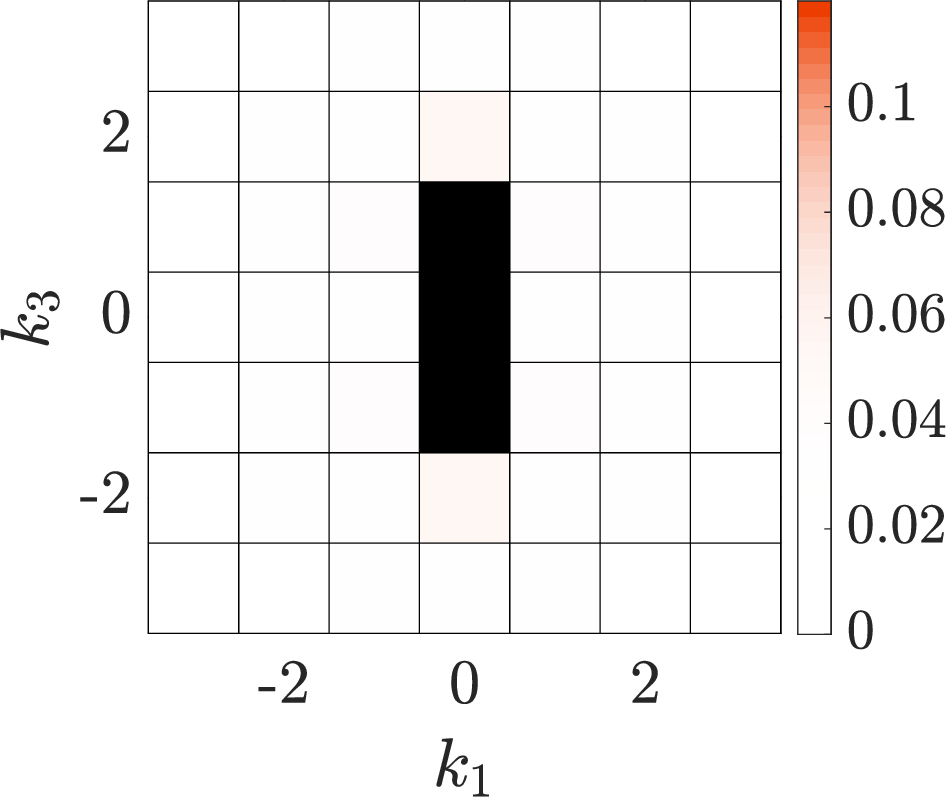}}
    \subfloat[]{
        \includegraphics[width=0.31\textwidth]{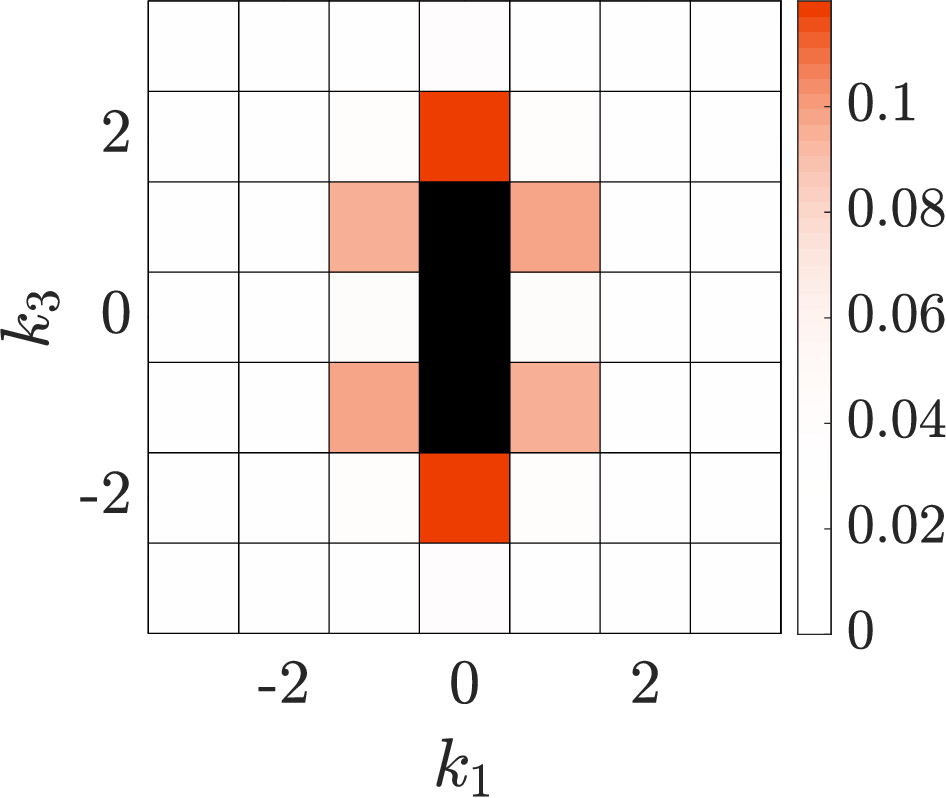}}
    \subfloat[]{
        \includegraphics[width=0.31\textwidth]{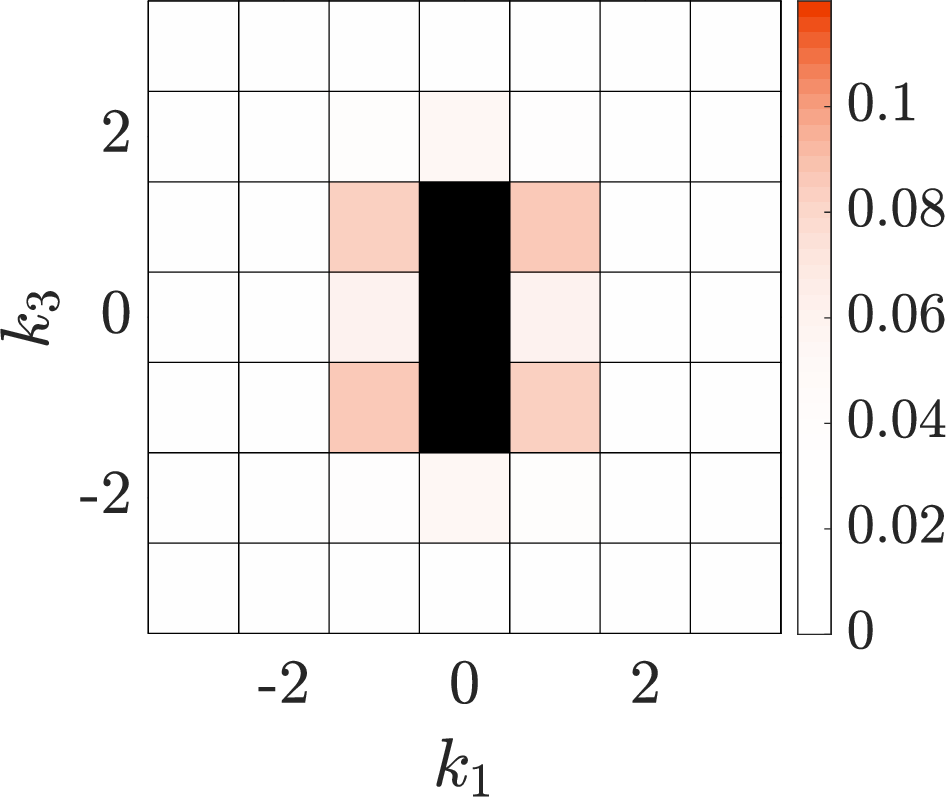}}
    \caption{Integrated streamwise spectral energy content $\overline{[|\hat{u_1}^{(k_1,k_3)}|^2]}/(2  u_\tau^2)$ for $\gamma = 2\%$ (top) and $\gamma = 10\%$ (bottom). The spectra shown are at times $t = 0 $ (left), $t = 0.6 \delta/u_\tau$ (middle) and $t = 1.2 \delta/u_\tau$ (right). The energy contained in the $(0,0)$, $(0,1)$ and $(0,-1)$--modes (black) are excluded for clarity.}
    \label{fig:SpectrumFixedTimes}
\end{figure}

\begin{figure}
    \centering
    \subfloat[]{
    \includegraphics[width=0.45\textwidth]{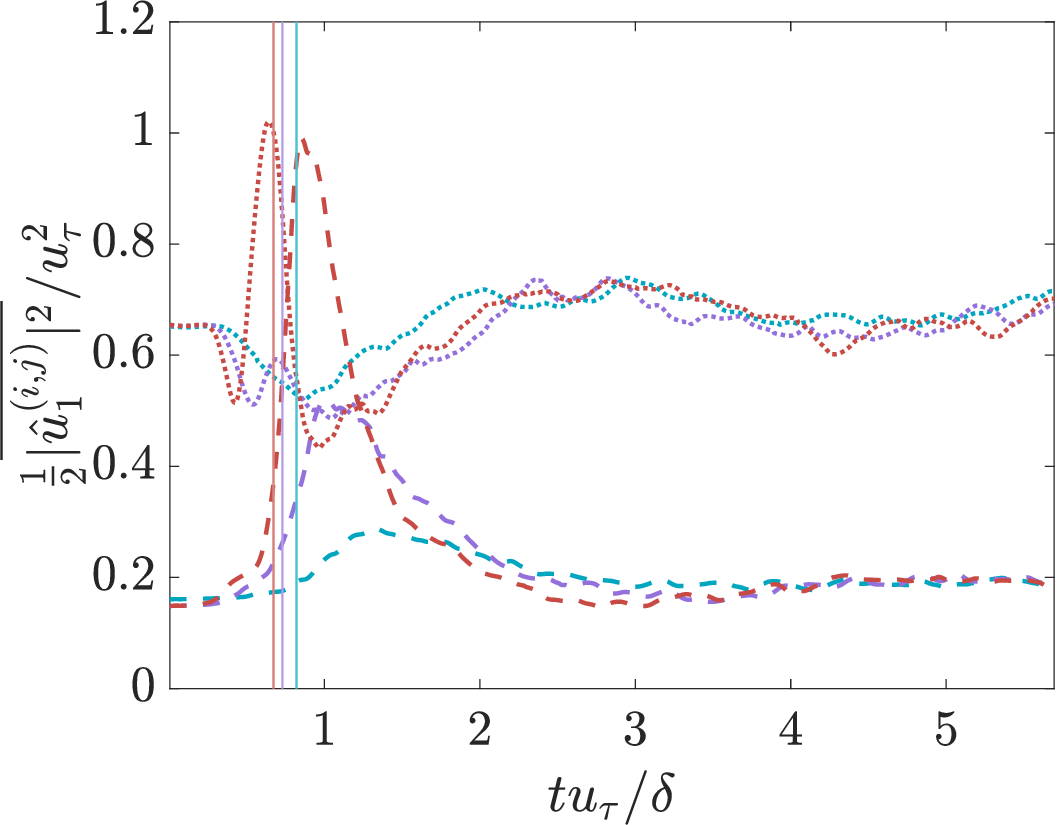}}
    \subfloat[]{
    \includegraphics[width=0.45\textwidth]{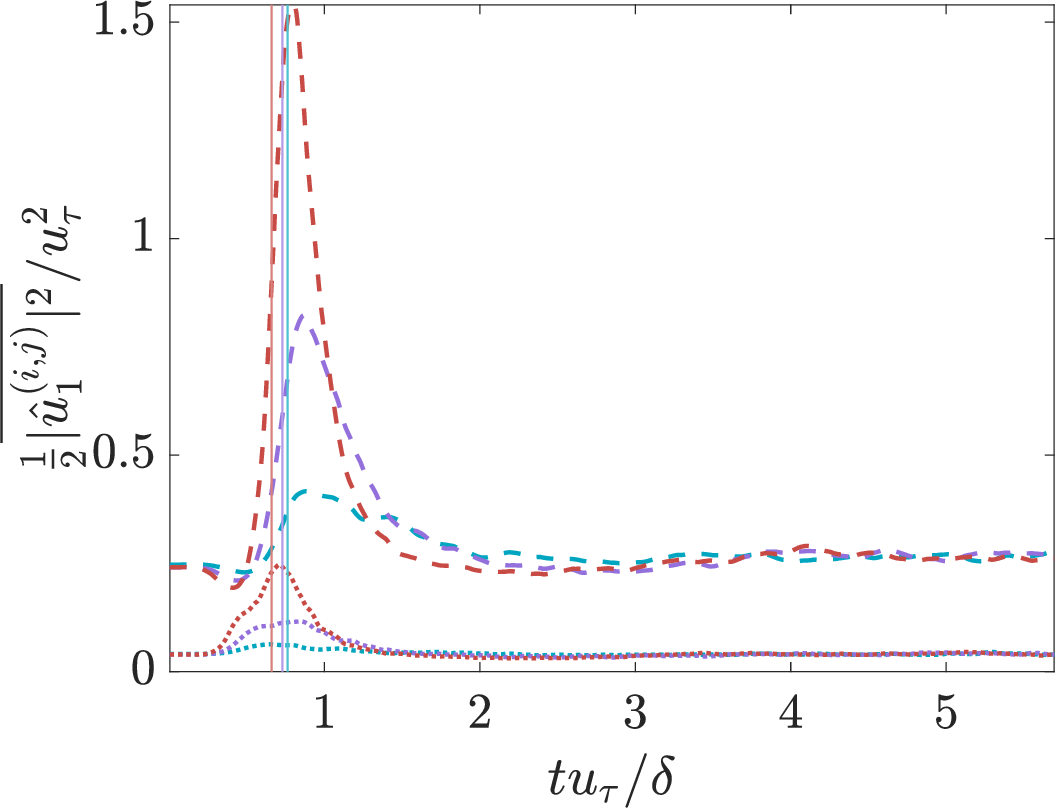}}
    
    \subfloat[]{
    \includegraphics[width=0.44\textwidth]{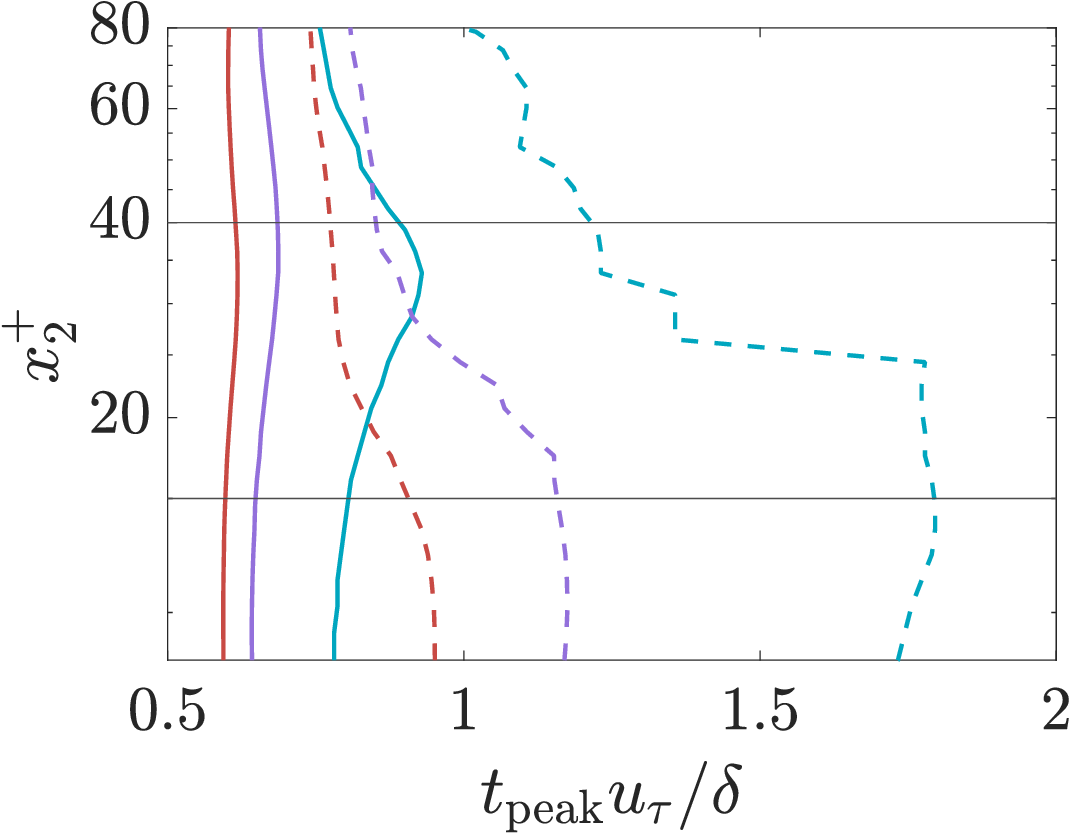}}
    \subfloat[]{
    \includegraphics[width=0.44\textwidth]{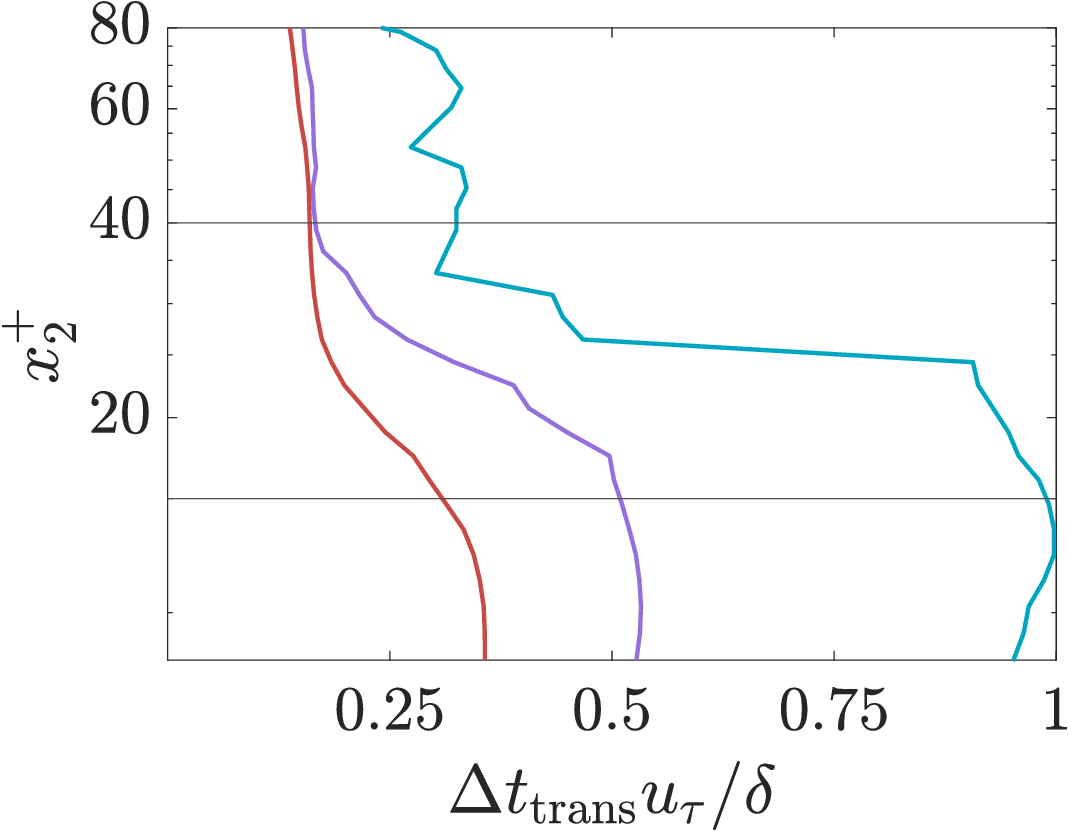}}
    
    \caption{Streamwise spectral energy content at (a) $x_2^+ \approx 16$  and (b) $x_2^+ \approx 39$  for the $(0,\pm 2)$--Fourier mode (dotted lines) and the $(\pm 1,\pm 1)$--Fourier mode (dashed lines).  
    (c) Energy peak times for the $(0, \pm1)$--mode (solid) and the smaller scales (dashed) cases. 
    (d) Timescale for the cross-scale energy transport. Colors indicate $\gamma = 2\%$  (blue), $\gamma \approx 5\%$ (purple), and $\gamma = 10\%$ (red). The vertical lines in (a) and (b) represent the local peaks of $|\hat u_1^{(0,1)}|$ The horizontal black lines in (c) and (d) represent $x_2^+ = 15$ and $x_2^+ = 40$, which delineate the buffer layer.}
    \label{fig:time_transfer_to_small_scales}
\end{figure}

\subsection{Optimality of resolvent forcing}
We compare the results of forcing using the principal resolvent forcing mode $\boldsymbol{\phi_1}$, the first suboptimal mode $\boldsymbol{\phi_3}$, and the random mode $\boldsymbol{\phi}_{\text{\bf{rand}}}$, \reviewerthree{using $\gamma = 5\%$}.
We see that the streak energy grows to higher peak when the minimal flow unit is forced by $\boldsymbol{\phi_1}$ than when forced by $\boldsymbol{\phi_3}$ (figure \ref{fig:optimality}). In both cases, the minimal flow unit is much more responsive compared to the case with random forcing. 
This suggests that resolvent analysis identifies a forcing structure to which the minimal flow unit is indeed sensitive, even when governed by the fully nonlinear Navier-Stokes equations. 
The advantage of the optimal forcing mode, however, is significantly reduced. The effective amplifications, $\sigma_\text{eff}$ in DNS forced by $\boldsymbol{\phi_1}$ and $\boldsymbol{\phi_3}$ differ only by a factor of $1.03$, whereas $\sigma_1/\sigma_3 = 2.16$.
%Though the initial growth of the streak energy is faster for the case forced by $\boldsymbol{\phi_1}$, the induced streak in both resolvent-forced cases triggers nonlinear effects at roughly the same time. 
Nonlinear effects quickly clip the transient growth of the induced streak and drain energy from the actuated mode before the differences between the responses to $\boldsymbol{\phi_1}$ and $\boldsymbol{\phi_3}$ can deepen.
 
We remind the reader that the mean profile is fixed by an additional forcing $\boldsymbol{\mathcal F}$ which removes the $(0, 0)$ contribution to $\partial {\boldsymbol u}/\partial t$ at every time step. Freezing the mean replenishes the energy in the $(k_1, k_3) = (0, 0)$ mode and preserves the energy transfer mechanism to the resolvent response mode. In other words, allowing the mean profile to vary may reduce the effectiveness of the forcing mode, since the mode is only optimal for the mean profile without forcing. 
This is observed in \citet{bae2021nonlinear}: subtracting the contribution of the resolvent forcing mode from the nonlinear term succeeds at reducing the turbulent kinetic energy initially, but this effect fades after $t u_\tau / \delta \approx 12$ as the mean flow is modified. 
However, our injected forcing has a short extent in time ($t u_\tau / \delta  \approx 1.5$), and we do not expect the initial linearly driven growth of the induced response mode to change much whether we fix the mean profile or not. Though not shown, the streak energy profile obtained for $\gamma = 5\%$ and an unfixed mean velocity is indeed very close to the what is shown in figure \ref{fig:transientStats}(a). 
%Changes to the results would depend on the nonlinear energy transfer to the $(0, 0)$--mode to change, which

\subsection{Spectra}

We compute the $x_2$--integrated streamwise and spanwise spectra 
for $\gamma = 2\%$ and $\gamma = 10\%$, at the initial time, right before the peak in streak energy ($t = 0.6 \delta/ u_\tau$), and during the streak energy decay ($t = 1.2 \, \delta/ u_\tau$). The results are shown figure \ref{fig:SpectrumFixedTimes}. Though not plotted, we note that the $(0, \pm 1)$--mode is the most dominant one across all times, accounting for $30\% - 70\%$ of the total turbulent energy over the entire simulated time horizon. 
For all forcing amplitudes -- though only the $\gamma = 2\%$ and $\gamma = 10\%$ cases are shown, the behaviour of the non-actuated modes is similar for all $\gamma$: the energy of the non-actuated modes grows during the decay of the $(0, 1)$--mode, with the $(0, \pm 2)$ and $(\pm 1, \pm 1)$--modes growing most significantly, which highlights their key role in exchanging energy with the actuated $(0, 1)$--mode. The share of total turbulent energy accounted for by the $(0, \pm 2)$ and $(\pm 1, \pm 1)$--modes is roughly constant across all forcing amplitudes, amounting to approximately $9-10\%$ and $7-8\%$ at their peaks, respectively. 
Some differences do exist across forcing amplitudes: the transient behaviour of the $(0, \pm 2)$--mode appears more sensitive to the value $\gamma$. For the lightly forced case, the energy of the $(0, \pm 2)$--mode continues to grow beyond $t_\mathrm{peak}$, while for the strongly forced case, the energy of the mode peaks soon after the peak in streak energy before decaying rapidly and ceding to the $(\pm 1, \pm 1)$--mode. This suggests that the $(0, \pm 2)$--mode tends to grow faster and peak earlier as the forcing amplitude increases. 
The particular sensitivity of the integrated energy of the $(0, \pm 2)$--mode to forcing amplitude can be partially explained by the fact that it is fed by the dyadic interaction involving the self-interaction of the actuated $(0, \pm 1)$--mode. The nonlinear energy transfer from the $(0, \pm 1)$--mode to secondary scales is discussed in more detail in the next section.

The instantaneous growth and decay of the two preferred secondary modes, the $(0, \pm 2)$-- and $(\pm 1, \pm 1)$--modes, at different wall-normal locations can be seen in figures \ref{fig:time_transfer_to_small_scales}(a, b).
Closer to the wall, the energy of the $(0, \pm 2)$--mode first decreases during the growth of the $(0, \pm 1)$--mode, then exhibits two peaks for the larger amplitude cases, one coinciding with the peak of the $(0, \pm 1)$--mode, and one much later at around $t u_\tau / \delta \approx 3$, as the energy of the mode reverts back to its initial state. The first peak, which is only visible for the higher amplitude cases, occurs earlier as $\gamma$ increases, while the later peak occurs roughly at the same time across forcing amplitudes. 
The behaviour of the $(\pm 1, \pm 1)$--mode is simpler and its energy peaks earlier as the forcing amplitude increases. 
Farther away from the wall, the energy peaks of both the $(0, \pm 2)$-- and  $(\pm 1, \pm 1)$--modes depend little on $\gamma$ and occur roughly at the same time for all forcing amplitudes, at $t u_\tau / \delta \approx 0.8$ and $t u_\tau / \delta \approx 1$, respectively. 
Thus, the energy of the $(0, \pm 2)$-- and $(\pm 1, \pm 1)$--modes is similarly influenced by $x_2$ and $\gamma$: while the forcing magnitude $\gamma$ affects the growth time scale of the secondary modes, causing their energy to peak earlier as it increases, the wall-normal location modulates the sensitivity of these modes to $\gamma$.

To better visualise the cross-scale energy transport time scales, we plot the streamwise energy peak times for the $(0, \pm1)$--mode and the smaller scales as a function of $x_2^+$ and for different forcing amplitudes (figure \ref{fig:time_transfer_to_small_scales}(c)). 
Across all wall-normal heights, increasing the forcing amplitude causes the energy of both the $(0, \pm1)$--mode and the smaller scales to peak earlier. For all forcing amplitudes, the peak times for the $(0, \pm1)$--mode and the smaller scales vary inversely with $x_2^+$. For the $(0, \pm1)$--mode, they are roughly constant with $x_2^+$ in the near-wall region, increase slightly in the buffer layer as we move farther away from the wall, and plateau in the outer region of the flow. For the non-actuated modes, the peak times are also constant in the near-wall region, but decrease dramatically within the buffer layer as $x_2^+$ increases, before levelling off in the outer region. The growth of the smaller scales is thus more sensitive to both forcing amplitude and distance to the wall than the actuated $(0, \pm 1)$--mode.

We define the timescale for cross-scale energy transport, $\Delta t_\mathrm{trans}$, as the time delay between the energy peaks of the $(0, \pm1)$--modes and the smaller scales; 
its dependence on $x_2^+$ is mostly determined by the energy peak time for the smaller scales (figure \ref{fig:time_transfer_to_small_scales}(d)). \reviewertwo{While the results are generally consistent with the established understanding of the energy cascade from larger to smaller scales, the trend of $\Delta t_\mathrm{trans}$ probes additional aspects of the energy transfer, mainly how the rate of the energy transfer to the secondary scales varies with the forcing amplitude and wall-normal height.}
Interestingly, though a larger forcing amplitudes accelerates the energy transport from the $(0, \pm 1)$--mode to smaller scales for all wall normal heights, the sensitivity of $\Delta t_\mathrm{trans}$ to forcing amplitude decreases as we move farther away from the wall. Indeed, in the outer regions of the flow, $\Delta t_\mathrm{trans}$ converges to a value of approximately $0.18 \, \delta / u_\tau $ for high forcing amplitude. 
The plots in figures \ref{fig:time_transfer_to_small_scales}(c, d) indicate two cross-scale energy transfer mechanisms: one for the near-wall region ($x_2^+ \leq 25$) which is highly dependent on $\gamma$, and one for the outer ($x_2^+ \geq 25$) region which is less dependent on $\gamma$. 
The outer region is already highly nonlinear, which allows for a rapid energy cascade from the actuated mode to smaller scales. Perturbing this region seems to have little effect on the time scale of this cross-scale energy transfer. Near the wall, linear mechanisms dominate; the $(0, \pm 1)$--mode, growing significantly under the action of the resolvent forcing term, greatly enhances the nonlinear interactions involving the mode and markedly changes the underlying energy transfer to secondary scales, which may explain the heightened sensitivity of $\Delta t_\mathrm{trans}$ to $\gamma$ in the near-wall region. 
%%%%%%% %%%%%%% %%%%%%% 
%The heightened sensitivity of $\Delta t_\mathrm{trans}$ to $\gamma$ in the near-wall region can be explained by the fact that the $(0, \pm 2)$--mode, which is fed by the self-interaction of the actuated $(0, \pm 1)$--mode, dominates close to the wall as shown in figure \ref{fig:time_transfer_to_small_scales}(a). The nonlinear energy transfer from the $(0, \pm 1)$--mode to secondary scales is discussed in more detail in the next section.
%%% TODO %%%%%%%%%%%%%%%%%%%%%%%%%%%%%
% should we mention scaling? to say that the near-wall region is more sensitive, we have to divide by gamma^1.44 but outer region is gamma^1.2 for collapse
%%%%%%%%%%%%%%%%%%%%%%%%%%%%%%%%%%%%%%

%%%%%%%%%%%%%%%%%%%%%%%%%%%%%%%%%%%%%%%%%%%%%%%%%%%%%%%%%
% Up until here: old article
%%%%%%%%%%%%%%%%%%%%%%%%%%%%%%%%%%%%%%%%%%%%%%%%%%%%%%%%%
\section{Nonlinear energy transfer}\label{sec:NLT}

The energy content of the secondary modes (figures \ref{fig:SpectrumFixedTimes}, \ref{fig:time_transfer_to_small_scales}) are the result of nonlinear interactions amongst all length scales. We wish to disentangle these interactions and focus on the nonlinear energy transfer that specifically drain energy from the actuated $(0, 1)$--mode. 
As in \citet{symon2021energy} and \citet{ding2025mode}, we represent the nonlinear energy transfer from a mode $(k_1, k_3)$ using the following term: 
\begin{equation}\label{eq:nonlinear_transfer}
\hat N(k_1, k_3) = -\hat u_i^{(-k_1, -k_3)} \widehat{\frac{\partial u_i u_j}{\partial x_j}}^{(k_1, k_3)}.
\end{equation}
%
%where $\hat u_i^{(k_1,k_3)}$ represents the Fourier transform of $u_i$ for streamwise and spanwise wavenumbers $2\pi k_1/L_1$ and $2\pi k_3/L_3$, respectively. 
We note that $(\cdot)^{(-k_1, -k_3)}$ refers to the complex conjugate of $(\cdot)^{(k_1, k_3)}$. The term $\hat N(k_1, k_3)$ satisfies
\begin{equation}
\begin{split}
    \int_0^{2\delta} \sum \limits_{k_1} \sum \limits_{k_3} \hat N(k_1, k_3) dy &= -\int_0^{2\delta} \widehat {u_i \frac{\partial u_i u_j}{\partial x_j}}^{(0, 0)} \\ 
    &= \frac{1}{L_1 L_3} \int_0^{2\delta} \int_0^{L_3} \int_0^{L_1} u_i \frac{\partial u_i u_j}{\partial x_j} dx_1 dx_3 dx_2 \\
    & = 0,
    \end{split}
\end{equation}
due to continuity and the no-penetration boundary conditions at the walls. \reviewertwo{This reflects the fact that the nonlinear transfer in channel flow does not add or remove energy from the system but simply redistributes it between scales \citep{tennekes1972first, pope2001turbulent}}. We can express $\hat N (k_1, k_3)$ as a sum of contributions from interacting scales:

\begin{equation} \label{eq:NM_equation}
\begin{split}
\hat N(k_1, k_3) &= - \sum \limits_{s_1} \sum \limits_{s_3} \hat u_i^{(-k_1, -k_3)} \widehat {\frac{\partial u_i}{\partial x_j}}^{(s_1, s_3)} \hat u_j ^{(k_1 - s_1, k_3 - s_3)} %\\
%&=: \sum \limits_{s_1 } \sum \limits_{s_3} \hat M (k_1, k_3, s_1, s_3). 
\end{split}
\end{equation}
We refer to an individual contribution to the sum as $\hat M^{(k_1, k_3)} (s_1, s_3)$, defined below

\mycomment{
\begin{equation}\label{eq:M_eq}
\begin{split}
    \hat M (k_1, k_3, s_1, s_3) &:=  - \hat u_i^{(-k_1, -k_3)} \widehat {\partial_j u_i}^{(s_1, s_3)} \hat u_j ^{(k_1 - s_1, k_3 - s_3)} 
    %\\
    %&- \hat u_i^{(-k_1, -k_3)} \widehat {\partial_j u_i}^{(-s_1, s_3)} \hat u_j ^{(k_1 + s_1, k_3 - s_3)} 
    \end{split}
\end{equation}}

\begin{equation}\label{eq:M_xz}
\begin{split}
    \hat M^{(k_1, k_3)}(s_1, s_3) = - 2\Real \left \{ \hat u_i^{(-k_1, -k_3)} \widehat {\frac{\partial u_i}{\partial x_j}}^{(s_1, s_3)} \hat u_j ^{(k_1 - s_1, k_3 - s_3)} \right \} .
    \end{split}
\end{equation}
The term $\hat M^{(k_1, k_3)}(s_1, s_3)$ represents the energy transfer from the $(k_1, k_3)$--mode to the $(s_1, s_3)$--mode and satisfies the following properties:

\begin{equation} \label{eq:property_1}
    \int_0^{2\delta} \hat M^{(k_1, k_3)}(s_1, s_3) dx_2 = - \int_0^{2\delta} \hat M^{(s_1, s_3)}(k_1, k_3) dx_2,
\end{equation}
\begin{equation} \label{eq:property_2}
    \int_0^{2\delta} \hat M^{(k_1, k_3)}(k_1, k_3) dx_2 = 0.
\end{equation} 
\reviewerone{These properties are also a consequence of the incompressibility condition and the boundary conditions. For example, to obtain equation \eqref{eq:property_1}, consider $\hat u_i^{(-k_1, -k_3)} \widehat {\partial u_i/\partial x_j}^{(s_1, s_3)} \hat u_j ^{(k_1 - s_1, k_3 - s_3)}$ and $\hat u_i^{(s_1, s_3)} \widehat {\partial u_i/\partial x_j}^{(-k_1, -k_3)} \hat u_j ^{(k_1 - s_1, k_3 - s_3)}$ which, along with their conjugates, constitute the quantities $\hat M^{(k_1, k_3)}(s_1, s_3)$ and $\hat M^{(s_1, s_3)}(k_1, k_3)$, respectively. These terms satisfy the following:
\begin{equation}
    \begin{split}
 & \int_0^{2\delta} \left ( \hat u_i^{(-k_1, -k_3)} \widehat {\frac{\partial u_i}{\partial x_j}}^{(s_1, s_3)} \hat u_j ^{(k_1 - s_1, k_3 - s_3)} + \hat u_i^{(s_1, s_3)} \widehat {\frac{\partial u_i}{\partial x_j}}^{(-k_1, -k_3)} \hat u_j ^{(k_1 - s_1, k_3 - s_3)} \right ) dx_2 \\
&= \int_0^{2\delta} \mathrm{i} (s_1-k_1) \hat u_i^{(-k_1, -k_3)} \hat u_i ^{(s_1, s_3)} \hat u_1 ^{(k_1 - s_1, k_3 - s_3)}  \\
& \qquad \qquad \qquad + \frac{\partial}{\partial x_2} \left( \hat u_i^{(-k_1, -k_3)} \hat u_i ^{(s_1, s_3)} \right )\hat u_2 ^{(k_1 - s_1, k_3 - s_3)}  \\
& \qquad \qquad \qquad \qquad \qquad  + \mathrm{i} (s_3 - k_3) \hat u_i^{(-k_1, -k_3)} \hat u_i ^{( s_1, s_3)} \hat u_3 ^{(k_1 - s_1, k_3 - s_3)} dx_2 \\
&= - \int_0^{2\delta}   \hat u_i^{(-k_1, -k_3)} \hat u_i ^{(s_1, s_3)} \left ( \mathrm{i}(k_1-s_1) \hat u_1 ^{(k_1 - s_1, k_3 - s_3)} \right )\\ 
& \qquad \qquad \qquad  + \hat u_i^{(-k_1, -k_3)} \hat u_i ^{(s_1, s_3)} \left ( \frac{\partial}{\partial x_2} \hat u_2 ^{(k_1 - s_1, k_3 - s_3)} \right ) \\
& \qquad \qquad \qquad \qquad \qquad  +  \hat u_i^{(-k_1, -k_3)} \hat u_i ^{(s_1, s_3)} \left ( \mathrm{i} (k_3 - s_3) \hat u_3 ^{(k_1 - s_1, k_3 - s_3)} \right ) dx_2 \\
& = 0. 
\end{split}
\end{equation}
}
\reviewerone{Note, $\mathrm{i} = \sqrt{-1}$ and $(\cdot)_i$ denotes a spatial direction. To obtain the wall-normal derivative term in the second equality, we integrate by parts and apply the no-penetration boundary condition; we enforce continuity for $\hat u_j^{(k_1 -s_1, k_3- s_3)}$ to obtain the final result.} Equation \eqref{eq:property_1} implies that the scale-to-scale energy transfer is conservative; equation \eqref{eq:property_2}, a consequence of equation \eqref{eq:property_1}, implies that it purely captures the energy transferred from one scale to a different scale and excludes self-interactions. We finally define $ \Delta \hat M^{(k_1, k_3)}$ as $\hat M^{(k_1, k_3)} - \hat M_0^{(k_1, k_3)}$, where $\hat M_0^{(k_1, k_3)}$ is defined as equation \eqref{eq:M_xz} using the unforced velocity field $\boldsymbol{u_0}$.

\begin{figure}
    \centering
    \subfloat[]{\includegraphics[width=0.5\linewidth]{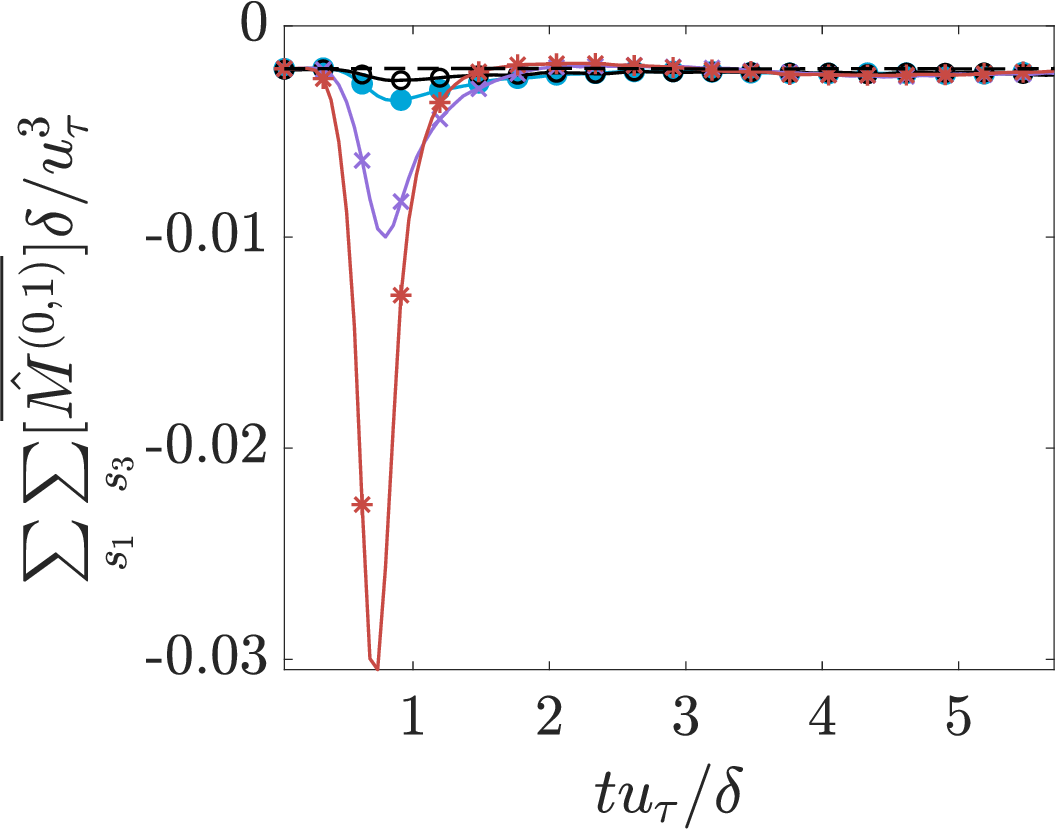}}
    \subfloat[]{\includegraphics[width=0.5\linewidth]{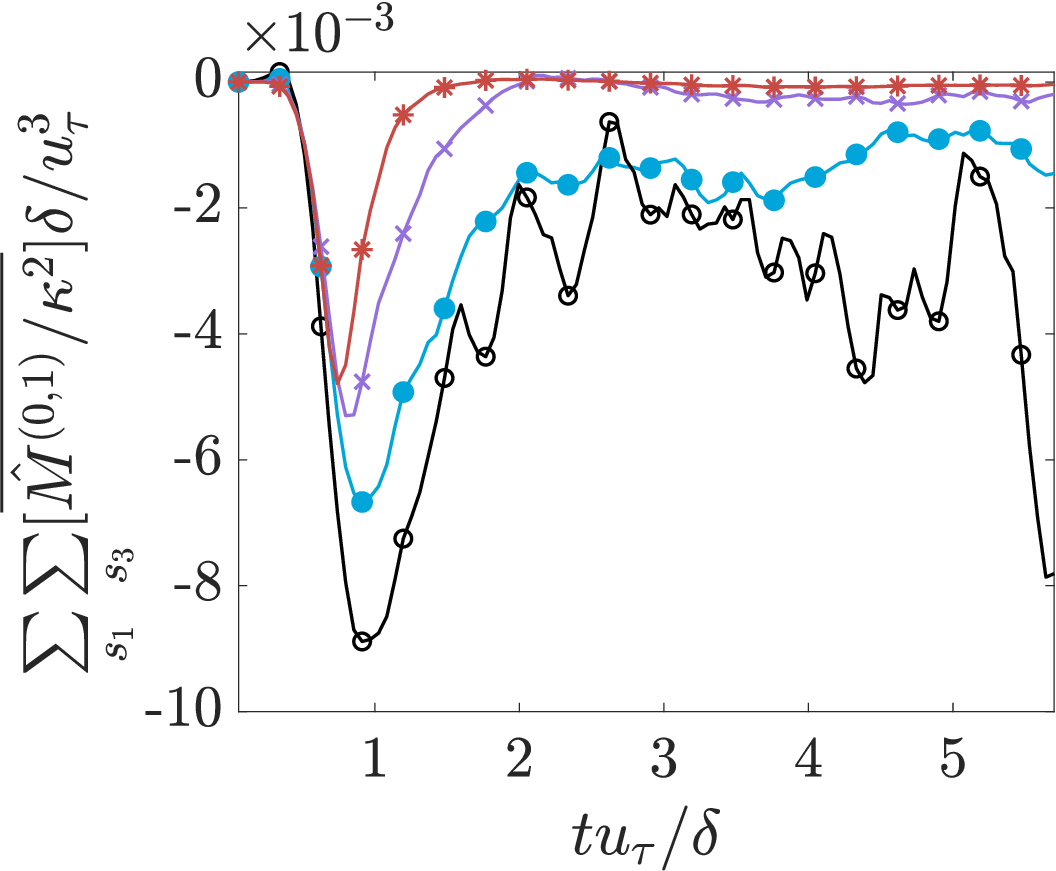}}
    \caption{(a) Integrated nonlinear energy transfer from the $(0,1)$--mode to the non-forced modes; (b) integrated nonlinear energy transport from the $(0,1)$--mode normalised by the forcing magnitude $|\kappa|^2$.
    %; and (c) magnitude peaks of the integrated nonlinear energy transfer. 
    The cases plotted are $\gamma = 1\%$ (black $\circ$), $2\%$ (cyan $\bullet$), $5 \%$ (purple $\times$) and $10 \%$ (red $*$). 
    %The dashed line in (c) represents the $0.78 |\gamma|^1.6$ trend
    }
    \label{fig:integrated_transfer}
\end{figure}

\begin{figure}
    \centering
    \subfloat[]{\includegraphics[width=0.33\linewidth]{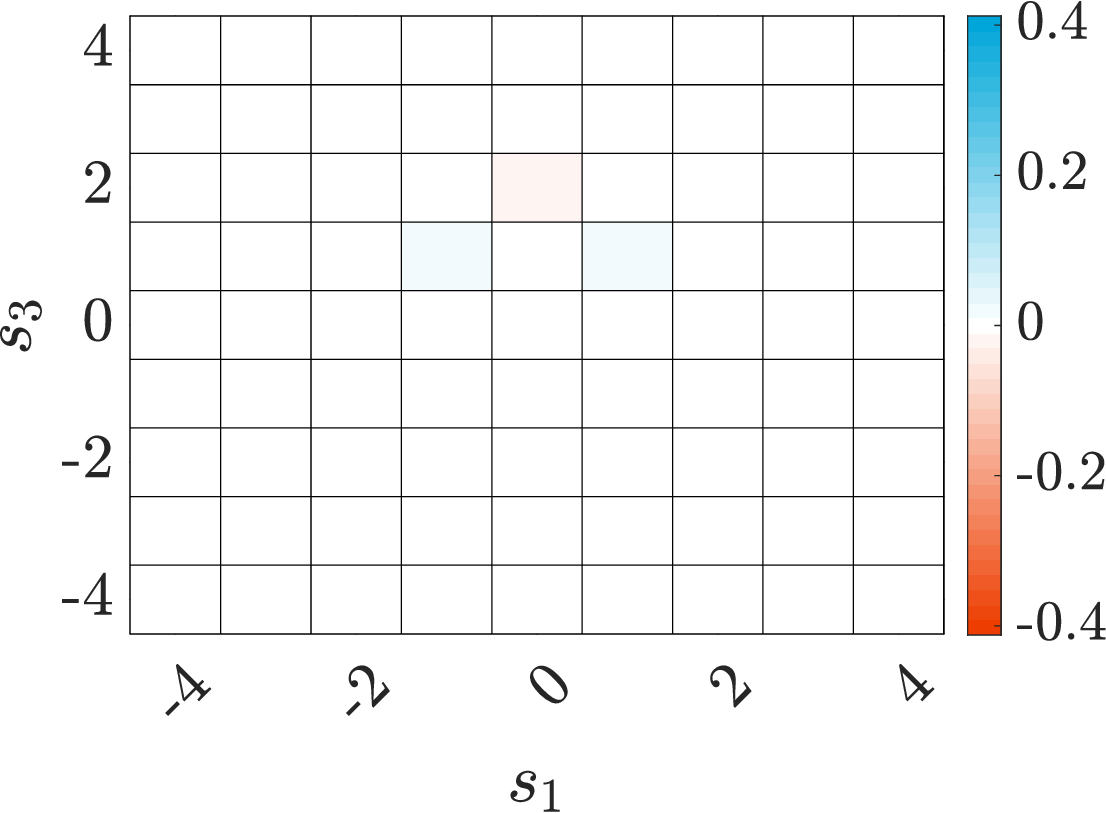}}
    \subfloat[]{\includegraphics[width=0.33\linewidth]{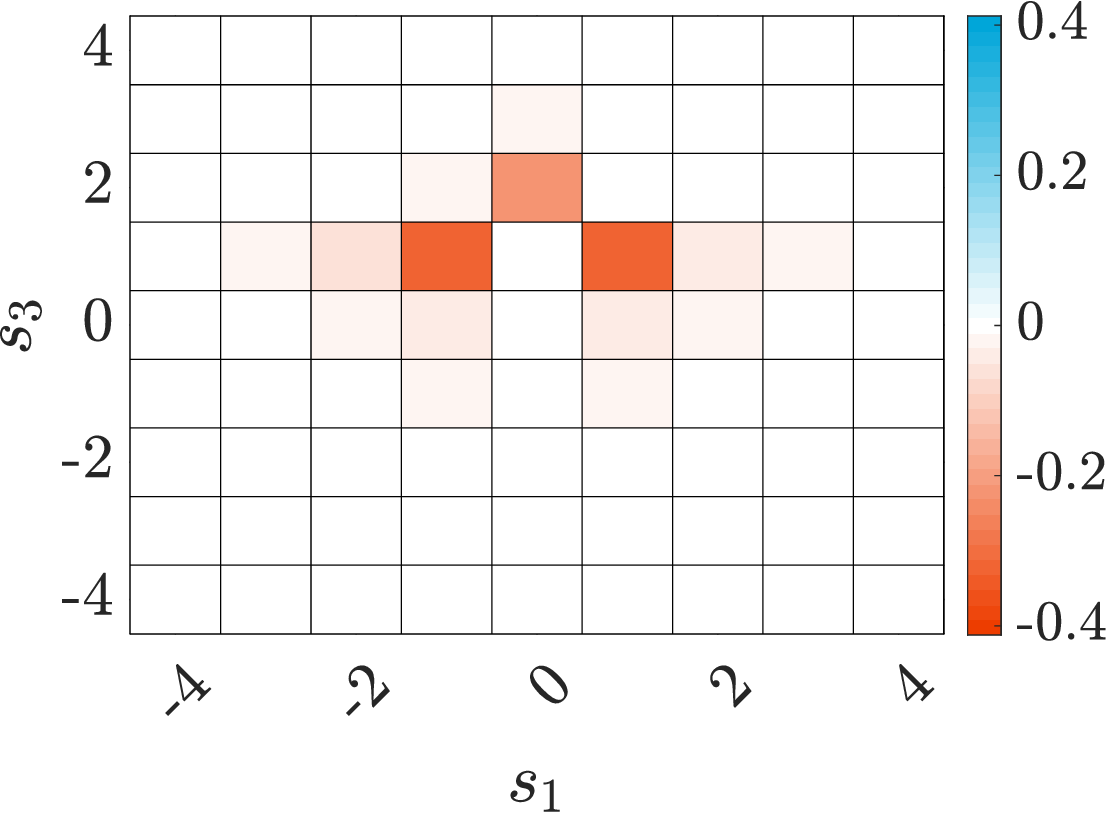}}
    \subfloat[]{\includegraphics[width=0.33\linewidth]{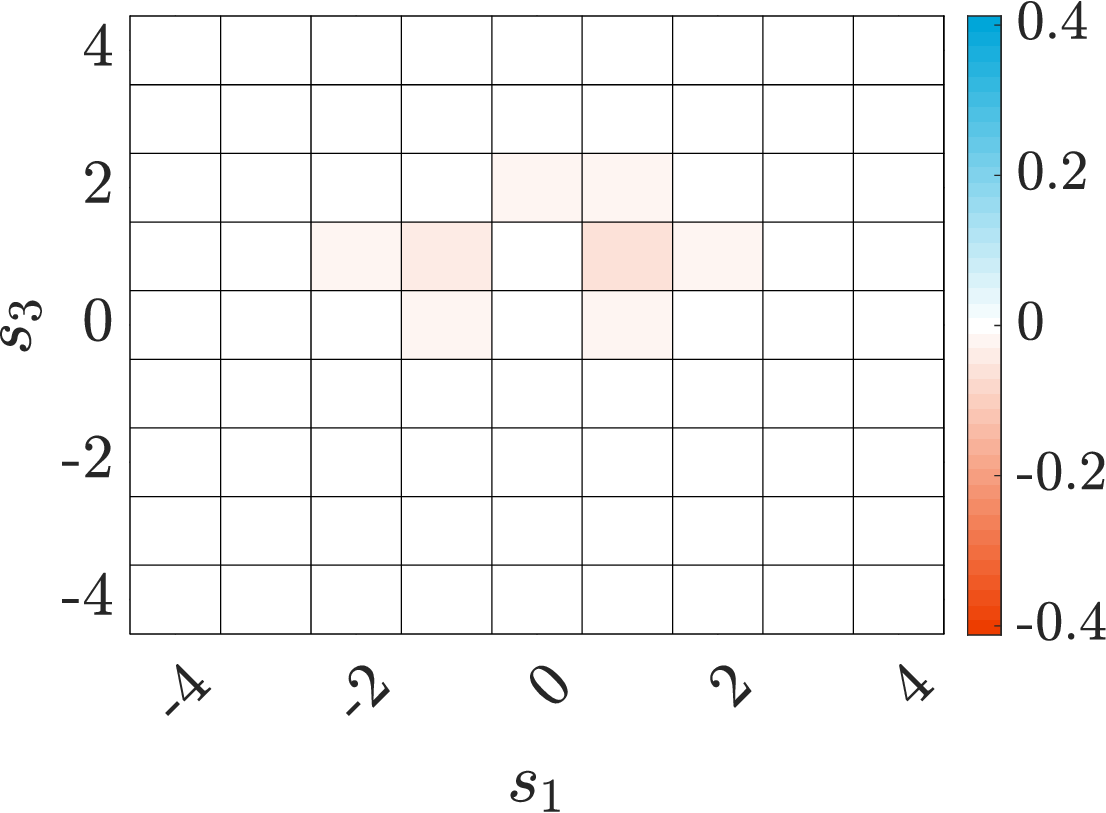}}
    
    \subfloat[]{\includegraphics[width=0.33\linewidth]{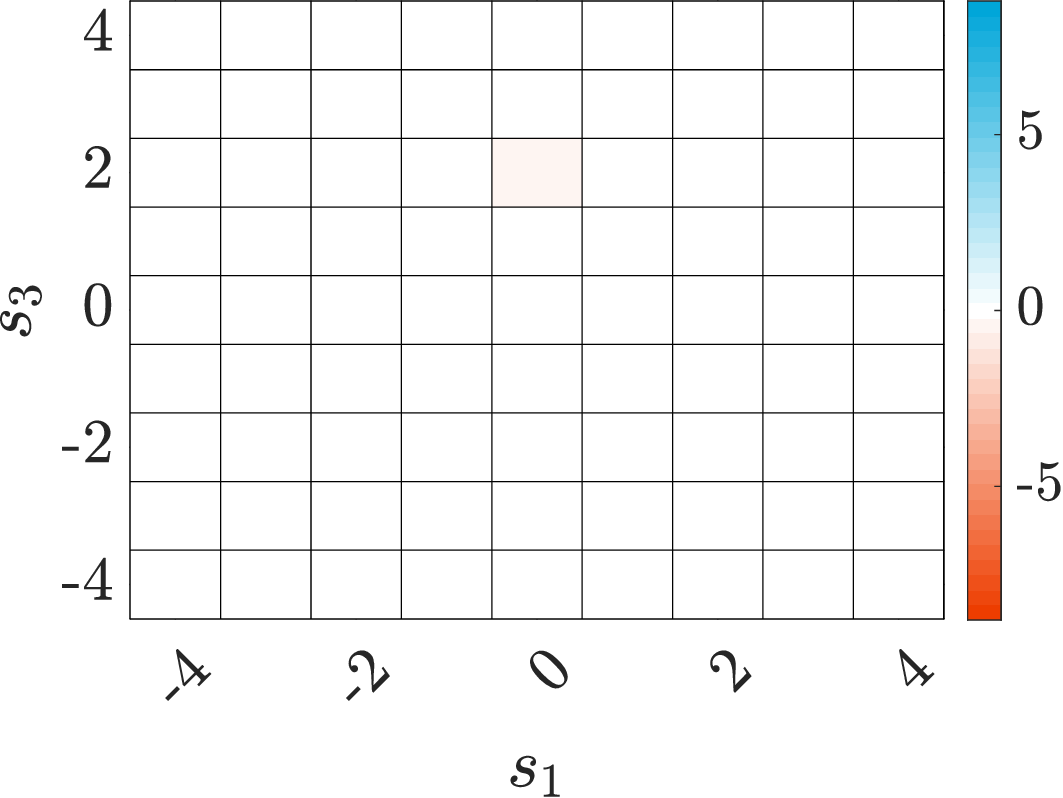}}
    \subfloat[]{\includegraphics[width=0.33\linewidth]{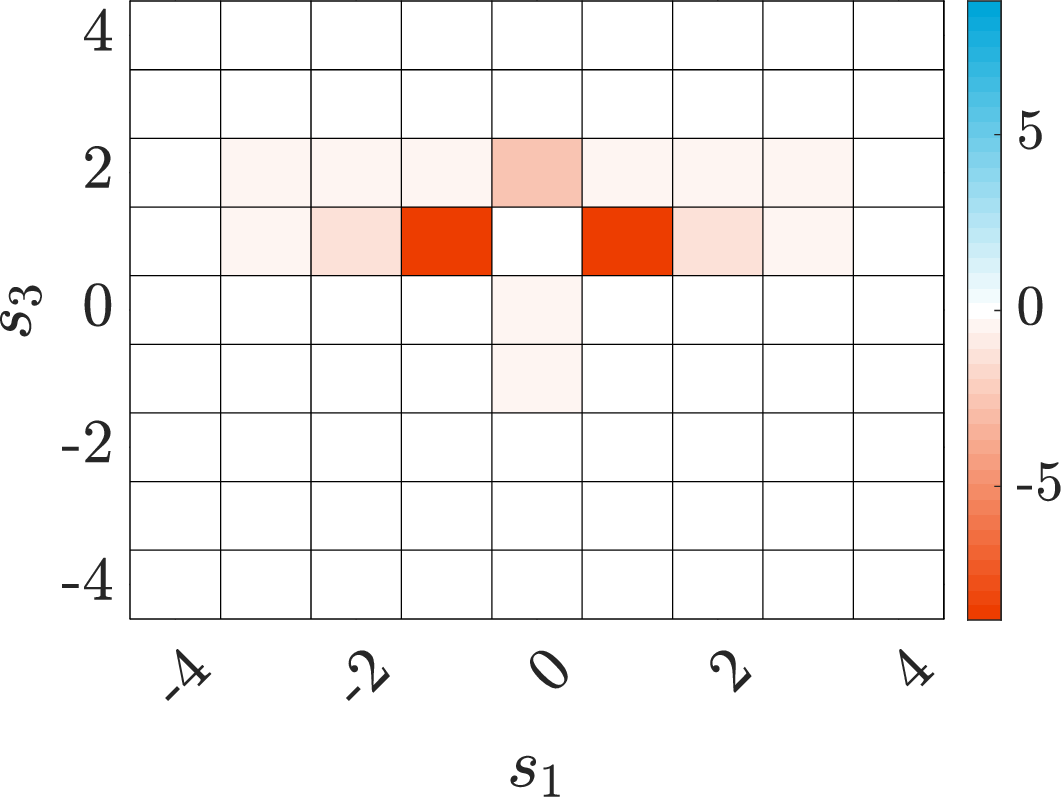}}
    \subfloat[]{\includegraphics[width=0.33\linewidth]{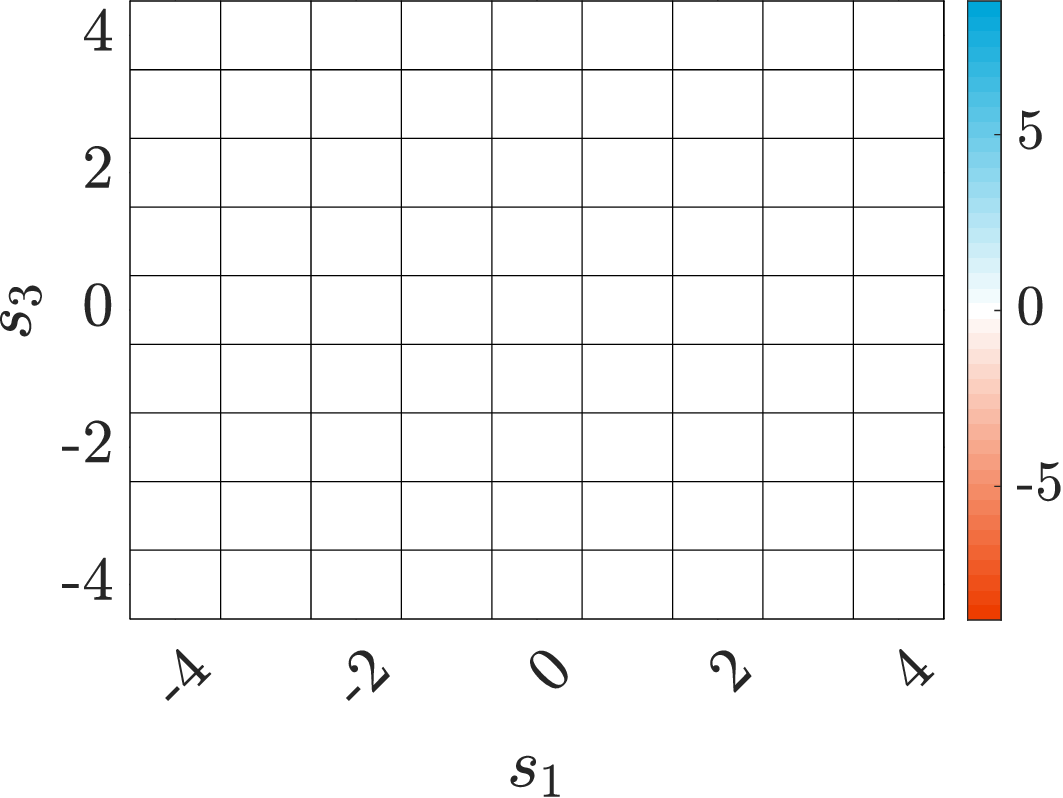}}
    \caption{Nonlinear energy transfer $\Delta \hat M^{(0, 1)} (s_1, s_3)$ for $\gamma = 2\%$ (a-c) and $\gamma = 10\%$ (d-f), at $tu_\tau / \delta = 0.35$ (a, d), $0.74$ (b, e), $2$ (c, f).}
    \label{fig:transfer_spectra}
\end{figure}

\begin{figure}
    \centering
    \subfloat[]{\includegraphics[width=0.5\linewidth]{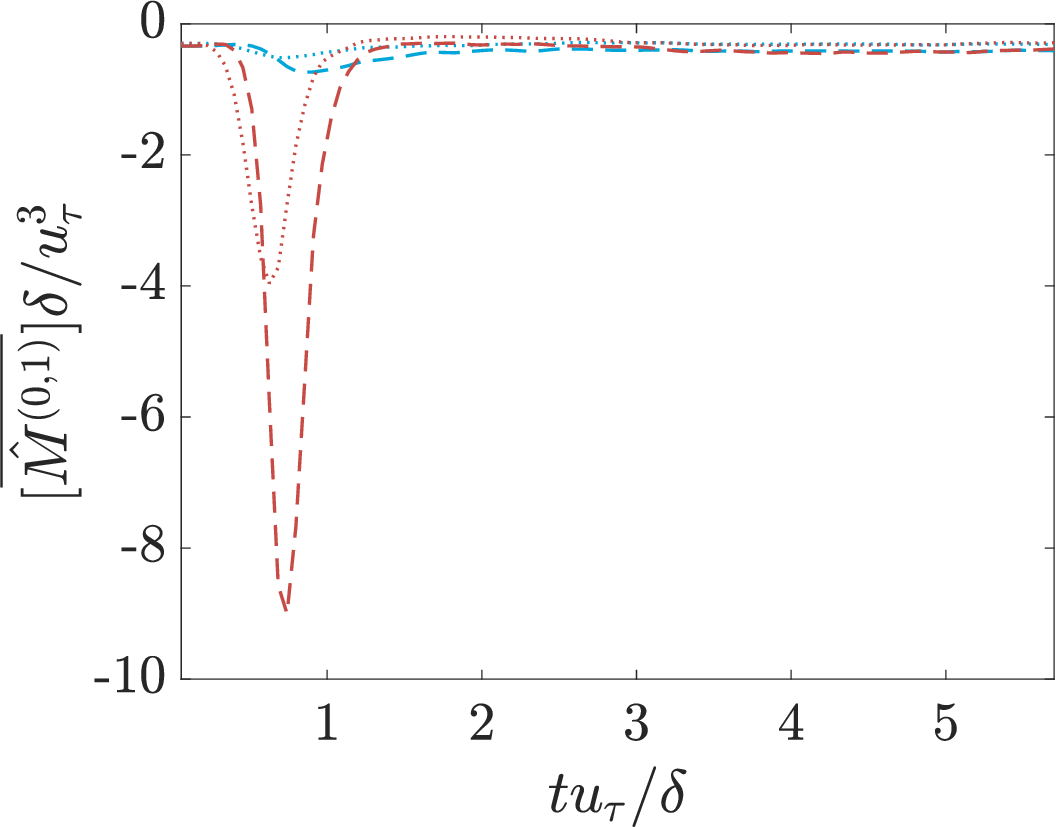}}
        \subfloat[]{\includegraphics[width=0.5\linewidth]{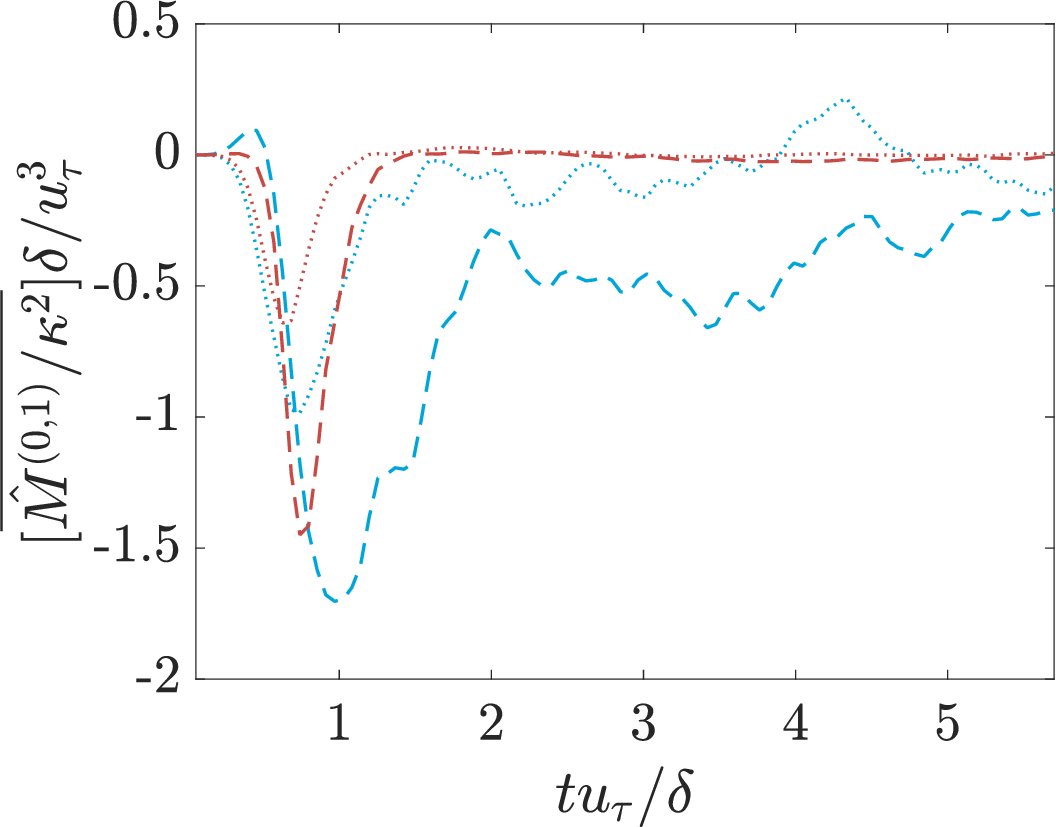}}
    \caption{(a) Integrated nonlinear energy transfer from the $(0,1)$--mode to the $(0, 2)$--mode ($...$) and the $(1,1)$--mode ($--$); (b) integrated nonlinear energy transfer from the $(0,1)$--mode normalised by the forcing magnitude $|\kappa|^2$. The cases plotted are $2\%$ (cyan) and $10 \%$ (red).}
    \label{fig:integrated_transfer_02_11}
\end{figure}

\begin{figure}
    \centering
    \subfloat[]{\includegraphics[width=0.5\linewidth]{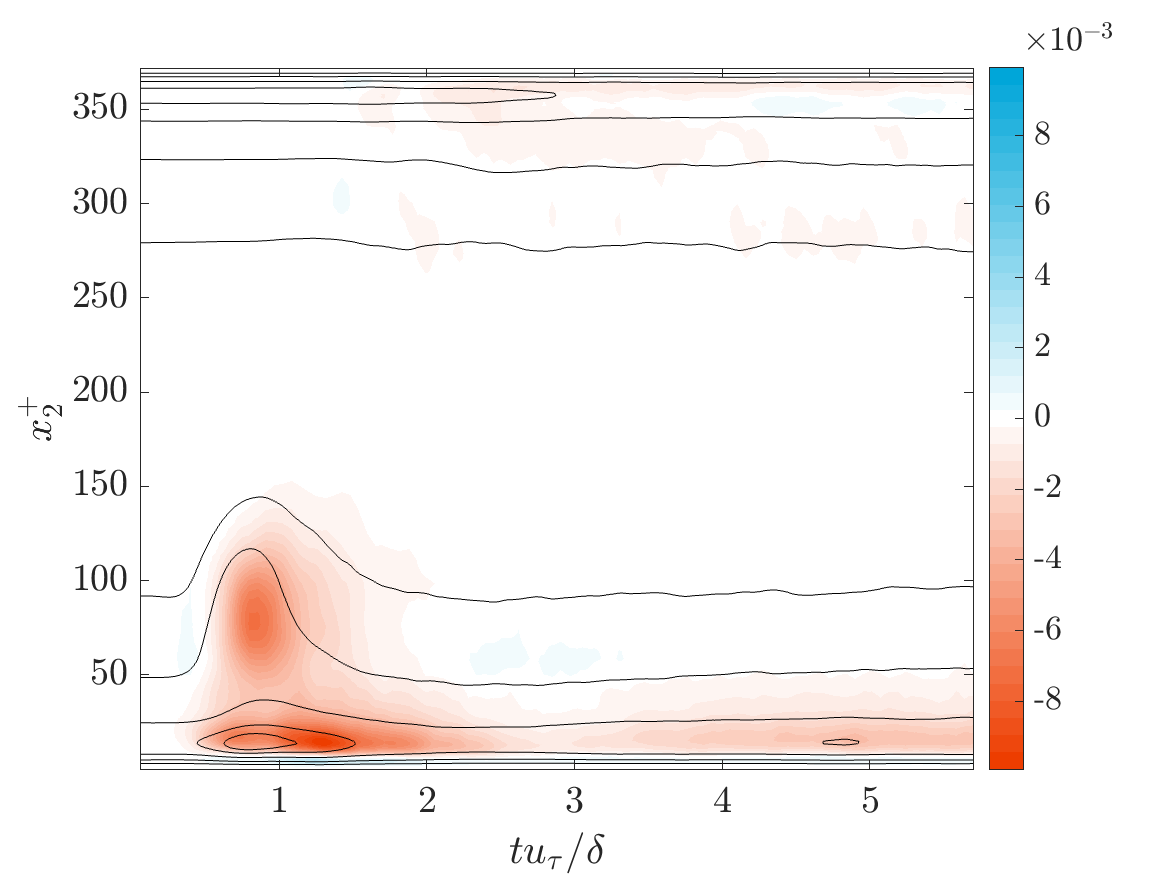}}
        \subfloat[]{\includegraphics[width=0.5\linewidth]{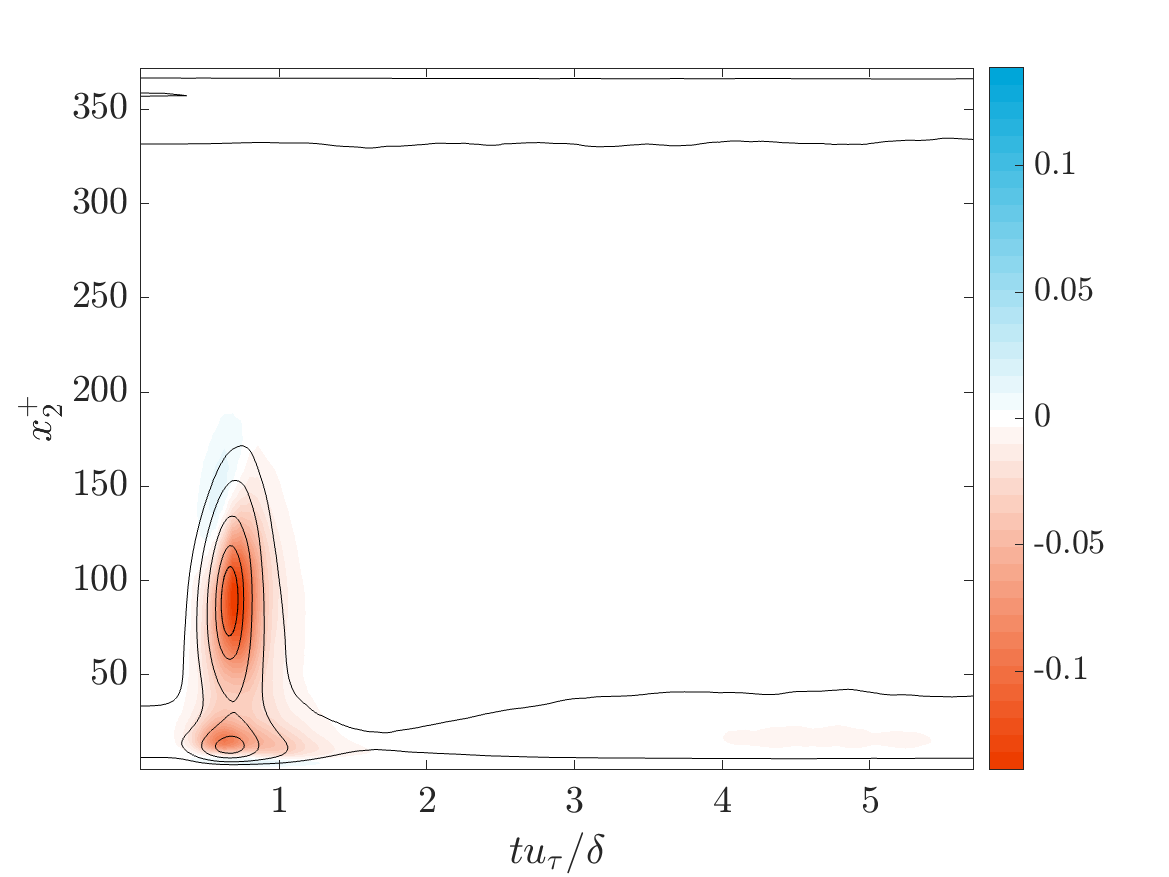}}
    \caption{(a) Total nonlinear energy transfer from the $(0,1)$--mode, \emph{i.e.} $ \sum_{s_1} \sum_{s_3}  \overline {\Delta \hat M^{(0, 1)}(s_1, s_3)}  \delta / u_\tau ^3 $. The cases plotted are for (a) $\gamma = 2\%$ and (b) $\gamma = 10 \%$. The black lines are contours of $\overline{|\hat u_1^{(0, 1)}|^2/2}$ and represent $10\%, \; 25 \%, 50\%, 75\%$ and $90\%$ of the maximum value across $x_2$ and $t$.}
    \label{fig:transfer_YT_plane}
\end{figure}

\begin{figure}
    \centering
    \subfloat[]{\includegraphics[width=0.5\linewidth]{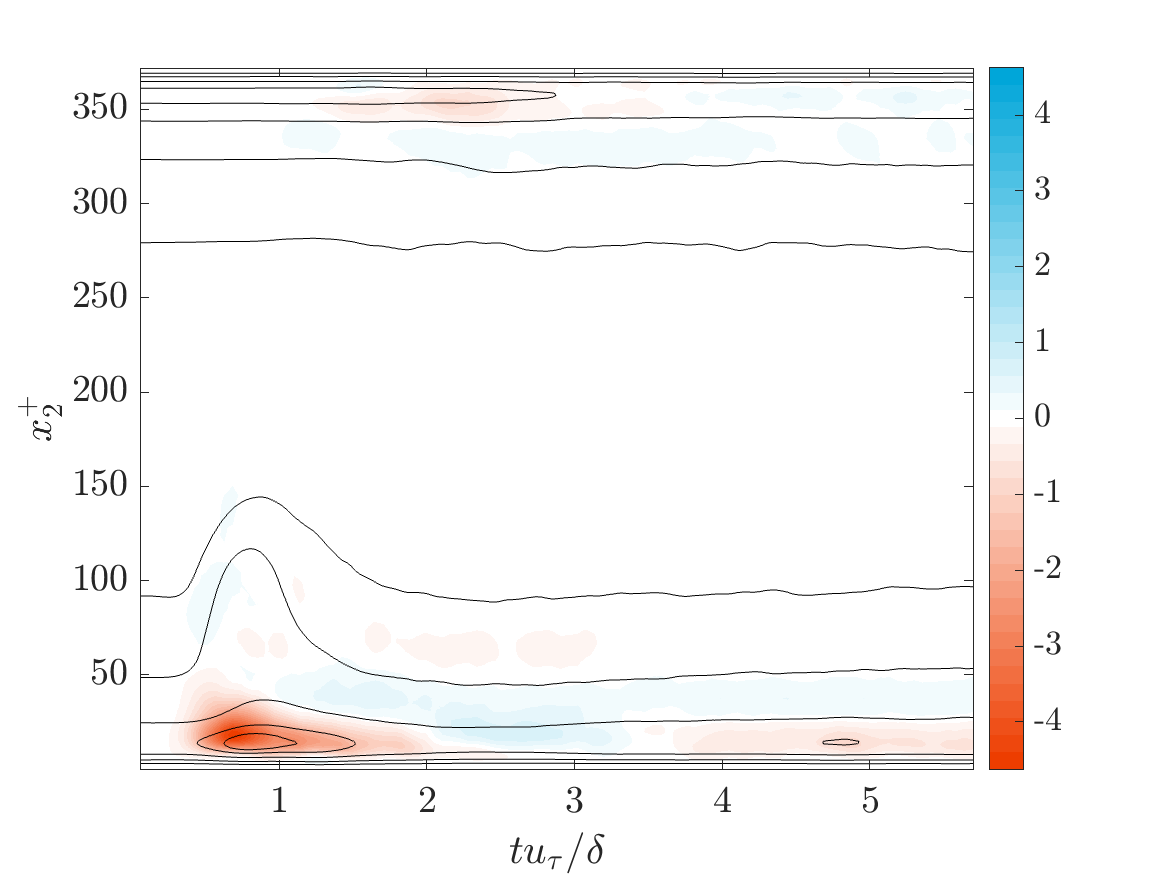}}
    \subfloat[]{\includegraphics[width=0.5\linewidth]{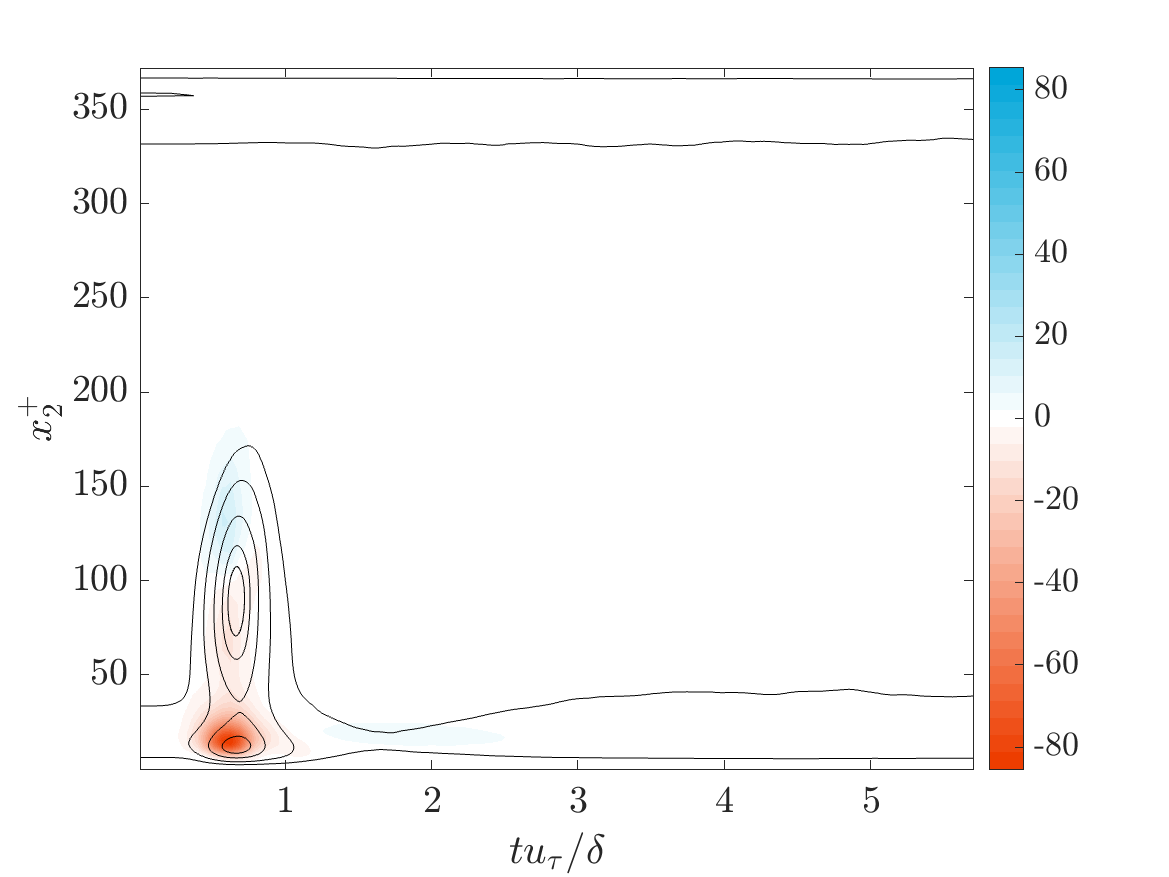}}

    \subfloat[]{\includegraphics[width=0.5\linewidth]{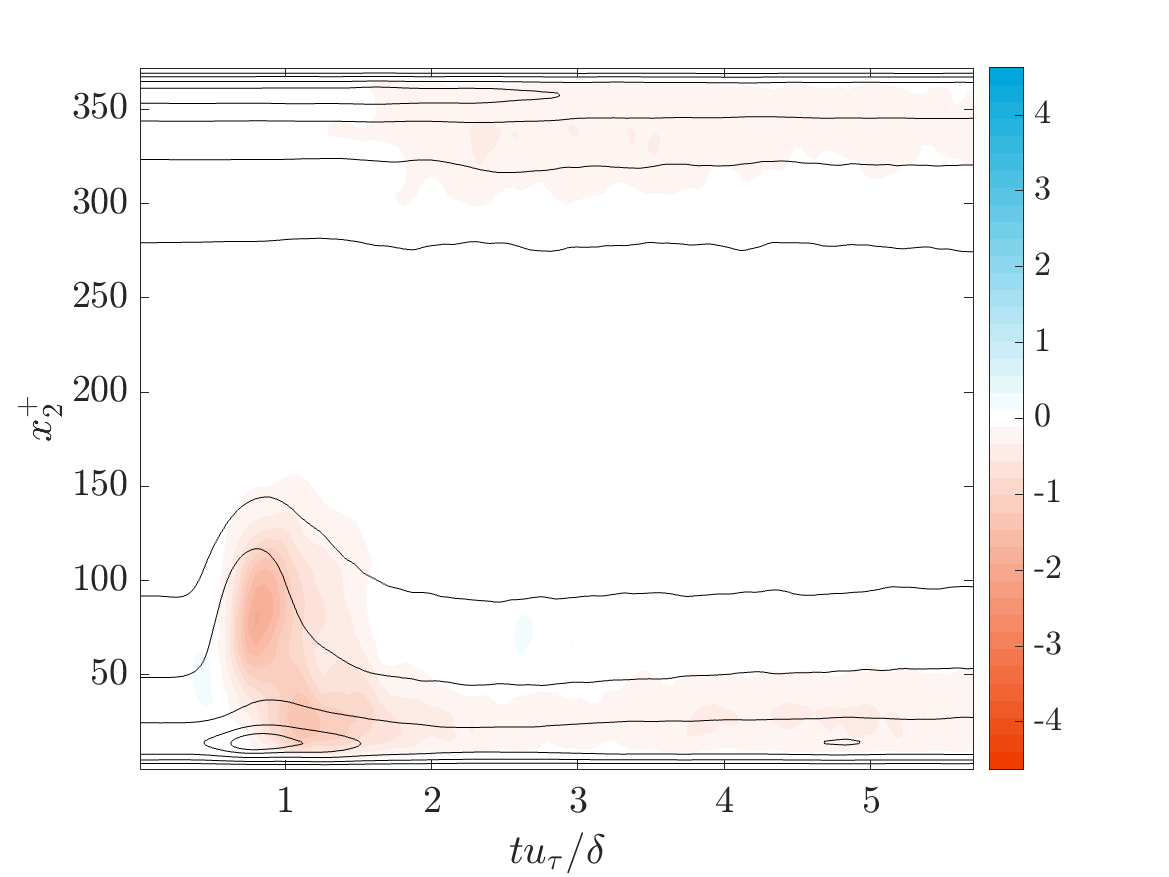}}
    \subfloat[]{\includegraphics[width=0.5\linewidth]{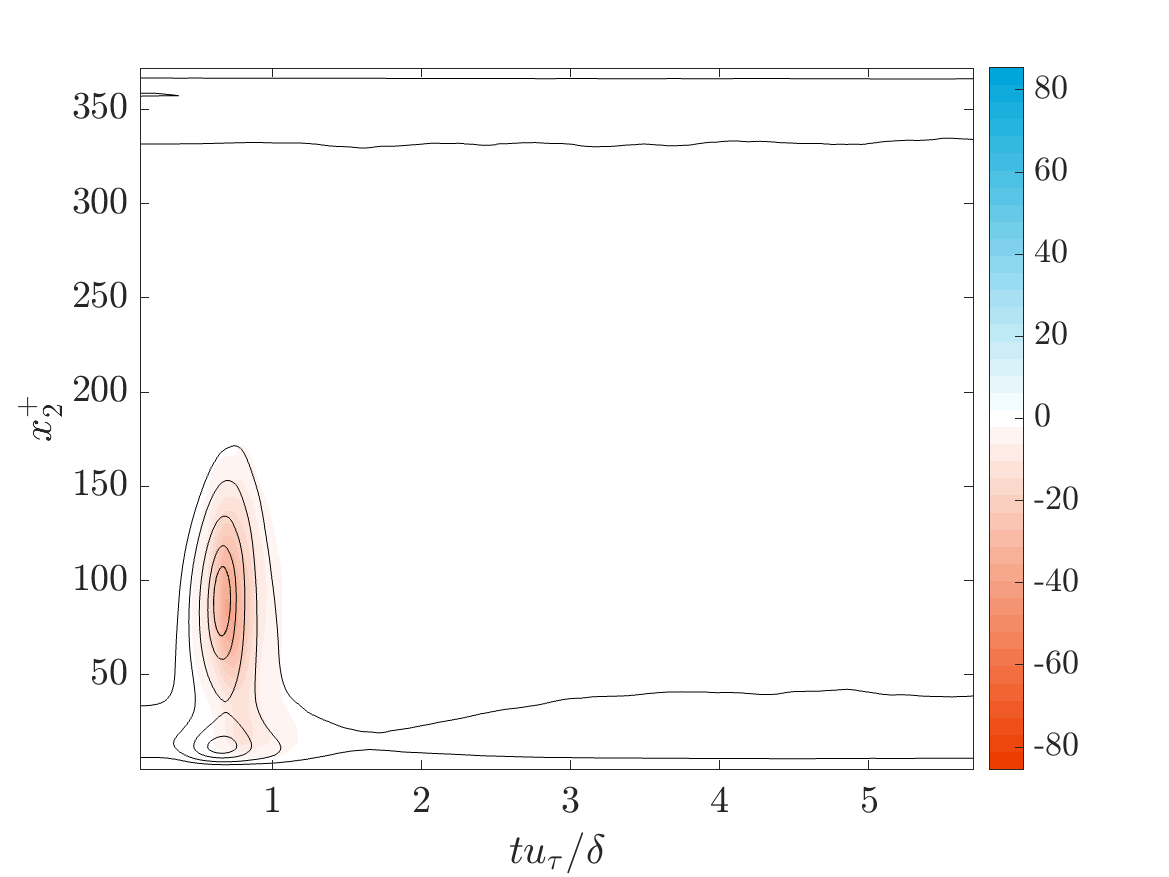}}
    \caption{
    (a, b) $\overline {\Delta \hat M^{(0,1)}(0, 2)}  \delta / u_\tau ^3 $ and (c, d) $\overline {\Delta \hat M^{(0,1)}(1, 1)}  \delta / u_\tau ^3 $ in the $t-x_2$ plane.
    The cases plotted are for (a,c) $\gamma = 2\%$ and (b,d) $\gamma = 10 \%$. The black lines are contours of $\overline{|\hat u_1^{(0, 1)}|^2/2}$ and represent $10\%, \; 25 \%, 50\%, 75\%$ and $90\%$ of the maximum value across $x_2$ and $t$.}
    \label{fig:transfer_YT_plane_11_02}
\end{figure}

\begin{figure}
    \centering
    \subfloat[]{\includegraphics[width=0.5\linewidth]{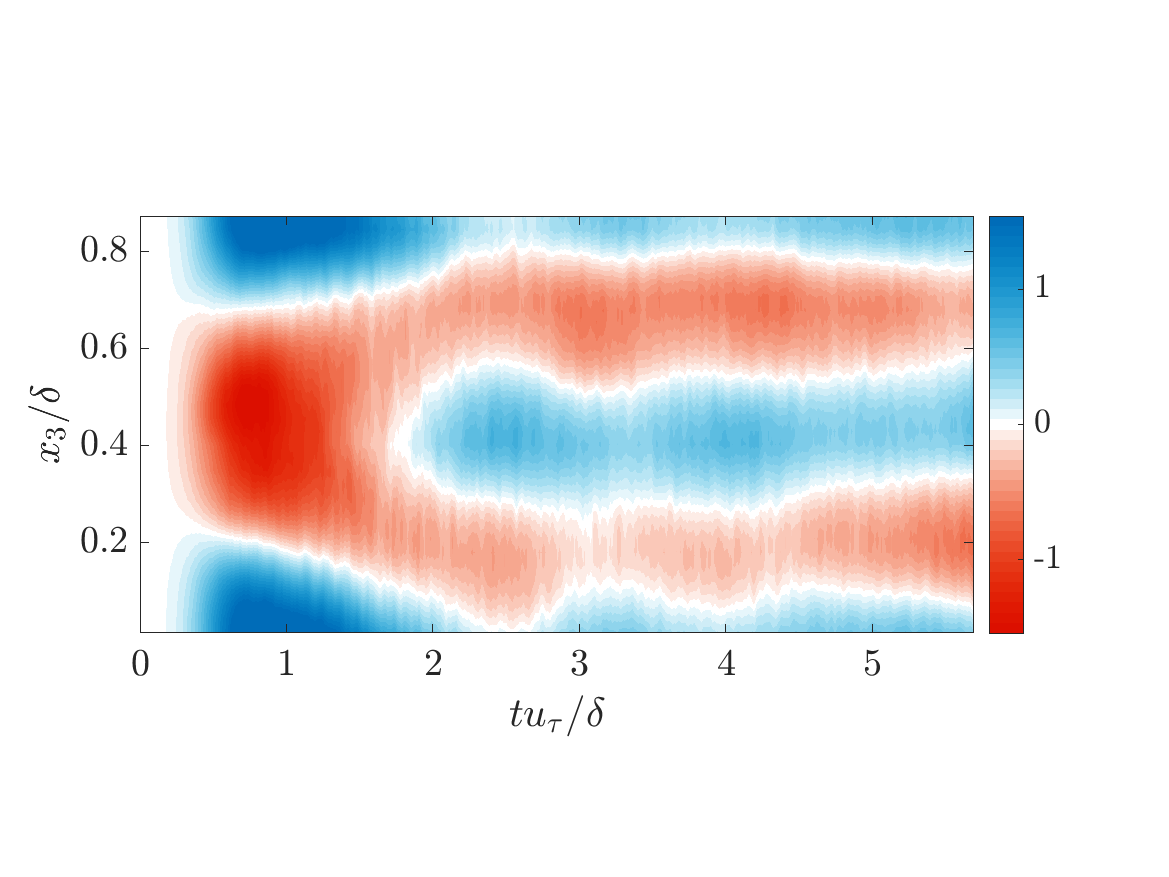}}
    \subfloat[]{\includegraphics[width=0.5\linewidth]{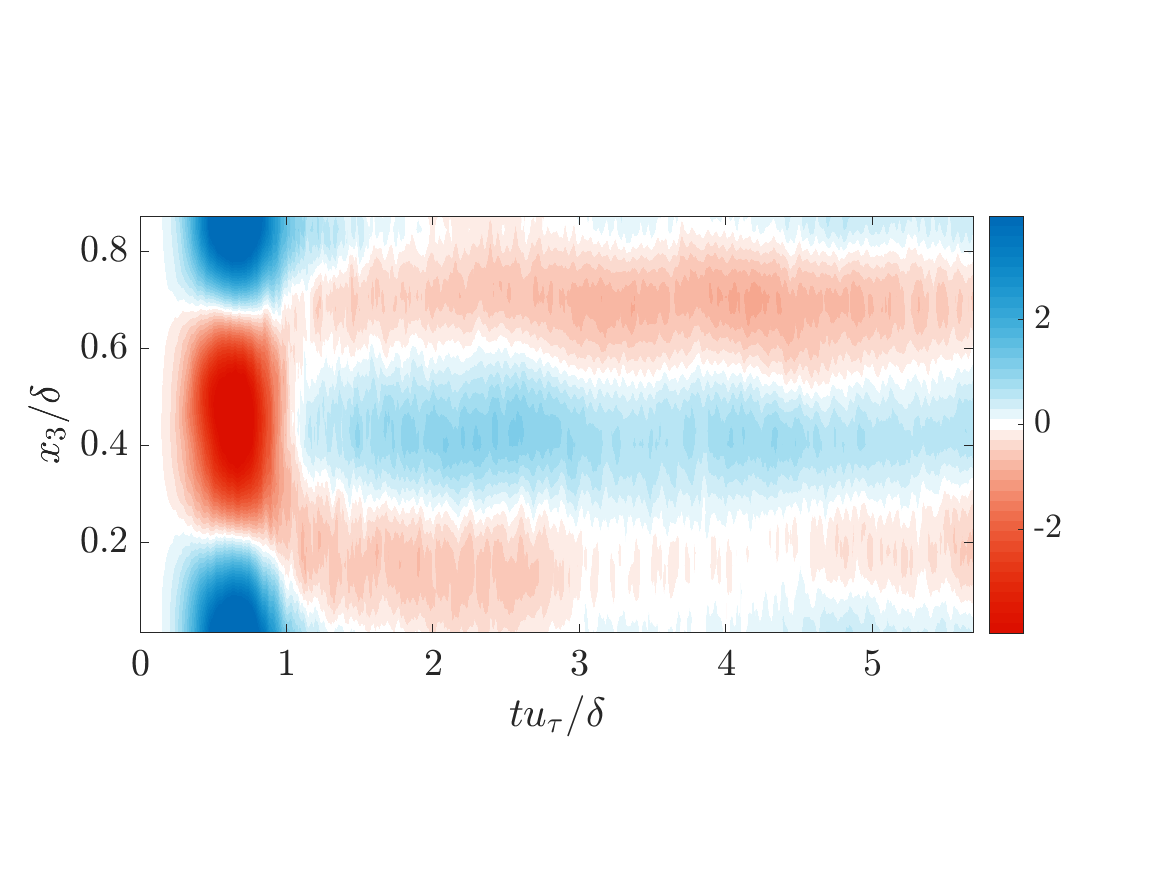}}

    \subfloat[]{\includegraphics[width=0.5\linewidth]{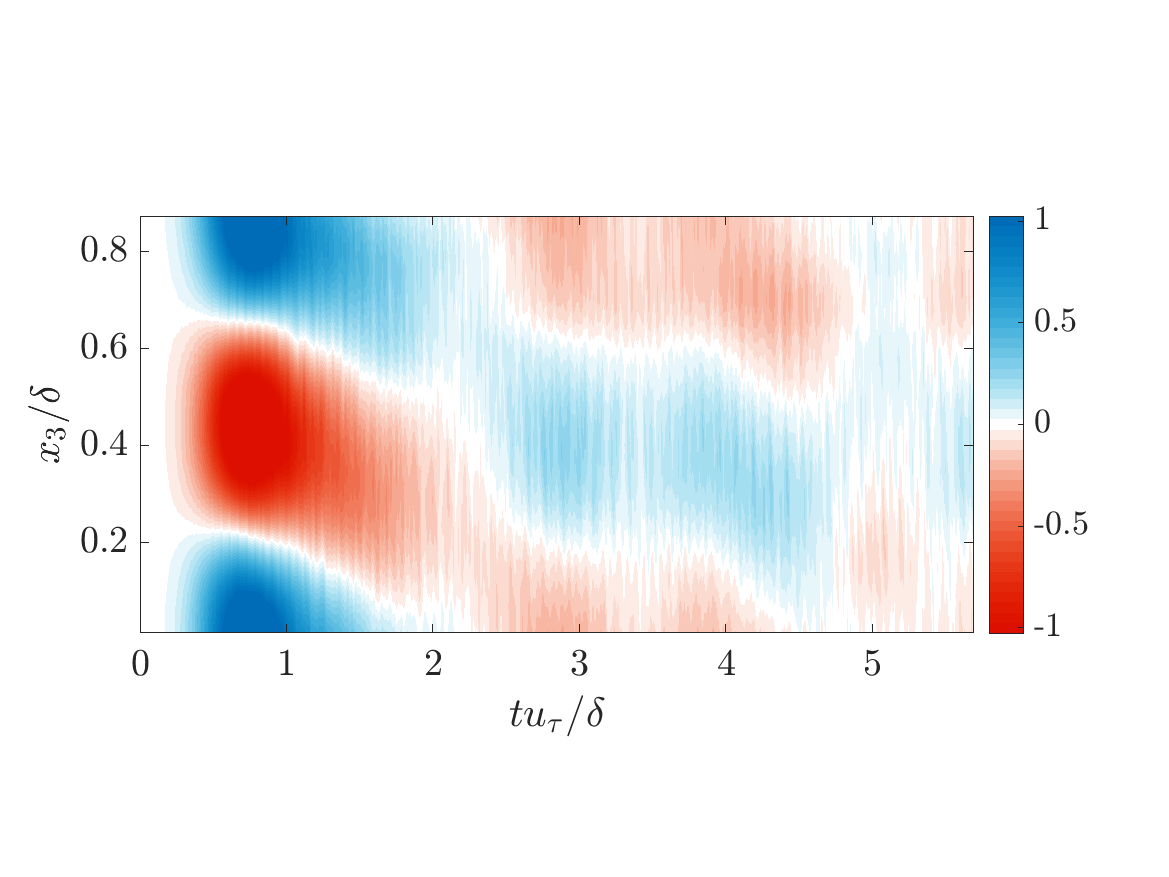}}
    \subfloat[]{\includegraphics[width=0.5\linewidth]{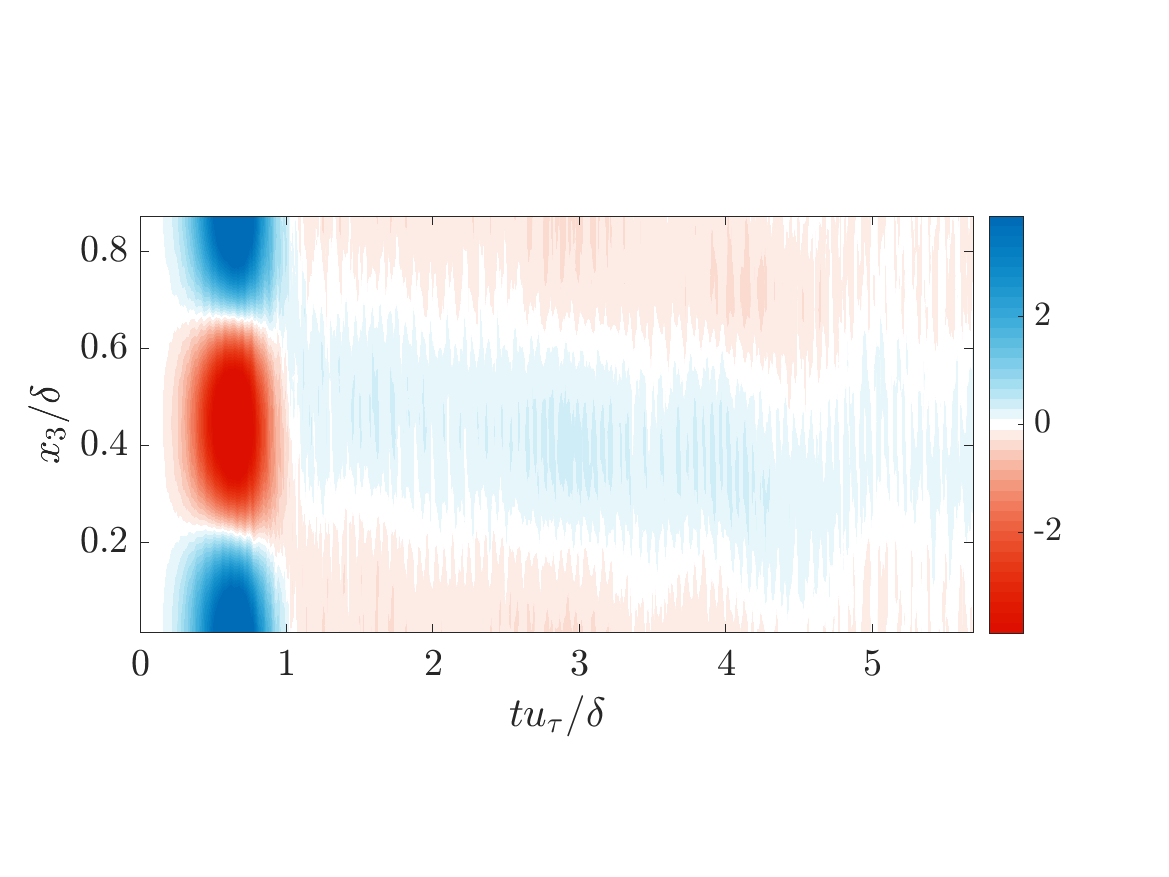}}
    \caption{Ensemble-averaged streamwise velocity deviation $\overline{\Delta u_1}$ for (a, c) $\gamma = 2\%$ and (b, d) $\gamma = 10 \%$. The results are for $x_1 = 0$ and two wall-normal heights: (a,b) $x_2^+ = 16$ and (c, d) $x_2^+ = 75$. Before ensemble averaging, the $(0, 1)$--mode for each ensemble member is multiplied by $\mathrm{e}^{-i\angle \kappa}$ to ensure their phase alignment.}
    \label{fig:split_breakup}
\end{figure}

\subsection{Interacting modes}
Figure \ref{fig:integrated_transfer} shows the total nonlinear transfer from the $(k_1, k_3) = (0, 1)$ mode integrated over the wall-normal direction. 
The integrated transfer is negative and confirms that nonlinear effects are indeed draining energy from the actuated scale. 
The trends in the plot echo the results in figure \ref{fig:transientStats}(a): the larger the forcing term, the larger the integrated transfer to unforced scales. The total energy transfer tends to peak during the decay of the streak, approximately $0.06 - 0.09 \; \delta / u_\tau$ later than the streak energy.
Considering the total energy transfer normalised by forcing magnitude $|\kappa|^2$, we see that the plots collapse for all values of $\gamma$, for $t u_\tau / \delta < 0.8$. The initial growth rate is the same across all forcing intensities and its magnitude scales with $|\gamma|^2$. The energy transfer mechanism at early times thus seems to be similar for all cases, regardless of forcing magnitude.
The larger the forcing, however, the earlier the energy transfer reaches its maximum and starts its decay. At later times, additional multiscale effects, more prominent for the strongly forced cases, likely take over and help drain energy more efficiently from the $(0, 1)$--mode. 

In figure \ref{fig:transfer_spectra}, we plot the streamwise-spanwise spectra for $\hat M^{(0, 1)}( s_1, s_3)$ at different times. Only wavenumbers corresponding to $ s_1, s_3 \in [-4, 4]$ are shown for clarity; the remaining wavenumbers interact negligibly with the $(0,1)$--mode. We observe that $\hat M^{(0, 1)} (s_1, s_3) \approx \hat M^{(0, 1)} (-s_1, s_3)$. Since the target mode extends the entire length of the channel, the streamwise phase of the interacting modes matters little once the results are averaged over the ensemble of simulations.  
The figure reveals that the nonlinear energy transfer is dominated by energy transfer to the $(0, 2)$ and $(1, 1)$--modes. This pattern is consistent across forcing intensities, though only the cases for $\gamma = 2\%$ and $\gamma = 10\%$ are shown. The qualitatively similar results across forcing magnitudes highlight the privileged role of the $(0, 2)$-- and $(1, 1)$--modes at exchanging energy with the actuated mode. Since the transfer to the $(0, 2)$--mode results from the self-interaction of the actuated mode, the importance of this particular scale is not surprising. %The transfer to the $(1, 1)$--mode can be considered as constituting a ``transverse cascade", which has been observed to play a role in other shear flows.
Figure \ref{fig:integrated_transfer_02_11} shows the $x_2$--integrated energy transferred to the $(1, 1)$ and $(0, 2)$ modes only. As in figure \ref{fig:integrated_transfer}(b), the initial growth rate of the magnitude-normalised quantities across forcing amplitudes, with the plot for $\gamma = 10\%$ peaking and starting its decay earlier than for $\gamma = 2\%$. 
Another notable observation is that, across forcing amplitudes, the energy transfer to the $(0, 2)$--mode occurs at a faster time-scale than the transfer to the $(1, 1)$--mode.

In figure \ref{fig:transfer_YT_plane}, we study the spatial distribution of the nonlinear energy transfer term $\hat M^{(0, 1)}$. We observe that, for $\gamma = 2\%$ and $\gamma = 10\%$, the additional nonlinear transfer to other scales due to the induced streak is centred on two wall-normal locations, $x_2^+ \approx 16$ located in the buffer layer, and  $x_2^+ \approx 75$ located in the outer region. This is consistent across all forcing magnitudes. 
The presence of two hubs of energy transfer echo the results in figure \ref{fig:time_transfer_to_small_scales}(d), which similarly shows two distinct regions each characterised by a time scale of energy transfer from the $(0, 1)$--mode to smaller scales.

In figure \ref{fig:transfer_YT_plane_11_02}, we plot the nonlinear energy transfer to the $(0, 2)$-- and $(1, 1)$--modes, for $\gamma = 2\%$ and $\gamma = 10\%$. The figure reveals that the transfer to the $(0, 2)$--mode accounts for the transfer at $x_2^+ \leq 25$, while the transfer to the $(1, 1)$--mode accounts for the transfer at $x_2^+ > 25$. In the buffer layer, the forced $(0, 1)$--mode thus tends to transfer its energy to a mode that is twice periodic in the spanwise direction; the induced streak splits into two branches, which can be seen in figure \ref{fig:split_breakup}(a, b). In the outer region, the nonlinear energy transfer favours the $(1, 1)$--mode, suggesting that the induced streak breaks up along the streamwise direction. This is seen in figures \ref{fig:split_breakup}(c, d), which show the streak meandering but not splitting into two branches as for $x_2^+ = 16$. 
We note that the transfer from the $(0, 1)$--mode to the $(0, 2)$--mode is due to the self-interaction of the actuated mode, and the term $\hat u^{(0, -1)}$ is squared in the expression for $\hat M ^{(0, 1)}(0, 2)$ (equation \ref{eq:M_xz}).
This along with the fact that the energy transfer to the $(0, 2)$--mode is predominant in the near-wall region explains the increased sensitivity of the energy transfer time scale $\Delta t_\mathrm{trans}$ to $\gamma$ for $x_2^+ \leq 25$ (figure \ref{fig:time_transfer_to_small_scales}(d)). 
More specifically, a higher of $\gamma$ disproportionately intensifies the transfer of energy to the $(0, 2)$--mode, causing the nonlinear transfer of energy near the wall to occur earlier -- closer to the focus of $\hat M^{(0, 1)}(0, 2)$ -- and decreasing $\Delta t_\mathrm{trans}$.

\subsection{Mechanism of nonlinear energy transfer}
\begin{figure}
    \centering
    \subfloat[]{\includegraphics[width=0.47\linewidth]{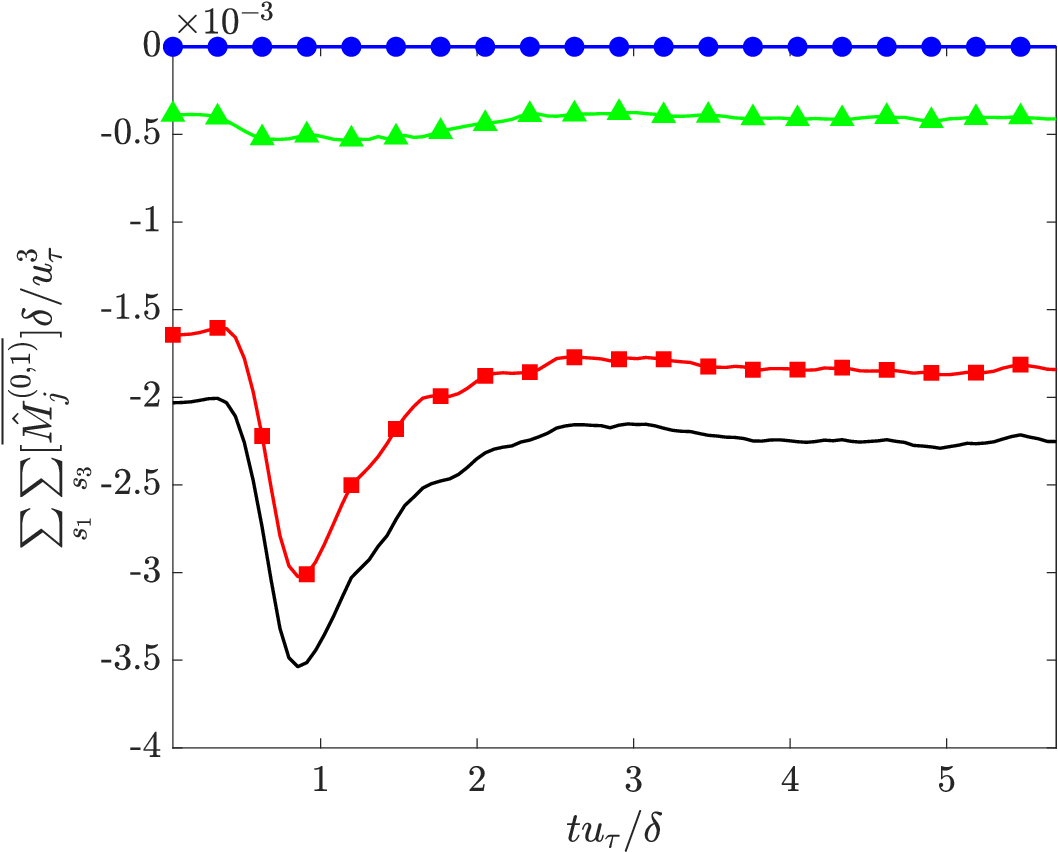}}
    \hspace{1mm}
    \subfloat[]{\includegraphics[width=0.47\linewidth]{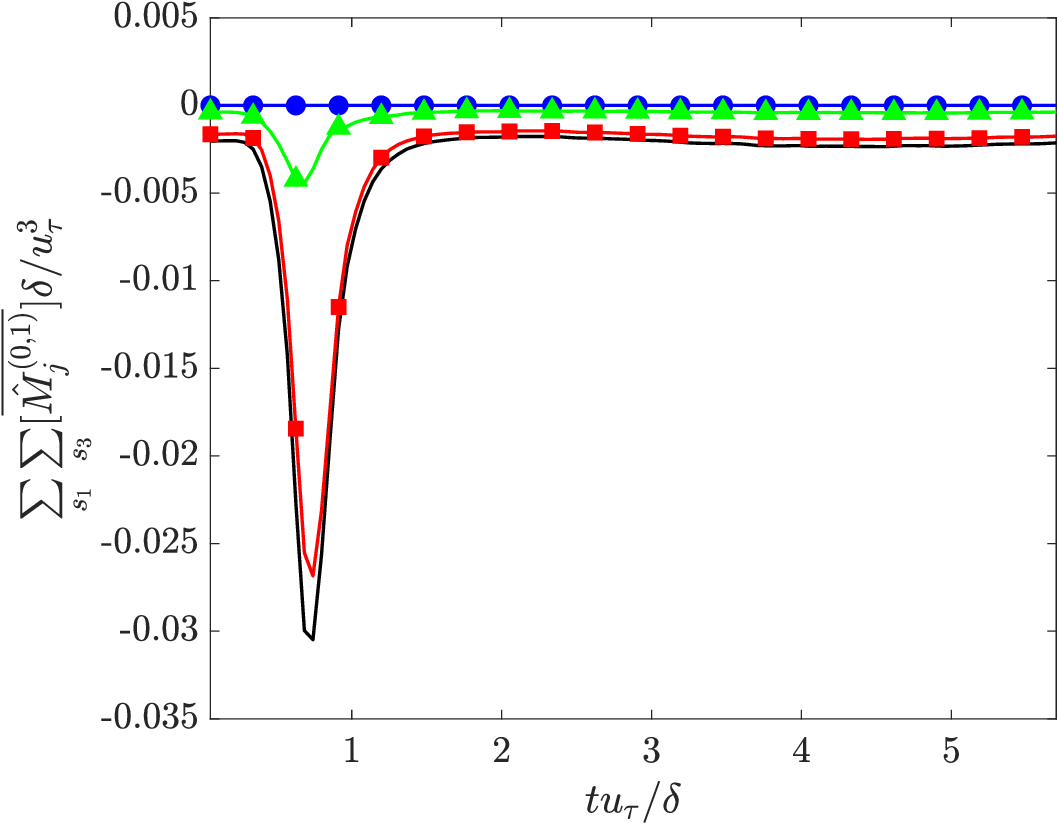}}
    
    \subfloat[]{\includegraphics[width=0.47\linewidth]{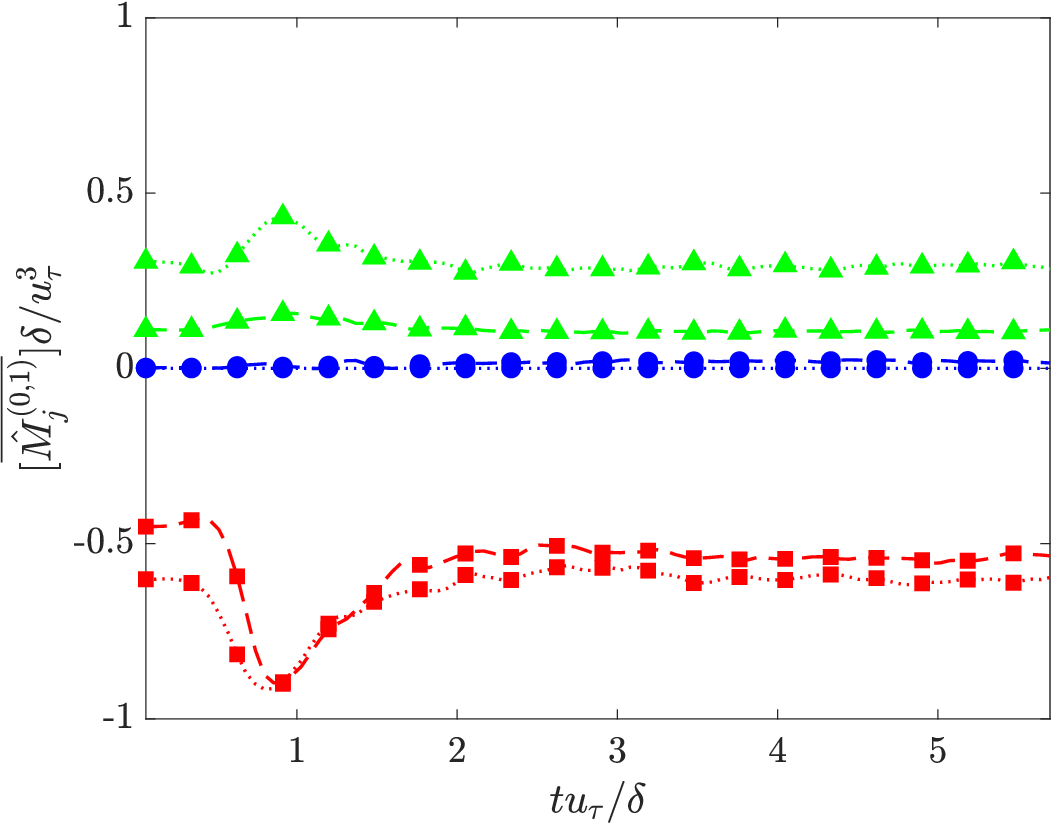}}
    \hspace{1mm}
    \subfloat[]{\includegraphics[width=0.47\linewidth]{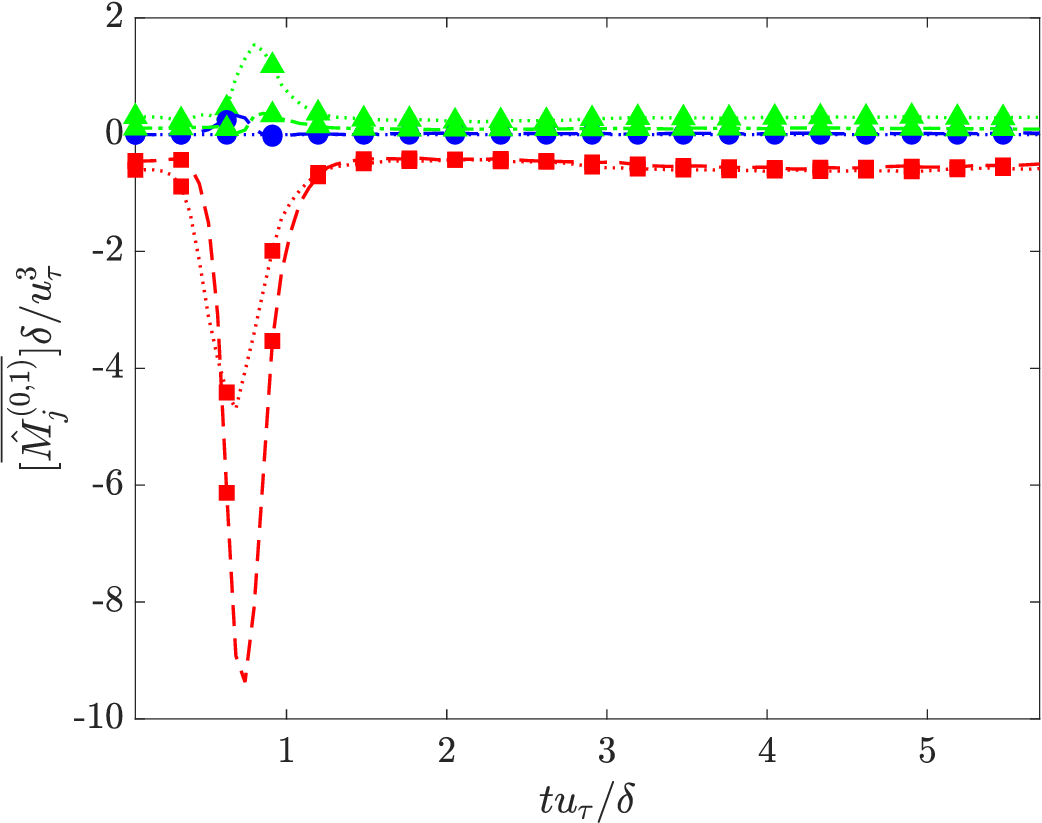}}
    \caption{Nonlinear energy transfer from the $(0,1)$--mode to unforced modes, broken down by streamwise (blue $-\bullet$), wall-normal (\reviewertwo{green} $-\blacktriangle$) and spanwise (\reviewertwo{red} $-\blacksquare$) contributions. The cases plotted are (a, c) $\gamma = 2\%$ and (b, d) $\gamma = 10\%$. Plots (a) and (b) represent the contributions to the total nonlinear energy transport from the $(0,1)$--mode, with the black line representing the sum of the contributions; plots (c, d) represent the transfer to the $(0, 2)$-- and $(1, 1)$--modes, denoted by $(\cdots)$ and $(--)$, respectively.}
    \label{fig:integrated_transfer_by_component}
\end{figure}

\begin{figure}
    \centering
    \subfloat[]{\includegraphics[width=0.5\linewidth]{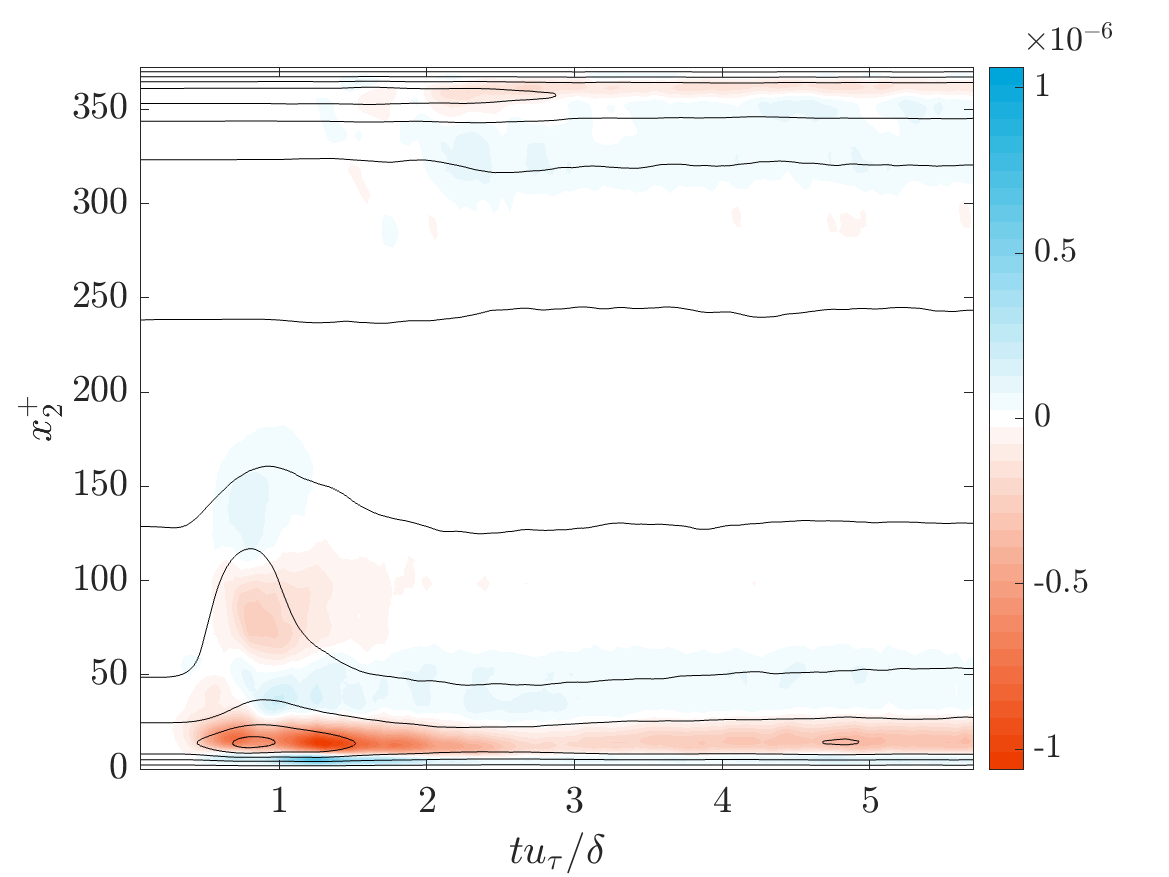}}
    \subfloat[]{\includegraphics[width=0.5\linewidth]{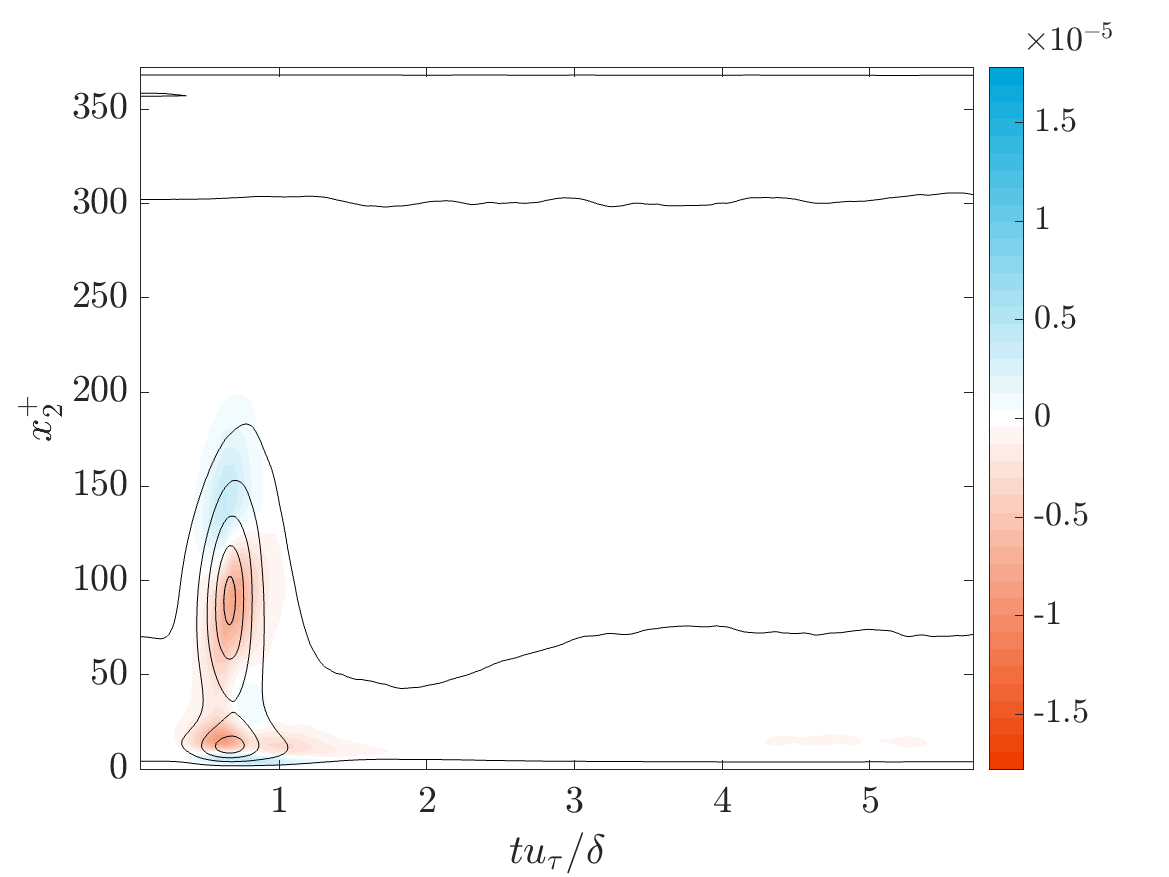}}

    \subfloat[]{\includegraphics[width=0.5\linewidth]{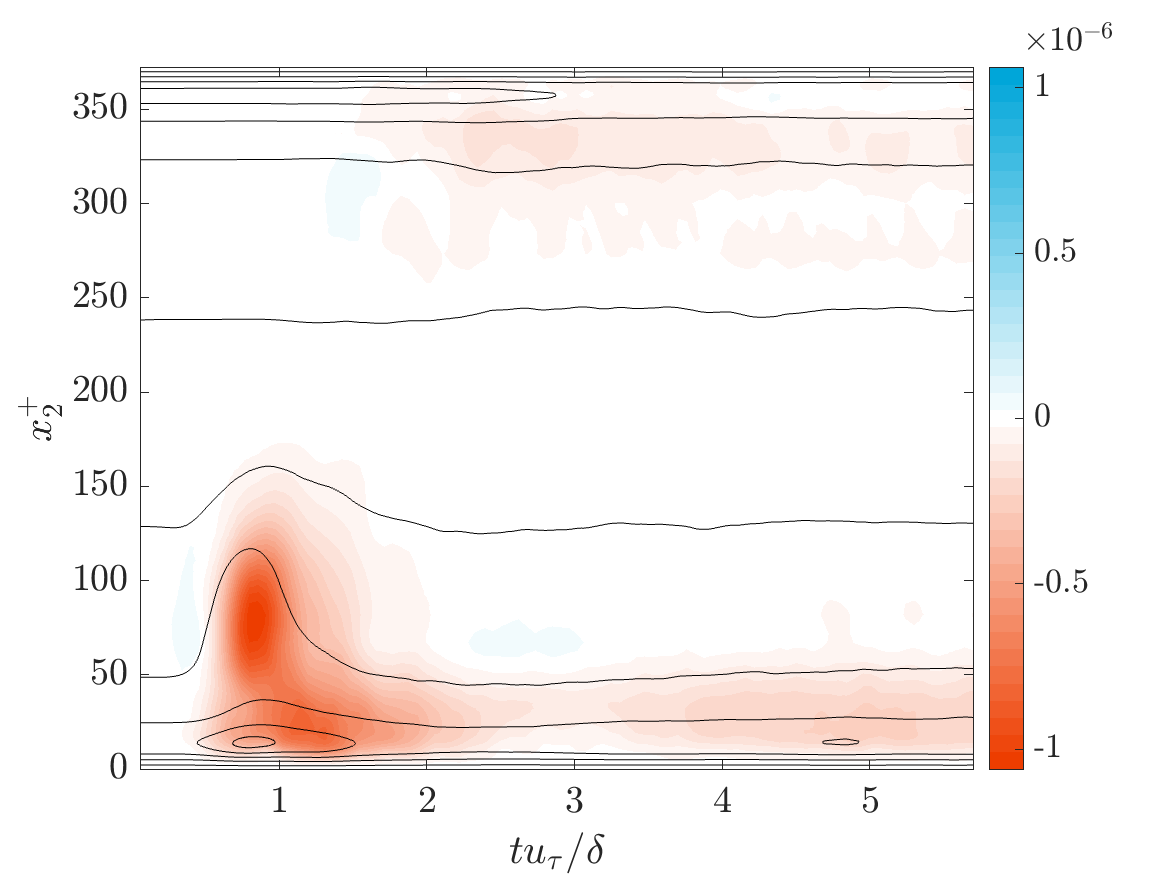}}
    \subfloat[]{\includegraphics[width=0.5\linewidth]{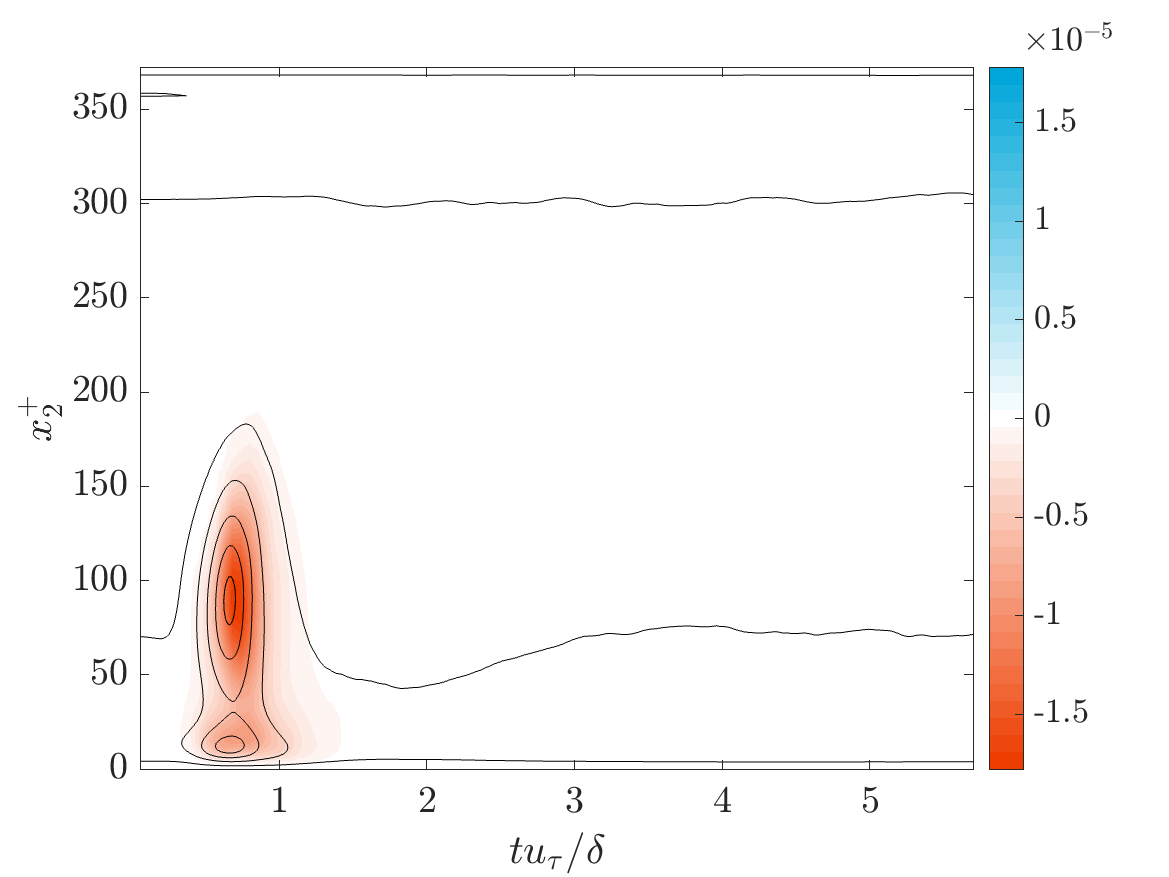}}
    \caption{Nonlinear energy transfer from the $(0,1)$--mode split by NLT$_2$ and NLT$_3$ contributions:
    (a, b) $\overline {\Delta \hat M^{(0,1)}_2}  \delta / u_\tau ^3 $ and (c, d) $\overline {\Delta \hat M^{(0,1)}_3}  \delta / u_\tau ^3 $. The cases shown are for (a,c) $\gamma = 2\%$ and (b,d) $\gamma = 10 \%$. The black lines are contours of $\overline{|\hat u_1^{(0, 1)}|^2/2}$ and represent $10\%, \; 25 \%, 50\%, 75\%$ and $90\%$ of the maximum value across $x_2$ and $t$.}
    \label{fig:transfer_YT_plane_componentwise}
\end{figure}
To shed light on the mechanism of this nonlinear energy transfer, we split the nonlinear transfer $\hat M^{(0, 1)}(s_1, s_3)$ into the terms that, respectively, reflect the contributions of the streamwise, spanwise and wall-normal nonlinear advection. Specifically, in order to study the nonlinear energy transfer from a mode $(k_1, k_3)$ due to streamwise nonlinear advection -- henceforth referred to as NLT$_1$ ("NonLinear Transfer, $x_1$"), we define $\hat M_{1}^{(k_1, k_3)}(s_1, s_3)$ as:

\begin{equation}\label{eq:M_xz_oneComponent}
    \hat M_{1} ^{(k_1, k_3)}(s_1, s_3) :=  - 2 \Real \left \{ \hat u_i^{(-k_1, -k_3)} \widehat {\frac{\partial u_i}{\partial x_1}}^{(s_1, s_3)} \hat u_1 ^{(k_1 - s_1, k_3 - s_3)} \right \}.
\end{equation}
We similarly define $\hat M_{2}$ and $\hat M_{3}$ to study NLT$_2$ and NLT$_3$, the nonlinear energy transfer due to wall-normal and spanwise gradients self-advection, respectively. 

Figure \ref{fig:integrated_transfer_by_component} shows that the energy transfer due to the coupling with the spanwise velocity component dominates the total energy transfer to other scales, even before the growth of the injected mode and after the mode completely decays for $t u_\tau / \delta > 2$. During the growth of the injected mode, the NLT$_3$ term grows significantly more than the transfer due to self-advection in the streamwise and wall-normal directions. This demonstrates the unique role of NLT$_3$ in transferring energy to secondary scales and restoring the system to its unforced state. 
This behaviour is consistent across forcing amplitudes, though only $\gamma = 2\%$ and $\gamma = 10\%$ are shown. The dominance of the NLT$_3$ contributions also holds for the $(0, 2)$ and $(1, 1)$--modes individually (figures \ref{fig:integrated_transfer_by_component}(c, d)). We do note that the NLT$_2$ is relatively important for the $(0, 2)$--mode. The induced streak thus tends to shed its energy to secondary modes via spanwise self-advection. We note that these results match those found in \citet{markeviciute2024threshold}, which computes a streak structure via optimal transient growth analysis then studies the growth of secondary instabilities added to the streak. Authors find that suppressing the pushover mechanism, \emph{i.e.} the spanwise advection of the secondary perturbation due to the streak, prevents the growth of these secondary instabilities. Thus, both this work and \citet{markeviciute2024threshold} agree that coupling via spanwise gradients dominate the interaction between the streak and secondary modes.

Additionally, by considering the wall-normal variations (figure \ref{fig:transfer_YT_plane_componentwise}), we see that spanwise self-advection is the primary pathway across all wall-normal heights by which the forced $(0, 1)$--mode sheds its energy. This is consistent across all forcing magnitudes, though only the cases for $\gamma=2\%$ and $\gamma = 10\%$ are shown.  
For the lightly forced case $\gamma = 2\%$, we do note a significant NLT$_2$ contribution at $x_2^+ \approx 18$, (figure \ref{fig:transfer_YT_plane_componentwise}(a)), but as $\gamma$ increases, the NLT$_2$ contribution is quickly overtaken by the NLT$_3$. 
\reviewerone{The NLT$_3$ term thus has a stronger dependence on $\gamma$ than the NLT$_2$ term, especially close to the wall; this suggests its key role in the nonlinear interactions which grow rapidly with $\gamma$.}
%We note that $x_2^+ \approx 18$ corresponds to the section of the wall-normal grid dominated by the transfer to the $(0, 2)$--mode, \emph{i.e.} the triadic interaction most sensitive to  $\gamma$ due to the self-interaction of the actuated mode.
The growth of secondary instabilities due to push-over occurs locally at both foci of nonlinear energy transfer, and both the $(1, 1)$-- and $(0, 2)$--modes benefit from this energy transfer mechanism.

\subsection{Quasi-linear approximation} \label{quasilinear}
\begin{figure}
    \centering
    \subfloat[]{\includegraphics[width=0.5\linewidth]{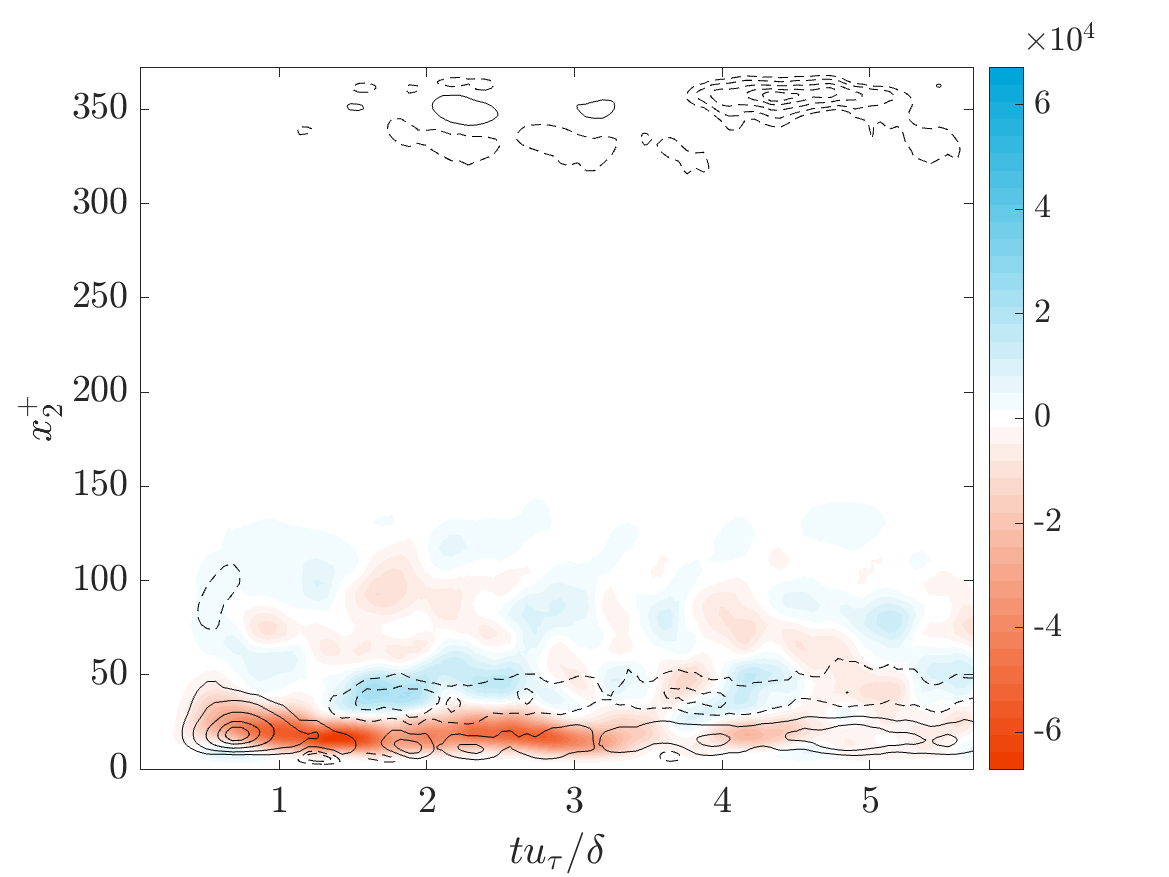}}
    \subfloat[]{\includegraphics[width=0.5\linewidth]{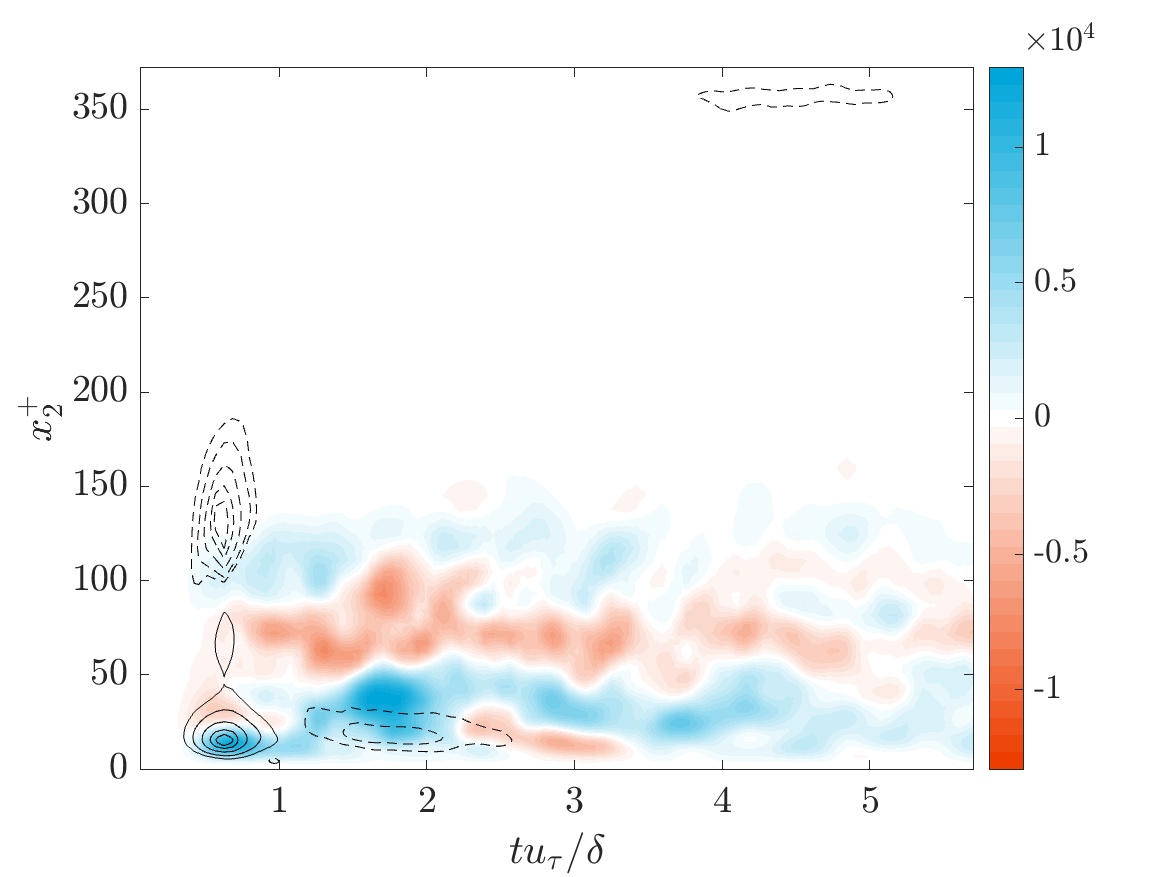}}
    
    \subfloat[]{\includegraphics[width=0.5\linewidth]{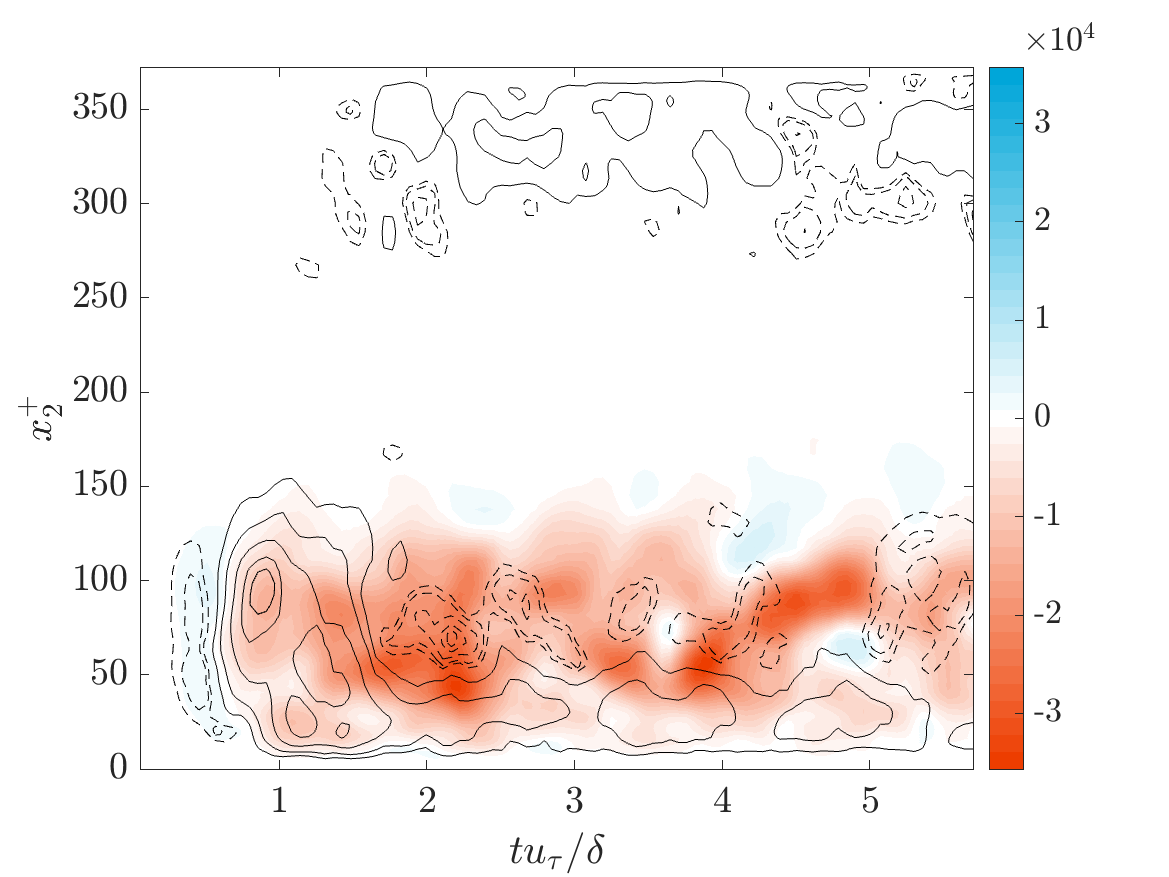}}
    \subfloat[]{\includegraphics[width=0.5\linewidth]{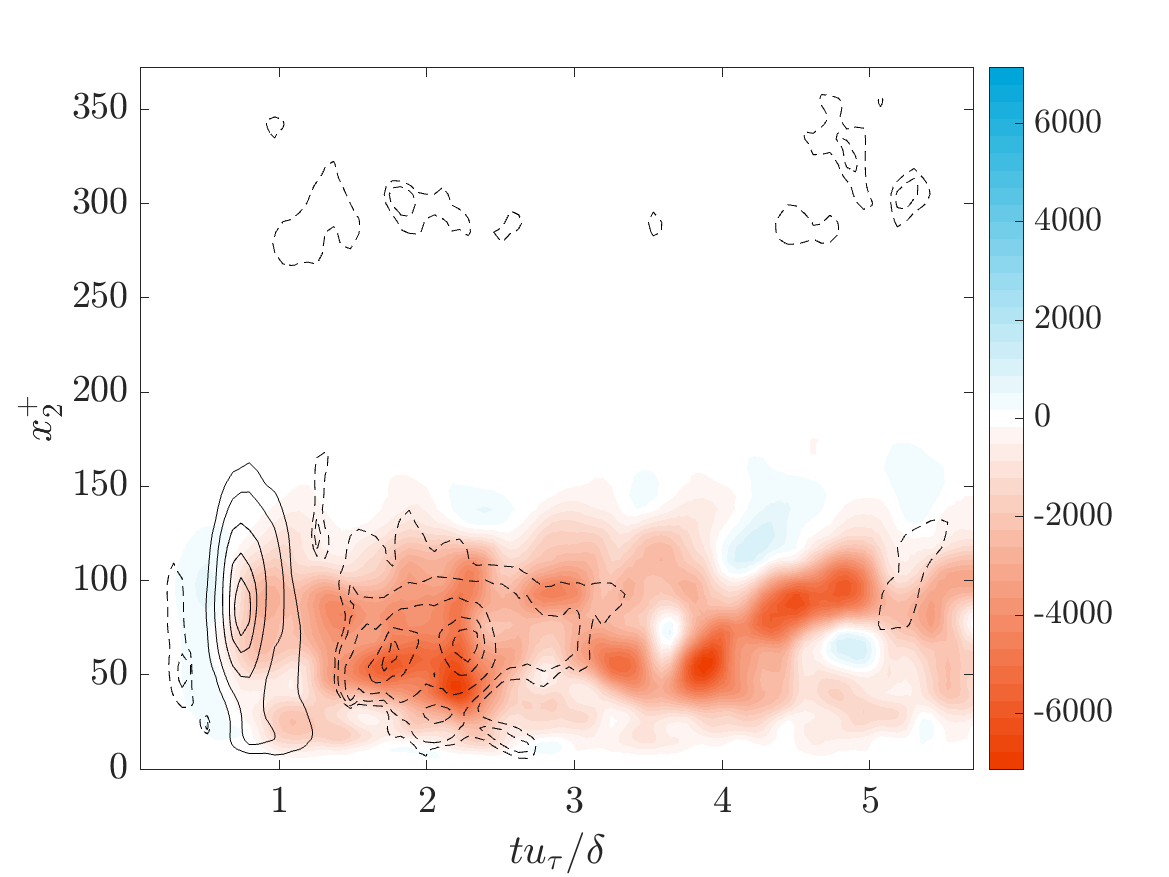}}
    \caption{ Quasi-linear approximation $\widetilde M^{(0,1)}$ for the nonlinear energy transfer from the $(0,1)$--mode, normalised by the forcing magnitude $|\kappa|^2$.  Figures (a) and (b) correspond to the transfer to the $(0,2)$--mode, and figures (c) and (d) correspond to the transfer to the  $(1,1)$--mode. The cases plotted are for (a, c) $\gamma = 2\%$ and (b, d) $\gamma = 10\%$. The black lines represent contours of $\hat { M}^{(0,1)}$ and correspond to $10\%$, $25\%$, $50\%$, $75\%$ and $90\%$ of the minimum (most negative) value (--), or $10\%$, $25\%$, $50\%$, $75\%$ and $90\%$ of the maximum (most positive) value ($--$).}  \label{fig:UpPsi_YT_plane}
\end{figure}

\begin{figure}
    \centering
    \subfloat[]{\includegraphics[width=0.5\linewidth]{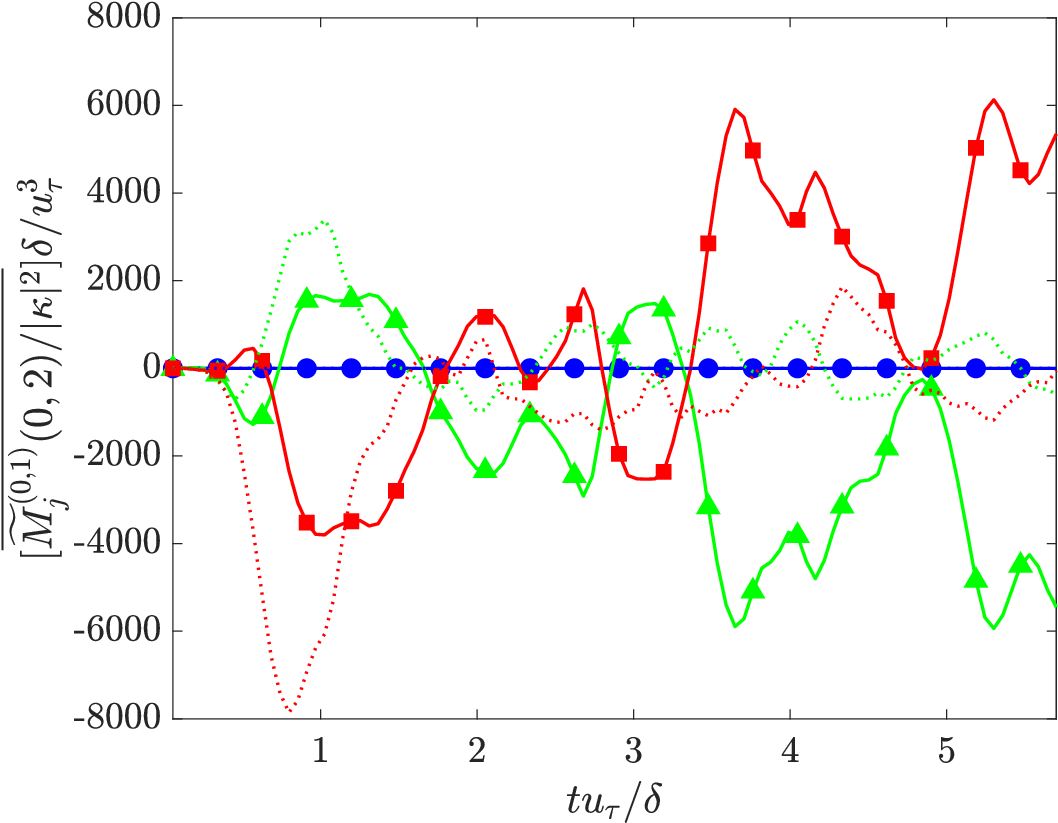}}
    \subfloat[]{\includegraphics[width=0.5\linewidth]{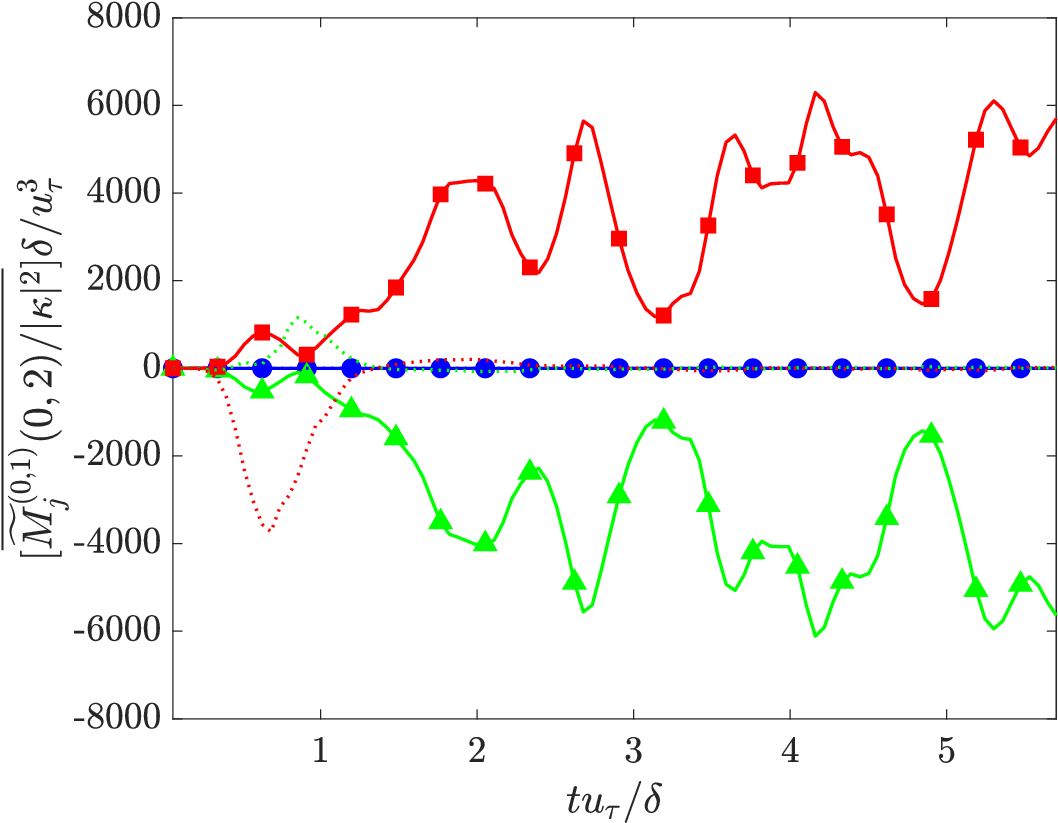}}
    
    \subfloat[]{\includegraphics[width=0.5\linewidth]{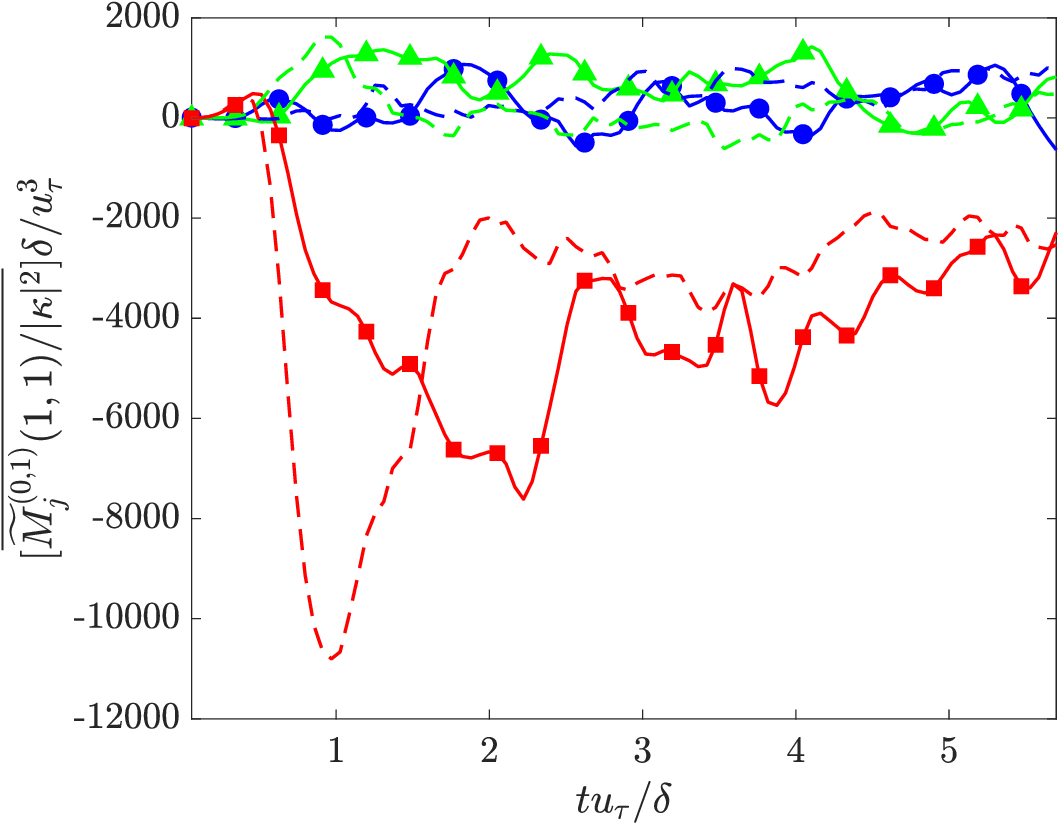}}
    \subfloat[]{\includegraphics[width=0.5\linewidth]{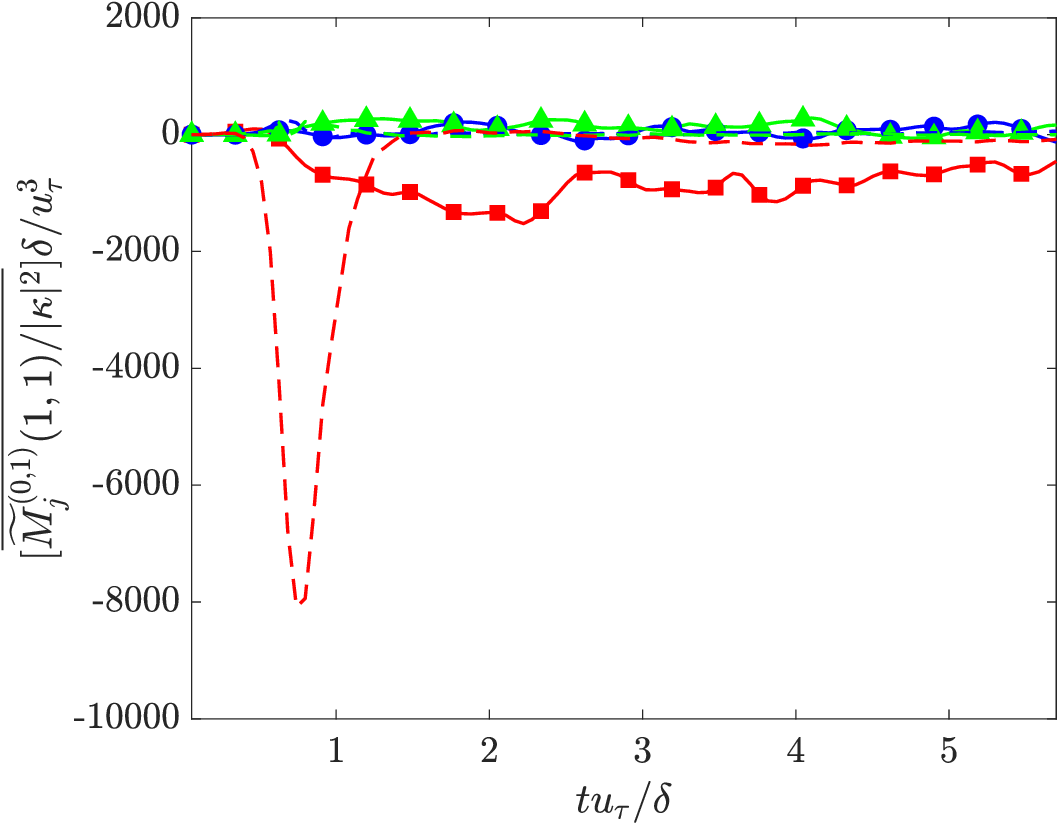}}
    \caption{  Nonlinear energy transport from the $(0,1)$--mode to the (a, b) $(0,2)$--mode and the (c, d) $(1,1)$--mode, broken down by streamwise (blue $-\bullet$), wall-normal (green $-\blacktriangle$) and spanwise (red $-\blacksquare$) contributions, and normalised by forcing magnitude. The solid lines correspond to the nonlinear energy transfer in the quasi-linear model $[\overline{\widetilde{ M}^{(0,1)} / |\kappa|^2}]$, while the dashed or dotted lines correspond to $[\overline{\hat{ M}^{(0,1)} / |\kappa|^2}]$, the results from DNS (also shown in figure \ref{fig:integrated_transfer_by_component}). The cases plotted are (a, c) $\gamma = 2\%$ and (b, d) $\gamma = 10\%$.}  \label{fig:breakdown_by_component_uppsi}
\end{figure} 
%

% rewrite this
In this section, we seek to determine whether the instantaneous behaviours of the nonlinear energy transfer can be simply modelled by the interactions of the principal response mode with the ``background" turbulence -- that is, the turbulence frozen in its state at $t = 0$.
By neglecting subsequent nonlinear feedback, the velocity field after the injection of the mode can be represented as $\boldsymbol{u}_\psi := \boldsymbol{u_0} + \kappa \sigma_1 \boldsymbol{\psi_1} \mathrm{exp}(\mathrm i 2 \pi / L_3) + \kappa^* \sigma_1 \boldsymbol{\psi_1}^* \mathrm{exp}(-\mathrm i 2 \pi / L_3) $. Using this field, we define the following nonlinear energy transfer term for the quasi-linear field analogously to equation \eqref{eq:M_xz}:

\begin{equation}
\begin{split}
    \widetilde { M} ^{(k_1, k_3)}(s_1, s_3) &:=  - 2 \Real \left \{ \hat u_{\psi, i}^{(-k_1, -k_3)} \widehat {\frac{\partial u_{\psi, i}^{(s_1, s_3)}}{\partial x_j}} \hat u_{\psi, j} ^{(k_1 - s_1, k_3 - s_3)} \right \}
    \end{split}
    \label{eq:Mpsi_eq}
\end{equation}
where $u_{\psi, i}$ is the $i^\mathrm{th}$ component of $\boldsymbol u_\psi$. 
This term captures the nonlinear energy transfer from the $(k_1, k_3)$--mode to the $(s_1, s_3)$--mode due to interactions of the injected mode with the background turbulence and ignoring the self-interactions of $\boldsymbol u_\psi $ at every time-step. We can thus view the term $\widetilde M^{(k_1 , k_3)} $ as a quasi-linear model of the nonlinear energy transfer. We define the quantities representing the NLT$_1$, NLT$_2$ and NLT$_3$ contributions for the $\boldsymbol u_\psi$ field as:
\begin{align}
\label{eq:Mpsi_xz_oneComponent_complex}
    \widetilde { M}_{1} ^{(k_1, k_3)}(s_1, s_3) :=  - 2 \Real \left \{\hat u_{\psi, i}^{(-k_1, -k_3)} \widehat {\frac{\partial u_{\psi, i}}{\partial x_1}}^{(s_1, s_3)} \hat u_{\psi, 1} ^{(k_1 - s_1, k_3 - s_3)} \right \}.
\end{align}
We respectively dub these terms QLT$_1$, QLT$_2$ and QLT$_3$ (``QuasiLinear Transfer, $x_i$") to distinguish them from NLT$_1$, NLT$_2$ and NLT$_3$.
Figure \ref{fig:UpPsi_YT_plane} shows the quasi-linear estimates for the nonlinear energy transfer to the $(0,2)$-- and $(1, 1)$--modes, normalised by forcing amplitude. 
\reviewertwo{For the lightly forced case of $\gamma =2\%$, the quasi-linear model predicts the regions of energy loss to the $(0, 2)$-- and $(1, 1)$-- well}, especially for $t < 1 \delta/u_\tau$ (figures \ref{fig:UpPsi_YT_plane}(a, c)). For $\gamma =10\%$, the quasi-linear interactions \reviewertwo{fail to predict the region of energy transfer to the $(0, 2)$--mode, but perform better for the $(1, 1)$--mode (figures \ref{fig:UpPsi_YT_plane}(b, d)).} 

\reviewertwo{In an attempt to test whether the quasi-linear model can predict the general NLT$_i$ trends observed in the previous section----mainly the dominance of the NLT$_3$ contributions---we plot the QLT$_1$, QLT$_2$ and QLT$_3$ quantities in figure \ref{fig:breakdown_by_component_uppsi}}. 
We see that the quasi-linear estimates for the transfer to mode $(0,2)$ can roughly predict a strongly negative QLT$_3$ component for the lightly forced system at early times, but fail to do so for later times. For large forcing amplitudes, the quasi-linear mode performs poorly for all times, and even reverses the observed trends for the QLT$_2$ and QLT$_3$ components of the nonlinear transfer. 
\reviewers{The quasi-linear model indirectly assumes that the growth of the secondary modes interacting with the injected streak is slow enough so that the dominant nonlinear interactions remain due to the "background" unperturbed field. The failure of the quasi-linear model to explain the trends of the $(0, 2)$--mode suggests that the assumption of slow growth cannot be applied to this mode.
Thus, in the near-wall region where streak splitting occurs, we suspect that transient growth mechanisms are key to explaining the behaviour mode $(0, 2)$. One such mechanism is linearly driven transient algebraic growth, which is responsible for amplifying the injected resolvent mode at early times.}
In contrast, the quasi-linear estimates for the transfer to mode $(1,1)$ qualitatively match the behaviour of the DNS and reveal a strongly negative QLT$_3$ contribution and a weakly positive QLT$_2$ contribution. Outside the buffer layer, the instantaneous nonlinear interactions between the target $(0,1)$--mode and the background turbulence can qualitatively explain the streak break-up. For better quantitative accuracy, we suspect that transient growth effects are also important in the outer region.

\begin{figure}
    \centering
    \subfloat[]{\includegraphics[width=0.5\linewidth]{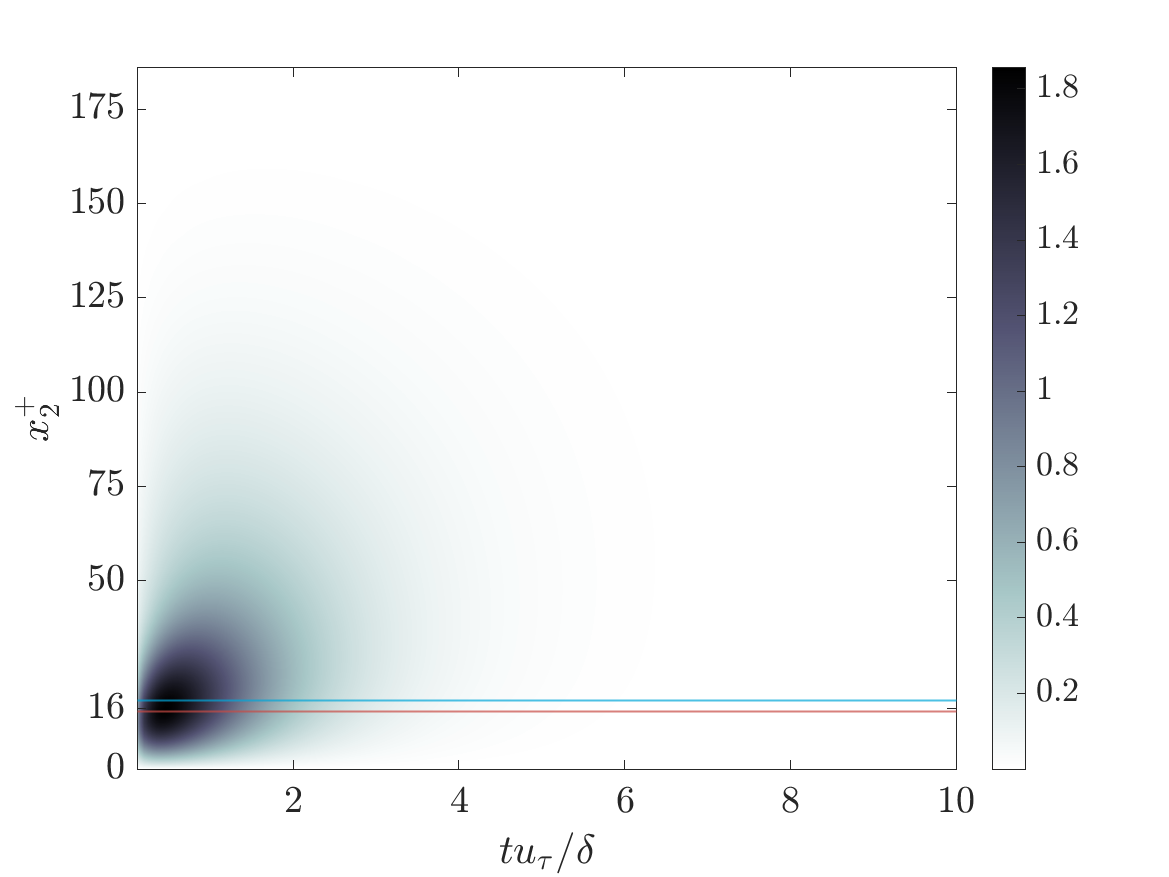}}
    \subfloat[]{\includegraphics[width=0.5\linewidth]{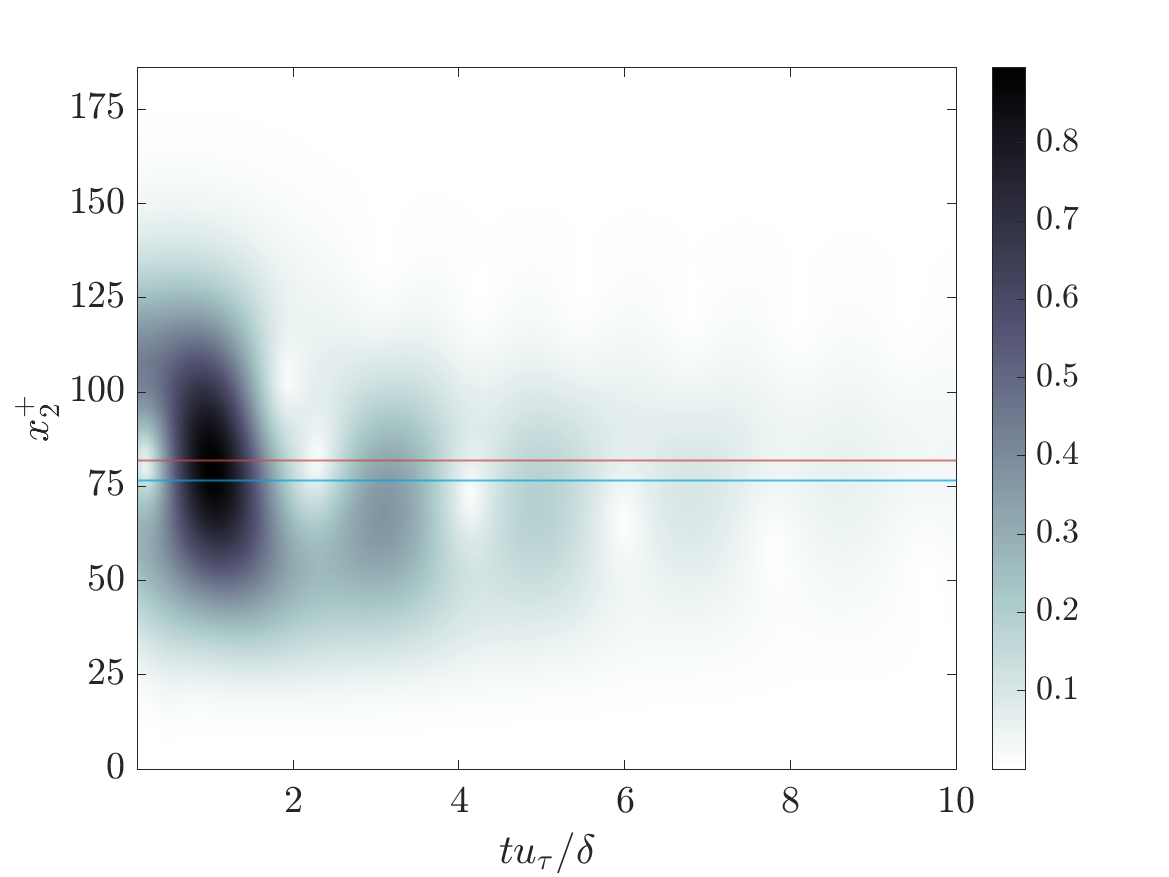}}

    \caption{Magnitude of the streamwise component of the principal resolvent response modes for (a) $(k_1, k_3) = (0, 2)$ and (b) $(k_1, k_3) = (1, 1)$. The modes are normalised so that their action pseudo-norm is 1. The horizontal lines represent the peak location of (a)  $\hat M^{(0, 1)}(0, 2)$ and (b)  $\hat M^{(0, 1)}(1, 1)$ for $\gamma = 2\%$ (blue) and $\gamma = 10\%$ (red).}  \label{fig:resolvent_modes_secondary}
\end{figure} 

In addition to studying the interaction of the transiently growing $(0, 1)$--mode with the background turbulence,
we consider the linear optimal growth of the $(0, 2)$-- and $(1, 1)$--modes. 
We repeat the wavelet-based resolvent analysis described in \S\ref{sec:resolvent} but with spatial parameters $(k_1, k_3) = (0, 2)$ and $(k_1, k_3) = (1, 1)$. We use the same spatial grid and choose $T = 44 \delta / u_\tau$ to allow the modes ample time to decay to zero, which occurs at $t u_\tau/\delta \approx 20$. The size of the temporal grid is chosen to be $N_t = 880$. 
Despite the fact that resolvent analysis captures a purely linear process, the $x_2$--location preferred by the principal resolvent response for $(k_1, k_3) = (0, 2)$ and $(k_1, k_3) = (1, 1)$ coincide with the foci of nonlinear interaction between the $(0, 1)$--mode, and the $(0, 2)$-- and $(1, 1)$--modes, respectively (figure \ref{fig:resolvent_modes_secondary}). 
The resolvent modes, which capture the maximal linear growth of the two scales considered, accurately predicts the locations of energy exchange with the actuated mode, and thus, transient growth via linear mechanisms can dictate the spatial structure of the nonlinear energy cascade.
Moreover, we observe that the principal resolvent response mode for $(k_1, k_3) = (0, 2)$ grows faster than the mode for $(k_1, k_3) = (1, 1)$, peaking earlier, which may explain the earlier peaks of $\hat M^{(0,1)}(0, 2)$ compared to $\hat M^{(0,1)}(1, 1)$ (figure \ref{fig:integrated_transfer_02_11}).

The results in figure \ref{fig:resolvent_modes_secondary} support works like \citet{huang2023spatio}, which studies the efficacy of individual dyadic interactions at exciting an energetic scale in the channel. The influence of individual dyadic contributions to the nonlinear term of the Navier-Stokes equations is measured via the projection of the contributions onto the principal resolvent forcing mode for the scale of interest, and dyadic interactions highly aligned with the forcing mode are deemed important contributors to turbulence. The underpinning assumption is that the response mode corresponding to the forcing mode indeed capture the behaviour of turbulence in the channel, which figure \ref{fig:resolvent_modes_secondary} suggests is true, even for secondary scales like the $(0, 2)$-- and $(1, 1)$--modes.
Figure \ref{fig:resolvent_modes_secondary} also helps explain the success of restricted nonlinear (RNL) models at replicating turbulent statistics \citep{thomas2014self, farrell2017statistical, gayme2019coherent}. 
In RNL models, the mean profile is governed by a modified version of the fully nonlinear Navier-Stokes equations where only a reduced subset of fluctuation scales contribute to the nonlinear advection terms, while the fluctuations obey the linearised Navier-Stokes equations about the mean profile. The location of the modes in figure \ref{fig:resolvent_modes_secondary} indeed suggest that linear mechanisms are enough to at least predict the correct spatial distribution of the nonlinear energy transfer to the two most important secondary modes involved in the energy cascade in this experiment.

%%%%%%%%%%%%%%%%%%%%%%%%%%%%%%%%%%%%%%%%%
\section{Conclusions}\label{conclusion}

In this work, we study the growth of time-localised resolvent modes in the minimal flow unit at $Re_\tau \approx 186$. We formulated resolvent analysis in a wavelet-basis in time that endows the resolvent modes with transient information, and obtained a linearly optimal time-localised forcing mode and its corresponding transient response mode.

Resolvent analysis ignores the feedback between the velocity fluctuations and the nonlinear terms; we thus tested the optimality of the resolvent forcing within a DNS of a minimal flow unit by numerically injecting the principal resolvent forcing mode into the flow at varying amplitudes. This allowed us to investigate the interactions between the transiently growing linear response mode and the nonlinear effects of the turbulent flow.
We compared the resulting flow to one forced by the first suboptimal forcing mode and another forced by a spatially-random forcing. The principal resolvent forcing produces a larger transient energy growth than the suboptimal mode, but the energy amplification is notably lower in both cases compared to the linearised case. Both systems were significantly more amplified than the random forcing case. 
In all cases, despite the fact that the injected forcing term is small compared to the initial nonlinearities, the amplification of the velocity perturbation due to linear mechanisms is significant enough that the simulated fields tracks the optimal linear response for a short time.

The nonlinearities of turbulence interrupt the initial algebraic energy growth driven by the linear dynamics of the flow. This is seen in all cases forced by the principal resolvent mode, and the amplitude of the resolvent forcing affects how closely the turbulent trajectory behaves like the optimal resolvent response mode. 
Across all forcing amplitudes, the initial growth phase is similar, and the systems peak at roughly the same time, but, the higher the forcing amplitude, the faster the decay of the system back to the unforced turbulent system. 
The more intense turbulence in the high-amplitude-forcing cases \reviewerthree{is} more effective at damping the effects of the initial forcing. 
\reviewerthree{We observe that the induced velocity perturbation field for the actuated Fourier mode matches the resolvent mode better, both in magnitude and spatiotemporal structure, within the near-wall region, where viscous effects are more prominent.}
%We observe that the forced DNS flow fields are closer to the resolvent response mode in the near-wall region, where viscous effects are more prominent. 
During the decay of the streak decays, the spectral energy content of the simulations becomes increasingly multi-scale due to a transfer of energy to the non-forced spatial scales. This cross-scale energy transfer is more prominent and occurs faster for the high-amplitude-forcing cases.

Nonlinear effects lead to streak breakdown, and by considering the nonlinear energy transfer from the induced streak to the non-actuated modes, two secondary modes are found to play a dominant role in draining the streak of its energy and curtailing its growth. The first mode, constant in the streamwise direction and twice periodic in the spanwise direction, corresponds to a splitting of the streak into two branches and dominates the energy transfer in the near-wall region, including the buffer layer. The second, once periodic in the streamwise and spanwise directions, corresponds to a streamwise break-up of the streak and dominates the energy transfer in the outer region. The branching mode receives energy from the actuated mode via the self-interaction of the injected streak, which explains the particular sensitivity of the near-wall region to forcing amplitude. Though a larger forcing amplitude accelerates the energy transfer from the actuated mode to smaller scales across all wall-normal heights, this effect is indeed more prominent in near the wall than in the outer region.

The streak interacts with these secondary modes mostly through an NLT$_3$ type energy transfer, \emph{i.e.} via the spanwise nonlinear self-advection term. We model the predicted nonlinear energy transfer by computing the interaction of the injected streak with the background turbulence of the initial conditions used, and find that these one-way interactions generally predict the trends for the streak breakup that dominates in the outer region. This crude model cannot, however, predict the dominance of the NLT$_3$ contribution that occurs during the streak splitting near the wall. We postulate that in the near-wall region and the buffer layer, transient growth phenomena are necessary to explain the behaviour of the $(0, 2)$--mode.
We compute the transiently growing principal resolvent modes 
for the two preferred secondary scales, and find that they are located exactly at the foci of nonlinear energy exchange with the actuated scale. The structures found using resolvent analysis predict the $x_2$--distribution of the nonlinear energy transfer to the $(0, 2)$-- and $(1, 1)$--modes in the nonlinear DNS, further proving that resolvent modes can be informative of turbulent flows despite using the linearised equations of motion. 

Testing the effectiveness of the principal resolvent forcing mode reveals valuable insights on the stability of the streak, the mechanism by which it sheds energy, and the sensitivity of this nonlinear energy transfer mechanism to forcing amplitude. 
To find structures more effective than the resolvent forcing mode at actuating the minimal flow unit, one could use nonlinear optimisation techniques \citep{kerswell2018nonlinear, leonid2023optimization}. These broadly aim to maximise the growth of kinetic energy within a user-defined time window, and enforce the satisfaction of the (nonlinear) Navier-Stokes equations as an optimisation constraint. 
Nevertheless, an important advantage of wavelet-based resolvent analysis is its computational efficiency and tractability.

%\begin{comment}
\section{Acknowledgments}
This work is supported in part by the European Research Council under the Caust grant ERC-AdG-101018287 and AFOSR grant FA9550-22-1-0109.

\section*{Declaration of Interests}
The authors report no conflict of interest.
%\end{comment}

\bibliographystyle{jfm}
\bibliography{references}
%\bibliography{bibliography}
\end{document}